\documentclass[singlecolumn,10pt]{article}

\usepackage{arxiv}

\usepackage[utf8]{inputenc} 
\usepackage[T1]{fontenc}    

\usepackage{url}            

\usepackage{amsfonts}       
\usepackage{nicefrac}       
\usepackage{microtype}      
\usepackage{graphicx}
\usepackage[authoryear,sort]{natbib}
\setcitestyle{authoryear,comma}  

\usepackage{amssymb}
\usepackage{amsmath}
\usepackage{amsthm}
\usepackage{tabularx}
\usepackage{booktabs} 
\usepackage{float}
\usepackage{makecell}
\usepackage{rotating}
\usepackage{xcolor} 
\usepackage[normalem]{ulem}
\usepackage{siunitx}
\usepackage{mathrsfs}  
\usepackage{mathtools}
\usepackage{enumitem}
\usepackage{romannum}
\usepackage[linktoc=page,
colorlinks=true,	
linkcolor=blue,
citecolor=blue,
urlcolor=blue]{hyperref}       

\usepackage{doi}

\usepackage{newtxtext}
\usepackage{newtxmath}

\theoremstyle{definition}
\newtheorem{lemma}{Lemma}
\newtheorem{proposition}[lemma]{Proposition}
\newtheorem{definition}{Definition}
\newtheorem{remark}{Remark}

\newcommand{\Set}[1]{\mathcal{#1}} 
\newcommand{\Setnum}[1]{\mathbb{#1}} 
\newcommand{\Group}[1]{\mathcal{#1}} 
\newcommand{\Ring}[1]{\mathcal{#1}} 

\newcommand{\PolRing}[3]{\mathbb{#1}\left[#2\right]^{\Group{#3}}}
\newcommand{\PolRingE}[3]{\mathbb{#1}\left[#2\right]^{#3}}

\newcommand{\rot}[2]{\te{Q}_{\ve{#1}}^{#2}}
\newcommand{\ve}[1]{\boldsymbol{#1}} 
\newcommand{\te}[1]{\mathbf {#1}} 
\newcommand{\tens}[2]{{}^{#2}\mathbf {#1}} 
\newcommand{\ts}[2]{\mathcal{L}^{#1}_{#2}} 
\newcommand{\tss}[1]{\mathcal{L}^{#1}_{\textrm{sym}}} 
\newcommand{\tsa}[1]{\mathcal{L}^{#1}_{\textrm{skw}}} 
\newcommand{\tsdetpos}{\mathcal{L}^{2}_{\textrm{det}>0}} 
\newcommand{\tsposdef}{\mathcal{L}^{2}_{\lambda>0}} 
\newcommand{\tr}{\operatorname{tr}} 
\newcommand{\ttu}[1]{_{\textrm{#1}}} 
\newcommand{\integg}[3]{\mathop{\int}_{ #1}^{ #2} \mathrm{d} #3 \:} 
\newcommand{\diffp}[2]{\frac{\partial #1}{\partial #2}} 
\newcommand{\point}{\text{ .}}
\newcommand{\commar}{\text{ ,}}
\newcommand{\inv}[1]{{}^{#1}I}
\newcommand{\invar}[2]{{}^{#1}{#2}}
\newcommand{\M}[1]{\uuline{\boldsymbol #1}} 
\newcommand{\V}[1]{\uline{\boldsymbol #1}} 
\newcommand{\aniso}[1]{\mathcal{A}_{#1}}
\newcommand{\Ms}{\mathscr{M}}
\newcommand{\Hs}{\mathscr{H}}
\newcommand{\rset}{\mathbb{R}}
\newcommand{\zset}{\mathbb{Z}}

\title{Construction of minimal integrity bases for  anisotropic hyperelasticity via structural tensors}

\author{
	Brain M. Riemer\\
	Institute of Solid Mechanics\\
	TU Dresden,
	01062 Dresden, Germany \\
	\And
	J\"{o}rg Brummund\\
	Institute of Solid Mechanics\\
	TU Dresden,
	01062 Dresden, Germany \\
	\And
	Karl A. Kalina\\
	Institute of Solid Mechanics\\
	TU Dresden,
	01062 Dresden, Germany \\
	\And
	Abel H. G. Milor\\
	Institute of Scientific Computing\\
	TU Dresden,
	01062 Dresden, Germany \\
	\And
	Franz Dammaß\\
	Institute of Solid Mechanics\\
	TU Dresden,
	01062 Dresden, Germany \\
	\And
	Markus K\"{a}stner\thanks{Corresponding author, email: \texttt{markus.kaestner@tu-dresden.de}.} \\
	Institute of Solid Mechanics\\
	TU Dresden, 
	01062 Dresden, Germany \\
}

\renewcommand{\shorttitle}{Construction of irreducible integrity basis for  anisotropic hyperelasticity via structural tensors}

\hypersetup{
pdftitle={Preprint\_RiemerEtAl\_2026},
pdfsubject={},
pdfauthor={RiemerEtAl},
pdfkeywords={},
}

\theoremstyle{definition}

\usepackage[font=small]{caption}

\begin{document}

\pagenumbering{arabic}

\maketitle

\begin{abstract}
	We present minimal integrity bases for all common anisotropies in hyperelasticity via the structural tensor concept, which can be used to formulate any algebraic invariant function in the elements of the respective bases. Hence, the provided minimal integrity bases are of great interest for formulating a concise but general anisotropic material model. Our work covers results for the 11 types of anisotropy that arise from the classical 7 crystal systems, as well as findings for 4 additional non-crystal anisotropies derived from the cylindrical, spherical, and icosahedral symmetry systems. By using well-known results from literature about structural tensors, isotropic invariants, and isotropic extension, functional bases are directly determined. A simple analytical-numerical approach is employed to identify polynomial relations between the invariants of these functional bases, thereby enabling the construction of functional bases of reduced cardinality. After that, we show that the determined reduced functional bases are also minimal integrity bases by identifying polynomial relations to known integrity bases from literature. Furthermore, fundamental concepts from invariant theory, including the Hironaka decomposition of invariant rings and the closely related Hilbert series, are employed to further validate the results. Alongside the presented findings, this article also aims to provide an introductory overview of the complex field of modeling anisotropic materials, especially for researchers with an engineering background.
\end{abstract}


\section{Introduction}
\label{sec:intro}
Materials such as composites or rolled metals are widely used in construction and engineering due to their direction-dependent mechanical responses, which arise from their internal structure. In continuum theory, accurately capturing this behavior requires a constitutive model that sufficiently reflects the material’s physical response. As \citet{truesdell_non-linear_1965} emphasized, formulating such models remains one of the central open problems in continuum mechanics. Many classical approaches~\citep{may-newman_constitutive_1998, holzapfel_new_2000, itskov_class_2004, holzapfel_determination_2005, ogden_introducing_2007, schroder_anisotropic_2008, chaimoon_anisotropic_2019, wollner_general_2023, Ciambella2024}, as well as modern machine learning-based techniques such as neural networks~\citep{linka_constitutive_2021, klein_polyconvex_2022, fuhg_learning_2022, tac_data-driven_2022, Linden2023, kalina_neural_2024,dammas_neural_2025,dammas_when_2025}, rely on invariants as foundational elements of the model. These invariants encode material symmetry or anisotropy according to Neumann’s principle, which states that the material symmetry must be preserved in the constitutive equations~\citep{neumann_vorlesungen_1885,Ebbing2010}. Finding the relationships between specific anisotropies and the corresponding invariants was an important focus of research in the second half of the 20th century, during which various types of invariants were identified and classified in the context of continuum theory. Below, we provide a brief overview of some of the work done in this field, mostly from the mechanics community.
\begin{remark}
	\label{rem:terminology}
	The authors would like to point out that the terminology used in invariant theory sometimes differs in the mathematics and mechanics community, even though the two refer to the same concept. 
	In particular, the terms irreducible integrity bases and irreducible functional bases are encountered in several publications of the mechanics community~\citep{smith_1963,Pennisi1987,Apel2004,Ebbing2010}. 
	In these works, a given set of invariants is termed irreducible, if no invariant in that set can be expressed as a polynomial or general function depending on the other elements of the set. 
	However, in mathematics, irreducibility is a property of polynomials. In particular, in invariant theory, irreducibility refers to homogeneous polynomial invariants of positive degree and means that such an invariant is not expressible as a product of two other non-constant invariant polynomials~\citep[p.~14]{froberg_introduction_1998}.
	Furthermore, an integrity basis such that none of its elements can be expressed as a polynomial in the others is denoted as minimal, not as irreducible~\citep[Definition 4.1]{olive_minimal_2017}.
	Likewise, a functional basis is called minimal if none of its elements can be expressed as a function of the others. This terminology is utilized throughout this paper. The most important terms used in this publication are defined in \ref{sec:glossar} and referenced at important points in the text. 
\end{remark}

\subsection{Invariants for anisotropic materials}
\label{sec:anisotropic_invariants}
\citet{smith_1958} initially computed invariants for the 32 point groups of the seven crystal systems leading to 11 out of 14 common anisotropies in hyperelasticity, which are presented in a so-called coordinate-dependent form. Shortly thereafter, \citet{smith_further_1962} revisited these results and demonstrated that the invariants originally formulated by \citet{smith_1958} form minimal integrity bases~(Def.~\ref{def:integrity_basis}). In subsequent works, this methodology was extended to include invariants for 1st and 2nd order tensors~\citep{smith_1963,smith_integrity_1964}, enabling the formulation of constitutive models for coupled field problems such as magneto-mechanics~\citep{eringenelectrodynamics, kiralEringen}. Although the invariants in \citet{smith_1963} were also classified as elements of minimal integrity bases, \citet{TAURINES2022167885} showed that the integrity basis associated with the cubic group $\Group{O}$ does not satisfy the criterion of minimality. Furthermore, \citet{smith_transversely_1982} also provided integrity bases for an arbitrary number of 1st order as well as symmetric and antisymmetric 2nd order tensors for all five groups related to the two cylindrical non-crystal systems (also denoted as transversely isotropic).

At the same time, numerous studies \citep{rivlin_further_1955,spencer_finite_1958,spencer_theory_1958,spencer_further_1959,smith_minimality_1960, spencer_invariants_1961, spencer_isotropic_1962, smith_isotropic_1965, wang_representations_1969-1, wang_representations_1969, smith_fundamental_1970, wang_new_1970, wang_corrigendum_1971, smith_isotropic_1971} have examined the determination of isotropic invariants.
This line of work extended the consideration from individual tensors to arbitrary sets of 1st and 2nd order tensors. While \citet{smith_minimality_1960, spencer_invariants_1961, spencer_isotropic_1962, smith_isotropic_1965} concentrated on minimal integrity bases, \citet{Boehler1977} provided a comprehensive summary of the developments of \citet{wang_representations_1969-1, wang_representations_1969, wang_new_1970, wang_corrigendum_1971, smith_fundamental_1970, smith_isotropic_1971} and has formulated a general framework for constructing functional bases~(Def.~\ref{def:function_basis}) of isotropic invariants involving arbitrary sets of 1st order tensors as well as symmetric and antisymmetric 2nd order tensors.
\citet{Pennisi1987} demonstrated that the formulation of \citet{Boehler1977} also yields a minimal functional basis for the isotropic case.

For the application to anisotropic cases, \citet{Lokhin1963} were the first to introduce the concept of structural tensors, which capture anisotropic properties through their invariance under specific symmetry transformations. Building on this idea, \citet{Zheng1993} showed that a single structural tensor is sufficient for capturing all symmetry properties of a specific group and provided explicit constructions for all relevant anisotropies. By incorporating these structural tensors as additional arguments into the constitutive functions (also known as isotropic extension~\citep{Xiao1996, Apel2004} or Rychlewski's theorem~\citep{zhang_structural_1990, itskov_tensor_2025, man_further_2026}), the results of \citet{smith_isotropic_1965, Boehler1977} could be extended to derive integrity and functional bases for anisotropies described by structural tensors up to 2nd order. However, this approach is limited by the fact that, for certain anisotropies such as hexagonal or cubic symmetry, structural tensors of order higher than two are required, whereas 2nd order tensors suffice for triclinic, monoclinic, orthorhombic, and cylindrical systems~\citep{xiao_2006}. The work by \citet{Xiao1996} addressed this limitation by introducing structural tensor functions restricted to at most 2nd order, thereby enabling the computation of invariants for the remaining relevant anisotropies via the approach of \citet{Boehler1977}. Despite the fact that these invariants form a functional basis, they suffer from the crucial drawback of consisting of a large number of invariants that are not independent of each other~\citep{Apel2004, Ebbing2010}. The polynomial and functional dependencies among the isotropic invariants of \citet{smith_isotropic_1965, Boehler1977} originate from the fact that the structural tensors are constant. Nevertheless, this form of invariants is currently often used in research. For practical reasons, the often large number of resulting invariants is usually reduced by only considering those up to a specified polynomial degree~\citep{Apel2004, Ebbing2010,kalina_neural_2024}.

Besides the works on integrity bases, numerous publications, such as \citet{boehler_simple_1979, xiao_general_1995, xiao_minimal_1996, xiao_anisotropic_1997, xiao_constitutive_1997, xiao_new_1998, Xiao_icosahedral, xiao_anisotropic_1998, xiao_scalar_1998, xiao_irreducible_1999, xiao_anisotropic_1999, xiao_irreducible_2000, xiao_irreducible_2000-1}, have focused on the determination of functional bases in the anisotropic case, which typically contain a fewer number of invariants than the corresponding integrity bases. Finally, it is worth noting that there also exist studies addressing invariants of tensors of higher order than two such as~\citet{betten_irreducible_1987, betten_irreduzible_1992, pennisi_third_1992, ghosh_generalized_2012, itin_quadratic_2016, olive_minimal_2017, desmorat_minimal_2021}.

\subsection{Objectives and contributions of this work}
\label{sec:objectives}
As highlighted in the brief literature review, numerous results already exist on integrity bases for various anisotropies, some of which are proven to be minimal~\citep{smith_further_1962}. However, many works in the mechanics community~\citep{Apel2004, Ebbing2010, kalina_neural_2024} use the concept of structural tensors in conjunction with the functional basis of \citet{Boehler1977}, leading to a functional basis containing many invariants that are not independent of each other.

To address this gap, we first construct functional bases using structural tensors, the structural function concept of \citet{Xiao1996}, and the isotropic invariants of \citet{Boehler1977}. Then the cardinality of these functional bases are reduced by finding polynomial relations between the isotropic invariants, respectively. Subsequently, we examine whether the reduced functional bases also constitute minimal integrity bases. Applying this method within the framework of hyperelasticity, we derive minimal integrity bases for all relevant anisotropies in finite strain elasticity using structural tensors, including the icosahedral symmetry group. In addition, alternative formulations of invariants based on different structural tensors are presented, along with the corresponding relationships between single and multiple structural tensors. 
As part of the validation process, reference is made to coordinate-dependent minimal integrity bases of \citet{smith_further_1962, smith_transversely_1982} as well as information provided by the Hilbert series and the structure of the corresponding invariant rings. In addition to the presented results, this article aims to provide an introductory overview of the complex field of modeling anisotropic materials, particularly for readers with a background in mechanics or engineering. For this reason, the article includes several boxes that not only highlight important key messages but also explain more complex concepts in a less formal way.

\paragraph{Organization of this work}
In Sect.~\ref{sec:fundamentals} we provide a brief introduction to strain and stress measures in continuum mechanics. Based on this, we give important requirements for hyperelastic potentials and summarize symmetry systems as well as point groups. Finally, this section introduces and compares the concepts of coordinate-dependent invariants and coordinate-free representation of invariants using structural tensors. 
In Sect.~\ref{sec:construction_of}, we describe how to construct functional bases and identifying polynomial relations between invariants by an analytical-numerical method. This is followed by two approaches that can be used to prove that a set of invariants forms a minimal integrity basis.
At the end of Sect.~\ref{sec:construction_of}, we illustrate the methodology using a monoclinic and cubic anisotropy as examples. Sect.~\ref{sec:complete_and_irr_sets} then presents the calculated minimal integrity bases for all anisotropies considered in this work. Finally, Sect.~\ref{sec:conclusions} summarizes the main advances for modeling the constitutive behavior of anisotropic materials and outlines directions for future research.

\paragraph{Notation}
Generally, sets and tuples are denoted by calligraphic symbols, e.g., $\Set{S}$, $\Set{T}$. Blackboard bold symbols are used for domains of numbers: $\Setnum{Z}$, $\Setnum{Q}$, $\Setnum{R}$ and $\Setnum{C}$ represent the integer, rational, real and complex numbers, respectively. Any domain of number with a condition as a subscript consists only of elements that satisfy that condition. For example, $\Setnum{Z}_{\geq0}:=\left\{z\in\Setnum{Z}\:|\:z \geq0\right\}$.
The space of $n$th order real tensors is denoted by $\Set{L}^n$. Fully symmetric or skew-symmetric $n$th order tensors are elements of $\tss{n}$ and $\tsa{n}$, respectively. Real 2nd order tensors with positive determinant $\det[\bullet]$ are elements of $\tsdetpos$, while positive definite real 2nd order tensors are elements of $\tsposdef$.
We denote 0th, 1st, and 2nd order tensors as $A$, $\ve{A}$ and $\te{A}$, respectively. To further simplify notation, we also use $\tens{A}{n}$ for a tensor of arbitrary order $n\in\Setnum{Z}_{\geq0}$. In index notation, we write $\tens{A}{n}=A_{i_1,i_2,\dots,i_n}\ve{e}_{i_1}\ve{e}_{i_2}\dots \ve{e}_{i_n}$ using lower case Latin letters, where the indices $i_1,\dots,i_n \in \{1, 2, 3 \}$, $\ve{e}_1, \ve{e}_2, \ve{e}_3$ are the standard basis vectors of $\mathbb R^3$ and Einstein's summation convention applies without restrictions if not stated differently.
The single, double and quartic contraction between tensors are denoted by $\cdot$, $:$ and $::$, respectively. For example, $\te{A}\cdot\te{B}=A_{ij}B_{jk} \,\ve{e}_i\ve{e}_k$, $\te{A}:\te{B}=A_{ij}B_{ji}$ and $\tens{A}{4}::\tens{B}{4}=A_{ijkl}B_{lkji}$.
If tensors are directly adjacent to each other, such as $\tens{A}{n}\tens{B}{m}=\tens{C}{n+m}$, this represents the dyadic product. In addition, the operator $\overset{m}{\otimes}\:\tens{A}{n}$ stands for the $m$-fold dyadic product of $\tens{A}{n}$, e.g., $\overset{3}{\otimes}\: \tens{A}{n}= \tens{A}{n}\tens{A}{n}\tens{A}{n}$, and $\tens{A}{n}^m$ denotes the $m$-fold single contractions between each $\tens{A}{n}$, e.g., $\tens{A}{n}^3 = \tens{A}{n}\cdot\tens{A}{n}\cdot\tens{A}{n}$. The trace operation of a 2nd order tensor or square matrix is denoted by $\tr[\bullet]$.\\
As common in literature, we generally do not explicitly distinguish between algebraic structures and the underlying sets. For example, the term symmetry group can stand for both the pair $(\Group{G}, \cdot)$ and the set $\Group{G}$ of tensors involved therein.
The special orthogonal group and the orthogonal group are denoted by $\Group{O}(3):=\Group{O}(3,\Setnum{R})$ and $\Group{SO}(3):=\Group{SO}(3,\Setnum{R})$, respectively.
$\Group{GL}_n(\Setnum{K})$ denotes the general linear group in $n$-dimension over a field $\Setnum{K}$.
For reasons of readability, the arguments of functions are often omitted within this work. 
\section{Fundamentals}
\label{sec:fundamentals}
This section provides a concise overview of the required continuum mechanical fundamentals and finite-strain hyperelasticity. It also introduces important crystal and non-crystal symmetry classes, as well as key concepts related to scalar-valued invariants for modeling elastic potentials.

\subsection{Kinematics and stress measures}
\label{sec:kinematics}
For finite deformations, a distinction is made between the reference configuration $\Set{B}_0 \subset \Setnum{R}^3$ at a reference time $\tau_0 \in\Setnum{R}$ and the current configuration $\Set{B}_\tau \subset \Setnum{R}^3$ at a time $\tau\geq \tau_0$. The smooth, bijective and orientation preserving motion mapping $\ve{\varphi}_\tau: \Set{B}_0 \to \Set{B}_\tau, \:\ve X \mapsto\ve{\varphi}_\tau\left(\ve{X}\right)$ maps each material point $\ve{X} \in \Set{B}_0$ from the reference configuration to the corresponding point $\ve{x}_\tau=\ve{\varphi}_\tau\left(\ve{X}\right) \in \Set{B}_\tau$ of the current configuration at time $\tau$.
In order to enable the calculation of derivatives w.r.t. time later on, we represent $\ve \varphi_\tau(\ve X)$ via a function of space and time in what follows, i.e., $\ve \varphi(\ve X,\tau)$ \citep[Sect.~2.2]{silhavy_mechanics_1997}.%
\addtocounter{footnote}{-1}%
The deformation gradient $\te{F} = \left(\nabla_{\ve{X}}\ve{\varphi}\right)^\top \in \tsdetpos$ with $\det[\te{F}] >0$, a so-called two-point tensor, is introduced as the central kinematic variable.\footnote{\label{foot:two-point}%
	Two-point 2nd order tensors are related to both reference and current configuration, e.g., the first coordinate of the deformation gradient $\te{F} = F_{kl} \, \tilde{\ve{e}}_k \, \ve{e}_l$ is related to $\Set{B}_\tau$ whereas the second coordinate is related to $\Set{B}_0$, cf. \citep[Sect.~1.4]{Marsden1984}, \citep[Sect.~2.4]{Holzapfel2000}. Similar to $\te F$, the 1st Piola-Kirchhoff stress tensor $\te P$ is also a two-point tensor.}
Here, $\nabla_X$ denotes the nabla operator acting on the function to its right; the subscript $X$ indicates that derivatives are taken with respect to the reference coordinates.
The deformation gradient $\te F = \te R \cdot \te U$ can be multiplicatively decomposed into unique rotational part $\te R\in\Group{SO}(3)$ and stretch part $\te U\in\tsposdef$.
From that, the right Cauchy-Green deformation tensor $\te{C} = \te{F}^\top \cdot \te{F} \in \tsposdef$, which is objective and positive definite, and the Green-Lagrange strain tensor $\te{E} = \frac{1}{2}\left(\te{C}-\te{1}\right) \in \tss{2}$ can be calculated. These two quantities are related to the reference configuration and are therefore often referred to as Lagrangian or material strain measures~\citep[Sect.~2.5]{Holzapfel2000}.

In addition to a variety of kinematic quantities, there are also several stress measures that can be introduced. Among the most important are the Cauchy stress tensor $\ve{\sigma} \in \tss{2}$, the 1st Piola-Kirchhoff stress tensor $\te{P} = J \ve{\sigma} \cdot \te{F}^{-\top} \in \ts{2}{}$ and the 2nd Piola-Kirchhoff stress tensor $\te{T} = J \te{F}^{-1} \cdot \ve{\sigma} \cdot \te{F}^{-\top} \in \tss{2}$.

For a much more comprehensive introduction to continuum mechanics, please refer to the textbooks of \citet{Marsden1984}, \citet{silhavy_mechanics_1997} and \cite{Haupt2000}.
\subsection{Conditions on hyperelastic potentials}
\label{sec:conditions}
An elastic constitutive model relates the deformation gradient to the stress induced at a material point. In hyperelasticity, this mapping is not defined directly, but via the scalar-valued  Helmholtz free energy
\begin{equation}
	\varPsi:\; \tsdetpos \to \Setnum{R}_{\geq 0} \; , \; \te{F} \mapsto \varPsi\left( \te{F} \right) \quad \text{and} \quad \te{P}=\frac{\partial \varPsi}{\partial \te{F}}\commar
	\label{qe:free_energy}
\end{equation}
which ensures thermodynamic consistency to be effectively demonstrated using the Clausius-Duhem inequality by definition of $\te{P}$ as the derivative with respect to $\te{F}$ \citep{Holzapfel2000,Haupt2000,Linden2023}. 
Other key principles in constitutive modeling include determinism, the principle of equipresence and local action, objectivity as well as material symmetry.\footnote{
	We recommend the reader \citet[p. 249-253]{lebon_understanding_2008}, see also \citet{rivlin_principles_1997}, for a critical discussion on the principles in constitutive modeling. Furthermore, we suggest \citet{frewer_more_2009} for an extensive review on the principle of objectivity.}

The latter two principles are given by $\varPsi\left( \te{Q}_1 \cdot \te{F} \cdot \te Q_2^\top \right) = \varPsi\left( \te{F}\right) \:\forall\:\te{Q}_1\in\Group{SO}(3), \te Q_2\in\Group{G}$, where $\Group{G}\subseteq\Group{O}(3)$ is the symmetry group of the material.
Thereby, setting $\te Q_2 = \te 1$ isolates the requirement of objectivity, that is, invariance of the material behavior with respect to rigid body motions of the current configuration $\Set{B}_\tau$. By setting $\te Q_1 = \te 1$, the requirement of material symmetry is isolated, i.e., invariance of the material behavior with respect to rigid body rotations $\te Q_2\in \Group{G}$ of the reference configuration $\Set{B}_0$.
To guarantee objectivity, the Helmholtz free energy is expressed as a function of a deformation or strain measure that is objective.
A common choice is $\varPsi: 
\tsdetpos \to \Setnum{R}_{\geq 0} \; , \; \te{F} \mapsto \varPsi\left( \te{C}(\te F) \right)$.
In the following, we will pursue this approach and simply write 
$\varPsi: \tsposdef \to \Setnum{R}_{\geq 0} \; , \; \te{C} \mapsto \varPsi\left( \te{C} \right)$.
As a consequence, merely the material symmetry condition in the form
\begin{equation}
	\varPsi\left( \te{Q} \cdot \te{C} \cdot \te{Q}^\top \right) = \varPsi\left( \te{C}\right) \:\forall\te{Q}\in\Group{G}  \subseteq\Group{O}(3) 
	\label{eq:material_symmetry}
\end{equation}
remains.
In general, only some functions of $\te{C}$ fulfill Eq.~\eqref{eq:material_symmetry}. These are called $\Group{G}$-invariant functions, or simply invariants.
Further constitutive conditions can be imposed on the elastic potential, such as the non-negativity of $\varPsi$, which are explained in detail in \citet{Holzapfel2000} or \citet{Linden2023}.
\subsection{Overview on crystal classes and non-crystal classes}
\label{sec:overview_crystal}
\begin{figure}[t]
	\centering
	\includegraphics[clip]{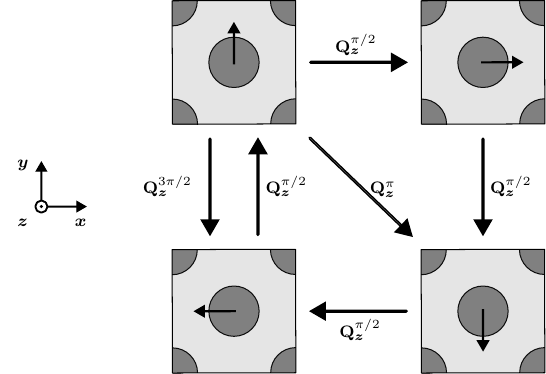}
	\caption{Concept of groups in the context of two-dimensional microstructures. In this illustration only rotations are considered. The set of transformations that keep the microstructure geometrically invariant is given by $\mathcal G=\left\{ \rot{z}{\pi/2}, \rot{z}{\pi}, \rot{z}{3\pi/2}, \te{1} \right\}$. This group can be generated in terms of a single generator: $\rot{z}{\pi/2}$ or $\rot{z}{3\pi/2}$. Note that more complex groups may not be generated by a single element.}	
	\label{fig:group_illustration}
\end{figure}

Many substances, both crystalline and non-crystalline, exhibit symmetry or geometric invariance with respect to certain rotations or roto-inversions.
Mathematically, a rotation or roto-inversion can be represented by an orthogonal 2nd order tensor $\te Q \in\Group{O}(3)$. The set of orthogonal tensors, whose corresponding rotation or roto-inversion leave the material geometrically unchanged, is called the material's symmetry group $\Group{G} \subseteq \Group{O}(3)$.
Therefore, we provide a brief overview of the symmetry systems relevant to this work.
\paragraph{Crystal systems}
A lot of materials, for example Diamond or Piypite, exhibit a special arrangement of their atoms, known as a crystal structure. The individual atoms are organized in a regular lattice, which consists of a periodic sequence of so-called unit cells. These cells are geometrically invariant to selected rotations represented by tensors $\rot{p}{\varphi}\in\Group{SO}(3)$ and roto-inversions $-\rot{p}{\varphi}\in\Group{O}(3)$, where $\ve{p}\in\ts{1}{}$ and $\varphi\in[0,2\pi)\subset\Setnum{R}$ denote the axis and angle of rotation, respectively.
\begin{center}
	\fbox{
		\begin{minipage}{0.9\textwidth}
			The material symmetry group $\Set{G}$, i.e., the set  of  geometry preserving transformations is  intrinsically defined by the respective unit cell or microstructure. Conversely, the corresponding set of transformations $\Set{G}$ also defines the symmetry of the microstructure. 
			In materials science, a symmetry group $\Set{G}$ is also called point group~\citep{voigt_lehrbuch_1910,Apel2004,Ebbing2010}. 
			
			Mathematically more strict, a group is a pair $(\Group{G}\subseteq\Group{O}(3), \cdot)$ consisting of two things~(Def.~\ref{def:group}, Rem.~\ref{rem:group_notation}): a set, in our case of some orthogonal tensors, and an operation, in our case the single contraction between these.\footnotemark\, 
			However, it is common to refer to both the set $\Group{G}$ and the pair $(\Group{G}\subseteq\Group{O}(3), \cdot)$ as the group, depending on the specific context.
			
			Furthermore, so-called generators~(Def.~\ref{def:generators}) are special elements of a group such that all other elements of the group can be expressed using, in our case, single tensor contraction between the generators. An illustration of these concepts can be found in Fig.~\ref{fig:group_illustration}.
		\end{minipage}
	}
\end{center}
\footnotetext{Independently of the specific microstructure, it can be verified that $(\Group{G}, \cdot)$ is indeed a group fulfilling the axioms of Def.~\ref{def:group} in straightforward manner: Single contraction of tensors is associative, the 2nd order identity tensor $\te{1}\in\tss{2}$ always is an element of $\Group{G}$ and $\rot{p}{\varphi} \in \Group{G}$ implies $\left(\rot{p}{\varphi}\right)^{-1} \in \Group{G}$.}
The seven crystal systems~\citep{voigt_lehrbuch_1910, Ebbing2010}: triclinic, monoclinic, rhombic, tetragonal, trigonal, hexagonal and cubic systems, are associated with 32 point groups. In addition to one possible set of generators that clearly describes a group, the point groups are often specified by the Schoenflies notation~\citep{schoenflies_krystallsysteme_1891} or Hermann-Mauguin notation~\citep{hermann_xvi_1929, mauguin_sur_1931}. 
Here, we use the Schoenflies notation along with the designation $\Group{G}_i$ used by \citet{Zheng1993} to distinguish between the considered groups. In the context of hyperelasticity, some group actions become equivalent due to the fact that the right Cauchy-Green deformation tensor $\te{C}$ remains invariant under additional inversion $-\te{1}$ of the transformation:
\begin{equation}
	\te{\tilde{C}} = \rot{p}{\varphi}\cdot\te{C}\cdot\left(\rot{p}{\varphi}\right)^\top = -\rot{p}{\varphi}\cdot\te{C}\cdot\left(-\rot{p}{\varphi}\right)^\top  
	\quad \forall \, \rot{p}{\varphi} \in\Group{SO}(3) \point
	\label{eq:C_transformation}
\end{equation}
Hence, transformations $\rot{p}{\varphi} \in\Group{SO}(3)$ and $-\rot{p}{\varphi}\in\Group{O}(3)$ are indistinguishable for $\te C$ if they share the same axis and angle of rotation. We highlight this by introducing a third notation $\aniso{i}$, cf. \citet{Zheng1993, Apel2004}, representing a specific anisotropy in hyperelasticity. 
\begin{remark}
	\label{rem:inversion_does_not_lead}
	In the case of even-order tensors that are attached at one single configuration, certain group actions become equivalent due to the fact that such tensors are invariant under inversion of the transformation, regardless of existing symmetries. Thus, transformations $\rot{p}{\varphi}$ and $-\rot{p}{\varphi}$ are indistinguishable for such tensors. This property, however, does generally not hold for tensors of odd order and two-point tensors such as the deformation gradient $\te F$, cf. Footnote~\ref{foot:two-point}.
\end{remark}
Tab.~\ref{tab:crystall_groups} provides an overview of the 32 crystallographic point groups, including the three notations used, a possible set of generators, and the group order (number of elements), denoted by $\left|\Group{G}\right|$.
It should be noted that, in addition to materials whose atomic structure leads to geometric invariance, materials like wood\footnote{Because of the non-straight grain direction in wood, the orientation of the symmetry transformations varies. However, for arbitrarily small volume elements, wood can still be classified as rhombic.}, rolled sheet metal, or even biological tissues can also be associated with one of the groups of the seven crystal systems.
\begin{table}[p]
	\centering
	\caption{The 32 finite point groups of the seven crystal systems, inspired by \citet{smith_1963,Zheng1993,Apel2004}. The Schoenflies symbols~\citep{schoenflies_krystallsysteme_1891} and the designations according to \citet{Zheng1993} are provided, along with the various types of anisotropy encountered in hyperelasticity.  $\ve{x}$, $\ve{y}$, $\ve{z}$ and $\ve{k}$ denote the $x$-, $y$-, $z$- and $[\ve{k}]=1/\sqrt{3}\left(-1,-1,-1\right)^\top$-axis.}
	\label{tab:crystall_groups}
	\renewcommand{\arraystretch}{1.2}
	\begin{small}
		\begin{tabularx}{0.92\textwidth}{p{2cm}p{3.4cm}p{1.6cm}p{2cm}p{2.2cm}p{0.6cm}p{2cm}}
			\toprule
			Crystal system & Point group & Schoenflies & Zheng/Spencer & Anisotropy type & $\left|\Group{G}\right|$ & Generators\\
			\midrule
			triclinic   & pedial 	  & $\Group{C}_1$                 & $\Group{G}_{1}$ & $\aniso{1}$ & 1 & $\te{1}$ \\
			& pinacoidal 	& $\Group{C}_i$ / $\Group{S}_2$ & $\Group{G}_{2}$ & $\aniso{1}$ & 2 & $-\te{1}$ \\
			\midrule
			monoclinic 	& sphenoidal 	& $\Group{C}_2$    & $\Group{G}_{4}$ & $\aniso{2}$  & 2 & $\rot{z}{\pi}$ \\
			& domatic 	  & $\Group{C}_{1h}$ & $\Group{G}_{3}$ & $\aniso{2}$ 	& 2 & $-\rot{z}{\pi}$ \\
			& prismatic     & $\Group{C}_{2h}$ & $\Group{G}_{5}$ & $\aniso{2}$ 	& 4 & $-\te{1},\rot{z}{\pi}$ \\
			\midrule
			rhombic & rhombic-disphenoidal 	& $\Group{D}_2$    & $\Group{G}_{7}$ & $\aniso{3}$  & 4 & $\rot{z}{\pi},\rot{x}{\pi}$ \\
			& rhombic-pyramidal 	  & $\Group{C}_{2v}$ & $\Group{G}_{6}$ & $\aniso{3}$ 	& 4 & $\rot{z}{\pi},-\rot{x}{\pi}$ \\
			& rhombic-dipyramidal   & $\Group{D}_{2h}$ & $\Group{G}_{8}$ & $\aniso{3}$ 	& 8 & $-\te{1},\rot{z}{\pi},\rot{x}{\pi}$ \\
			\midrule
			tetragonal  & tetragonal-pyramidal 	  & $\Group{C}_4$ & $\Group{G}_{10}$ & $\aniso{4}$ 	& 4 & $\rot{z}{\frac{\pi}{2}}$ \\
			& tetragonal-disphenoidal & $\Group{C}_{2i}$ / $\Group{S}_4$  & $\Group{G}_{9}$ & $\aniso{4}$ 	  & 4 & $-\rot{z}{\frac{\pi}{2}}$ \\
			& tetragonal-dipyramidal 	& $\Group{C}_{4h}$                  & $\Group{G}_{11}$ & $\aniso{4}$ 	  & 8 & $-\te{1},\rot{z}{\frac{\pi}{2}}$ \\
			& tetragonal-trapezohedral 	& $\Group{D}_4$    & $\Group{G}_{14}$ & $\aniso{5}$ 	& 8  & $\rot{z}{\frac{\pi}{2}},\rot{x}{\pi}$ \\
			& ditetragonal-pyramidal 	& $\Group{C}_{4v}$ & $\Group{G}_{13}$ & $\aniso{5}$ 	& 8  & $\rot{z}{\frac{\pi}{2}},-\rot{x}{\pi}$ \\
			& tetragonal-scalenohedral 	& $\Group{D}_{2d}$ & $\Group{G}_{12}$ & $\aniso{5}$ 	& 8  & $-\rot{z}{\frac{\pi}{2}},-\rot{x}{\pi}$ \\
			& ditetragonal-dipyramidal 	& $\Group{D}_{4h}$ & $\Group{G}_{15}$ & $\aniso{5}$ 	& 16 & $-\te{1},\rot{z}{\frac{\pi}{2}},\rot{x}{\pi}$ \\
			\midrule
			trigonal  & trigonal-pyramidal 	& $\Group{C}_3$                    & $\Group{G}_{16}$ & $\aniso{6}$ & 3 & $\rot{z}{\frac{2\pi}{3}}$ \\
			& rhombohedral 	   & $\Group{C}_{3i}$ / $\Group{S}_6$ & $\Group{G}_{17}$ & $\aniso{6}$ & 6 & $-\rot{z}{\frac{2\pi}{3}}$ \\
			& trigonal-trapezohedral  & $\Group{D}_3$     & $\Group{G}_{19}$ & $\aniso{7}$ 	& 6  & $\rot{z}{\frac{2\pi}{3}},\rot{x}{\pi}$ \\
			& ditrigonal-pyramidal 	  & $\Group{C}_{3v}$  & $\Group{G}_{18}$ & $\aniso{7}$ 	& 6  & $\rot{z}{\frac{2\pi}{3}},-\rot{x}{\pi}$ \\
			& hexagonal-scalenohedral & $\Group{D}_{3d}$  & $\Group{G}_{20}$ & $\aniso{7}$ 	& 12 & $-\rot{z}{\frac{2\pi}{3}},\rot{x}{\pi}$ \\
			\midrule
			hexagonal   & hexagonal-pyramidal 	  & $\Group{C}_6$    & $\Group{G}_{22}$ & $\aniso{8}$   & 6  & $\rot{z}{\frac{\pi}{3}}$ \\
			& trigonal-dipyramidal 	& $\Group{C}_{3h}$ & $\Group{G}_{21}$ & $\aniso{8}$ 	& 6  & $-\rot{z}{\frac{\pi}{3}}$ \\
			& hexagonal-dipyramidal 	& $\Group{C}_{6h}$ & $\Group{G}_{23}$ & $\aniso{8}$ 	& 12 & $-\te{1},\rot{z}{\frac{\pi}{3}}$ \\
			& hexagonal-trapezohedral & $\Group{D}_6$    & $\Group{G}_{26}$ & $\aniso{9}$ & 12 & $\rot{z}{\frac{\pi}{3}},\rot{x}{\pi}$ \\
			& dihexagonal-pyramidal   & $\Group{C}_{6v}$ & $\Group{G}_{25}$ & $\aniso{9}$ & 12 & $\rot{z}{\frac{\pi}{3}},-\rot{x}{\pi}$ \\
			& ditrigonal-dipyramidal  & $\Group{D}_{3h}$ & $\Group{G}_{24}$ & $\aniso{9}$ & 12 & $-\rot{z}{\frac{\pi}{3}},-\rot{x}{\pi}$ \\
			& dihexagonal-dipyramidal & $\Group{D}_{6h}$ & $\Group{G}_{27}$ & $\aniso{9}$ & 24 & $-\te{1},\rot{z}{\frac{\pi}{3}},\rot{x}{\pi}$ \\
			\midrule
			cubic   & tetartoidal 	    & $\Group{T}$   & $\Group{G}_{28}$ & $\aniso{10}$ & 12 & $\rot{z}{\pi},\rot{k}{\frac{2\pi}{3}}$ \\
			& diploidal 		& $\Group{T}_h$ & $\Group{G}_{29}$ & $\aniso{10}$ & 24 & $\rot{z}{\pi},-\rot{k}{\frac{2\pi}{3}}$ \\
			& gyroidal 		    & $\Group{O}$   & $\Group{G}_{31}$ & $\aniso{11}$ & 24 & $\rot{z}{\frac{\pi}{2}},\rot{x}{\frac{\pi}{2}}$\\
			& hextetrahedral 	& $\Group{T}_d$ & $\Group{G}_{30}$ & $\aniso{11}$ & 24 & $-\rot{z}{\frac{\pi}{2}},-\rot{x}{\frac{\pi}{2}}$\\
			& hexoctahedral 	& $\Group{O}_h$ & $\Group{G}_{32}$ & $\aniso{11}$ & 48 & $\rot{z}{\frac{\pi}{2}},-\rot{k}{\frac{2\pi}{3}}$\\
			\bottomrule
		\end{tabularx}
	\end{small}
\end{table}
\paragraph{Non-crystal systems}
Many materials, such as fiber-reinforced composites~\citep{ehret_polyconvex_2007,Kalina2023,chatterjee_role_2025, lauff_influence_2025} or magneto-rheological elastomers~\citep{bustamante_transversely_2010, Kalina2024}, exhibit geometric invariance under transformations that are not among the 32 crystallographic point groups. In this context, one speaks of non-crystal systems, of which there are infinitely many. Among the most important systems are the spherical (isotropic) and the cylindrical (transversal-isotropic), whose associated point groups have an infinite number of elements. 
However, not all groups associated with non-crystal systems contain an infinite number of elements. A notable example is the icosahedral system, which comprises two groups, each possessing a finite number of transformations.
As with the crystal systems, several groups coincide in hyperelasticity due to the transformation properties of the 2nd order tensor $\te{C}$, cf. Eq.~\eqref{eq:C_transformation}. Consequently, the five groups of the cylindrical system reduce to two distinct anisotropies, and one additional anisotropy each arises from the spherical and icosahedral systems, respectively. Again, we use the Schoenflies notation, the notation of \citet{Zheng1993} and the anisotropy type $\aniso{i}$ for the different groups of the non-crystal systems. It should be noted that \citet{Zheng1993} did not consider the spherical and icosahedral systems, which is why no notation is available in these cases.
The point groups of the considered non-crystal systems are listed in Tab.~\ref{tab:non_crystall_groups}.
\begin{table}[ht]
	\centering
	\caption[Non crystal systems]{Point groups of the cylindrical, spherical and icosahedral system, inspired by \cite{Apel2004}. The Schoenflies symbols~\citep{schoenflies_krystallsysteme_1891} and designations according to \citet{Zheng1993} are provided, along with the various types of anisotropy encountered in hyperelasticity. $\ve{x}$, $\ve{z}$ denote the $x$-, $z$-axis. $[\ve{g}]=1/\sqrt{\tau^2-2\tau+2}\left(\tau-1,0,1\right)^\top$ with the golden ratio $\tau=(1+\sqrt{5})/2$ and $[\ve{p}]=1/\sqrt{3}\left(1,1,1\right)^\top$.}
	\label{tab:non_crystall_groups}
	\renewcommand{\arraystretch}{1.2}
	\begin{small}
		\begin{tabularx}{0.86\textwidth}{p{2.7cm}p{1.8cm}p{2cm}p{2.2cm}p{0.6cm}p{1.2cm}p{2.1cm}}
			\toprule
			Non-crystal system & Schoenflies & Zheng/Spencer & Anisotropy type & $\left|\Group{G}\right|$ & Generators &\\
			\midrule
			cylindrical & $\Group{C}_\infty$ / $\Group{T}_1$     & $\Group{T}_{1}$ & $\aniso{12}$   & $\infty$ & $\rot{z}{\gamma},$         & $\gamma\in [0,2\pi)$ \\
			& $\Group{C}_{\infty h}$ / $\Group{T}_3$ & $\Group{T}_{3}$ & $\aniso{12}$   & $\infty$ & $-\te{1},\rot{z}{\gamma},$ & $\gamma\in [0,2\pi)$ \\
			& $\Group{D}_{\infty}$ / $\Group{T}_5$   & $\Group{T}_{5}$ & $\aniso{13}$   & $\infty$ & $\rot{z}{\gamma},\rot{x}{\pi},$         & $\gamma\in [0,2\pi)$ \\
			& $\Group{C}_{\infty v}$ / $\Group{T}_2$ & $\Group{T}_{2}$ & $\aniso{13}$   & $\infty$ & $\rot{z}{\gamma},-\rot{x}{\pi},$        & $\gamma\in [0,2\pi)$ \\
			& $\Group{D}_{\infty h}$ / $\Group{T}_4$ & $\Group{T}_{4}$ & $\aniso{13}$   & $\infty$ & $-\te{1},\rot{z}{\gamma},\rot{x}{\pi},$ & $\gamma\in [0,2\pi)$ \\
			\midrule
			spherical 	& $\Group{K}$ / $\Group{SO}(3)$  & - & $\aniso{14}$ & $\infty$ & $\rot{z}{\gamma},\rot{x}{\alpha},$         & $\gamma,\alpha\in [0,2\pi)$ \\
			& $\Group{K}_h$ / $\Group{O}(3)$ & - & $\aniso{14}$ & $\infty$ & $-\te{1},\rot{z}{\gamma},\rot{x}{\alpha},$ & $\gamma,\alpha\in [0,2\pi)$ \\
			\midrule
			icosahedral & $\Group{I}$     & - & $\aniso{15}$ & 60  & $\rot{g}{\frac{2\pi}{5}},\rot{p}{\frac{2\pi}{3}}$   & \\
			& $\Group{I}_h$   & - & $\aniso{15}$ & 120 & $\rot{g}{\frac{2\pi}{5}},-\rot{p}{\frac{2\pi}{3}}$  & \\
			\bottomrule
		\end{tabularx}
	\end{small}
\end{table}

A simple script to compute the group elements of the finite groups considered in Tab.~\ref{tab:crystall_groups}~and~{\ref{tab:non_crystall_groups}} using the respective generators can be found at \url{https://github.com/NEFM-TUDresden/anisotropic_hyperelasticity_integrity_bases}.
\subsection{Formulation of hyperelastic potentials with invariants}
\label{sec:modeling_of}
In continuum mechanics, material models are often formulated using deformation or strain tensors that ensure objectivity, see Sect.~\ref{sec:conditions}. To further ensure material symmetry and thus reflect the underlying atomistic or microscopic structure, the use of scalar-valued invariants is particularly appealing. These invariants are then used to formulate the elastic potential.
Therefore, we begin with a general definition of invariants within this framework. To this end, we first introduce the function
\begin{equation}
	\star : \Group{G} \times \bigcup_{n\geq0}\ts{n}{} \to \bigcup_{n\geq0}\ts{n}{} ,\quad
	\left\{
	\begin{array}{llll}
		\left(\te{Q},H\right) \mapsto \te{Q}\star H := H & n=0 \\
		\left(\te{Q},\ve{H}\right) \mapsto \te{Q}\star \ve{H} := \te{Q}\cdot\ve{H} = H_i \:\te{Q}\cdot\ve{e}_i & n=1 \\
		\left(\te{Q},\te{H}\right) \mapsto \te{Q}\star \te{H} := \te{Q}\cdot\te{H}\cdot\te{Q}^\top = H_{i_1i_2} \te{Q}\cdot\ve{e}_{i_1}\te{Q}\cdot\ve{e}_{i_2} & n=2 \\
		\left(\te{Q},\tens{H}{n}\right) \mapsto \te{Q}\star {}\tens{H}{n} := H_{i_1\dots i_n}\te{Q}\cdot\ve{e}_{i_1}\dots \te{Q}\cdot\ve{e}_{i_n} & n>2 \\
	\end{array} \right. , \quad \te{Q}\in\Group{G} \commar
	\label{eq:rayleigh}
\end{equation}
which provides a compact notation for describing the action of a group element $\te{Q}\in\Group{G}\subseteq\Group{O}(3)$ on tensors of arbitrary order, cf. \citet{Xiao1996, Apel2004}. 
In other words, $\star$ maps the pair of transformation $\te Q$ and an (untransformed) tensor $\tens{H}{n}$ to the transformed match of $\tens{H}{n}$.
To further shorten the notation, the transformation of several tensors, that are gathered in a tuple $\Set{H}=\left( \tens{H}{n_1}_1, \ldots, \tens{H}{n_k}_k \right)$, $k\in\Setnum{Z}_{\ge 1}$, by the same group element $\te{Q}\in\Group{G}$ is denoted by
\begin{equation}
	\te{Q}\star\Set{H} := \left( \te{Q}\star\tens{H}{n_1}_1, \dots, \te{Q}\star\tens{H}{n_k}_k \right) \point
	\label{eq:tensor_set_trafo}
\end{equation}
We will now introduce two important concepts: these are the terms $\mathcal G$-equivariance and $\mathcal G$-invariance.

\begin{definition}[Tensor-valued equivariant]
	\label{def:equivariant}
	Let $m\in\Setnum{Z}_{\geq 0}$, $k\in\Setnum{Z}_{\geq1}$ and $ \tens{E}{m} : \ts{n_1}{} \times \dots \times \ts{n_k}{} \rightarrow \ts{m}{}, \, \Set{H} \mapsto \tens{E}{m} (\Set{H})$ be a tensor-valued function of $\Set{H}=\left( \tens{H}{n_1}_1,  \dots, \tens{H}{n_k}_k \right)$ with $n_1, \dots, n_k \in \Setnum{Z}_{\geq0}$. The tensor-valued function $\tens{E}{m}(\Set{H})$ is called equivariant under the group $\Group{G}\subseteq\Group{O}(3)$, $\Group{G}$-equivariant or simply equivariant, if it satisfies the condition
	\begin{equation}
		\te Q \star \tens{E}{m}(\Set{H}) = \tens{E}{m}(\te{Q}\star\Set{H}) \quad \forall \: \te{Q}\in\Group{G}\quad \forall\: 
		\tens{H}{n_\lambda}_\lambda \in \ts{n_\lambda}{}\point
		\label{eq:equivariant_definition}
	\end{equation}
\end{definition}

\begin{definition}[Tensor-valued invariant]
	\label{def:tensor_invariant}
	Let $m\in\Setnum{Z}_{\geq 0}$ and $ \tens{I}{m} : \ts{n_1}{} \times \dots \times \ts{n_k}{} \rightarrow \ts{m}{}, \, \Set{H} \mapsto \tens{I}{m}(\Set{H})$ be a tensor-valued function of $\Set{H}=\left( \tens{H}{n_1}_1,  \dots, \tens{H}{n_k}_k \right)$, $n_1, \dots, n_k \in \Setnum{Z}_{\geq0}$ and $k\in\Setnum{Z}_{\geq1}$. The tensor-valued function $I(\Set{H})$ is called invariant under the group $\Group{G}\subseteq\Group{O}(3)$, $\Group{G}$-invariant or simply invariant, if it satisfies the condition
	\begin{equation}
		\tens{I}{m}(\Set{H}) = \tens{I}{m}(\te{Q}\star\Set{H}) \quad \forall \: \te{Q}\in\Group{G}\quad \forall\: 
		\tens{H}{n_\lambda}_\lambda \in \ts{n_\lambda}{}\point
		\label{eq:tensor_invariant_definition}
	\end{equation}
\end{definition}

\begin{definition}[Scalar invariant]
	\label{def:invariant}
	Let $ I : \ts{n_1}{} \times \dots \times \ts{n_k}{} \rightarrow \Setnum{R}, \, \Set{H} \mapsto I(\Set{H})$ be a scalar-valued function of $\Set{H}=\left( \tens{H}{n_1}_1,  \dots, \tens{H}{n_k}_k \right)$, $n_1, \dots, n_k \in \Setnum{Z}_{\geq0}$ and $k\in\Setnum{Z}_{\geq1}$. The scalar-valued function $I(\Set{H})$ is called invariant under the group $\Group{G}\subseteq\Group{O}(3)$, $\Group{G}$-invariant or simply invariant, if it satisfies the condition
	\begin{equation}
		I(\Set{H}) = I(\te{Q}\star\Set{H}) \quad \forall \: \te{Q}\in\Group{G}\quad \forall\: 
		\tens{H}{n_\lambda}_\lambda \in \ts{n_\lambda}{}\point
		\label{eq:invariant_definition}
	\end{equation}
\end{definition}
As scalar invariants are of particular relevance for formulating hyperelastic potentials, we will restrict our focus to this scenario throughout the paper. 
However,  we want to note that other authors also consider tensor-valued equivariants and/or invariants, e.g., \citep{zheng_theory_1994}.
We also want to note that equivariance and invariance are identical for the special case of scalar-valued functions.

\subsubsection{The concepts of integrity and functional bases}
We now discuss two important concepts that are central in invariant theory. These are introduced in a descriptive  manner in this section. 
Mathematically more precise definitions can be found in the glossary in \ref{sec:glossar}.

\begin{center}
	\fbox{
		\begin{minipage}{0.9\textwidth}
			For any group $\Group{G}\subseteq\Group{O}(3)$, there exists an infinite number of scalar-valued $\Group{G}$-invariants according to Def.~\ref{def:invariant}, which can be algebraic, rational, polynomial or homogeneous polynomial functions. 
			As shown by \citet{spencer_theory_1971}, it is useful to focus on homogeneous polynomial invariants. This is because algebraic invariants are solutions of algebraic equations with the coefficients being rational invariants. Furthermore, these rational invariants can be written as ratios of polynomial invariants, which can be expressed as a sum of homogeneous polynomial invariants. 
			This means that every $\mathcal G$-invariant function $f(\mathcal H)$ can be written as 
			\begin{align}
				f(\Set{H}) = g(I_1(\Set{H}), \ldots, I_n(\Set{H}))\; ,
			\end{align}
			with $I_1(\Set{H}), \ldots, I_n(\Set{H})$, $n\in\Setnum{Z}_{\geq1}$ being homogeneous polynomial invariants and $g$ an algebraic function (not necessarily polynomial).
		\end{minipage}
	}
\end{center}
Consequently, it becomes essential to identify a finite set of homogeneous polynomial invariants from which all other $\Group{G}$-invariant polynomials can be expressed as a polynomial function in the elements of that subset.

\begin{center}
	\fbox{
		\begin{minipage}{0.9\textwidth}
			A set $\{I_1,\ldots,I_n\}, n\in \Setnum{Z}_{\geq1}$, of homogeneous polynomial invariants such that any polynomial invariant can be written as a polynomial of $I_1,\ldots,I_n$ is referred to as an integrity basis~(Def.~\ref{def:integrity_basis}).\footnotemark\,
			If, in addition, none of the invariants in this set can be expressed as a polynomial of the others, the integrity basis is called minimal and its elements are called fundamental invariants (Def.~\ref{def:fundamental_invariant}).
			While a minimal integrity basis is not unique, the number of fundamental invariants with a given polynomial degree is always the same~\citep{spencer_theory_1971, goodman_symmetry_2009}\citep[Remark 4.2]{olive_minimal_2017}.
		\end{minipage}
	}
\end{center}
\footnotetext{The elements of an integrity basis are not restricted to be homogeneous polynomials. In fact, they are also allowed to be inhomogeneous. However, usually one focuses on homogeneous polynomial invariants, since this simplifies a lot of theory.}

Many algorithms deal with the task of determining minimal integrity bases, see for instance \citet{derksen_computational_2002, sturmfels_algorithms_2008}. However, there is another more general type of basis. 
\begin{center}
	\fbox{
		\begin{minipage}{0.9\textwidth}
			Functional bases~(Def.~\ref{def:function_basis}) consist of so-called separable invariants~(Def.~\ref{def:separating_set}) that can be used to distinguish between group orbits~\citep[Def. 4.5.]{olive_minimal_2017}. In practice, this means that every invariant function can be written as a function of the separable invariants~\citep{wineman_material_1964,pipkin_formulation_1959, pipkin_material_1963}. Unlike an integrity basis, the elements of a functional basis are not restricted to be polynomials. For compact groups, it also holds that every integrity basis is a functional basis, whereas the converse is generally false~\citep{zheng_theory_1994}~\citep[Appendix C]{abud_geometry_1983}. A functional basis in which none of its elements can be expressed as a function of another is called minimal. However, the cardinality of a minimal functional basis is not unique. One might find different minimal functional bases with larger or smaller cardinality~\citep[Remark 4.7]{olive_minimal_2017}.
		\end{minipage}
	}
\end{center}
We conclude that both integrity and functional bases are sufficient to express any polynomial $\Group{G}$-invariant as a function in their respective elements. The main difference is that, for an integrity basis, this has to hold true considering solely polynomial functions, while, for a functional basis, functions can be more general, e.g., rational or algebraic. 
Furthermore, even if no element of a minimal integrity basis can be expressed as a polynomial in the other elements of that basis, there might be a non-polynomial or functional dependence between the elements. This functional dependence may therefore be used to obtain a smaller set of invariants. By definition, such a set is no longer an integrity basis, but still a functional basis. Accordingly, it is often possible to find a functional basis with a smaller cardinality then an integrity basis.
\begin{center}
	\fbox{
		\begin{minipage}{0.9\textwidth}
			A set of invariants may be polynomially, functionally, or algebraically dependent or independent. 
			For example, the elements of minimal integrity bases are polynomially independent, while the elements of minimal functional bases are polynomially and even functionally independent. However, for both minimal integrity bases and minimal functional bases, the invariants may be algebraically dependent.
		\end{minipage}
	}
\end{center}
\subsubsection{Invariants in coordinate-dependent form}
\label{sec:approach_smith_rivlin}
In the context of hyperelasticity, \citet{smith_1958} were the first to compute integrity bases for all anisotropy types arising from the 32 point groups of the seven crystal systems. \citet{smith_further_1962} also proved later that the proposed invariants of \citet{smith_1958} are also minimal integrity bases.
This represents a significant advantage as it ensures that the number of invariants in terms of polynomial representation is minimal.
Similar, \citet{smith_transversely_1982} provided integrity bases for an arbitrary number of 1st order as well as symmetric and antisymmetric 2nd order tensors for all five groups related to the two cylindrical non-crystal systems (transversely isotropic). Therefore, also integrity bases for the two cylindrical anisotropies (transversely isotropic) are known. The minimal integrity bases determined by \citet{smith_further_1962} and \citet{smith_transversely_1982} are listed in the supplementary material.
Typically, these integrity bases are formulated by using theorems about symmetric polynomials \citep{smith_1958,smith_further_1962}.

\begin{figure}[t]
	\centering
	\includegraphics[clip]{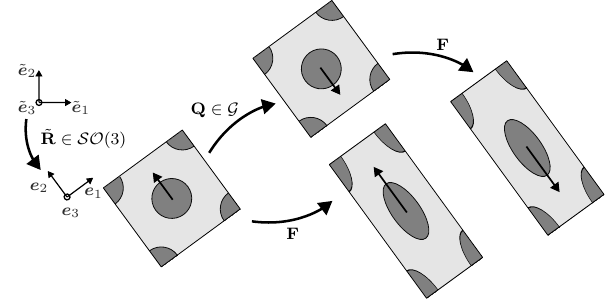}
	\caption{Principle of material symmetry using coordinate-dependent invariants. The use of invariants in coordinate-dependent form requires specification in a fixed reference frame, as the orientation of the transformation $\te{Q}\in\Group{G}$ is predefined. Therefore, a preliminary coordinate transformation using $\tilde{\te{R}}\in\Group{SO}(3)$ is necessary to ensure that the reference frame $\{\tilde{\ve{e}}_1, \tilde{\ve{e}}_2,\tilde{\ve{e}}_3\}$ coincides with that in which the invariants are formulated, i.e., $\ve{e}_i = \tilde{\te{R}} \cdot \tilde{\ve{e}}_i$ with $i\in\{1,2,3\}$.}	
	\label{fig:smith_invariants}
\end{figure}

Although these invariants are invariant under all transformations of a given group, as defined by Eq.~\eqref{eq:invariant_definition}, they also require a fixed orientation of the reference coordinate system. We refer to such invariants as coordinate-dependent invariants.
To apply this concept using Def.~\ref{def:invariant}, we define $\Set{H}=\left(\te{C}\right)$.\footnote{\label{foot:same_invariants}%
	Any suitable deformation or strain measure may be used instead of the right Cauchy-Green deformation tensor $\te{C}\in\tsposdef$, as long as it is a symmetric 2nd order tensor, which is not a two-point tensor. For example, the invariants of the Green-Lagrange strain tensor $\te{E}\in\tss{2}$ have exactly the same form as those formulated in $\te{C}$.} 
Before coordinate-dependent invariants can be used in a model, the given reference frame $\tilde{\ve e}_i \in \Setnum{R}^3$, $i\in\{1,2,3\}$, must be aligned with that in which the invariants are formulated, i.e.,  ${\ve{e}_i} = \tilde{\te R} \cdot \tilde{\ve{e}}_i \in \Setnum{R}^3$, $\tilde{\te{R}}\in\Group{SO}(3)$.
Thus, for the constitutive variable being the right Cauchy-Green tensor $\te{C} = \tilde C_{ij} \tilde{\ve e}_i \tilde{\ve e}_j =  C_{ij} \ve e_i \ve e_j$ we have its coordinate-dependent invariants given by
\begin{equation}
	I: \Setnum{R}^6 \to \Setnum{R},\quad (C_{11}, \ldots, C_{12}) \mapsto I(C_{11}, \ldots, C_{12}) 
	\quad\text{ with } \quad
	C_{ij} = \tilde R_{ia} \tilde R_{jb} \tilde C_{ab} \;.
	\label{eq:smith_needs_rotation}
\end{equation}
The necessary passive rotation~$\tilde{\te{R}}$ (coordinate transformation) in context of material symmetry is illustrated in Fig.~\ref{fig:smith_invariants}.

\begin{center}
	\fbox{
		\begin{minipage}{0.9\textwidth}
			By using the concept of coordinate-dependent invariants, we can represent the Helmholtz free energy based on $r\in \Setnum{Z}_{>0}$ invariants $I_\alpha(C_{11}, \ldots, C_{12})$, $\alpha\in\{1,\ldots,r\}$, as 
			\begin{align}
				\varPsi_\Group{G}:
				\Setnum{R}^6 \to \Setnum{R}_{\geq 0} \; , \; (C_{11}, \ldots,  C_{12})
				\mapsto
				\varPsi_\Group{G}(I_1(C_{11}, \ldots, C_{12}), \ldots, I_r(C_{11}, \ldots, C_{12})) \; .
				\label{eq:free_energy_coordinate}
			\end{align}
			From Eq.~\eqref{eq:free_energy_coordinate} it directly follows that the Helmholtz free energy formulated in this way is a $\mathcal G$-invariant function.
		\end{minipage}
	}
\end{center}

\subsubsection{Structural tensor concept}
\label{sec:structural_tensor_concept}
A completely different approach is the structural tensor concept, which is favored in modern work~\citep{schroder_invariant_2003,Apel2004,ehret_polyconvex_2007,ogden_introducing_2007,bustamante_transversely_2010,reese_using_2021,klein_polyconvex_2022,wollner_general_2023,Ciambella2024,kalina_neural_2024,holthusen_complement_2025} and can be traced back to the work of \citet{Lokhin1963}. Descriptively speaking, the idea is to extend the list of arguments of the potential $\varPsi$ by a tuple
\begin{equation}
	\Set{M}_\Group{G} = \left( \tens{M}{n_1}_1, \ldots,  \tens{M}{n_l}_l \right)
	\label{eq:structure_functions_set}
\end{equation}
of $l\in\Setnum{Z}_{\ge 1}$ structural tensors $\tens{M}{n_\mu}_\mu\in\ts{n_\mu}{}$, $\mu \in\{1,\ldots,l\}$, to obtain scalar-valued isotropic tensor functions. The structural tensors are Lagrangian tensors, i.e., attached to the reference configuration $\Set B_0$, and defined such that they describe the material's underlying symmetry.
For a given material, structural tensors are constant, that is, they are independent of motion and time.
\footnote{As is common practice, we can define the push-forward of the structural tensors to the current configuration $\Set{B}_\tau$ using the rotational part $\te R\in\Group{SO}(3)$ of $\te F = \te R \cdot \te U$, i.e., by $\te R \star \tens{M}{n_\mu}_\mu$. This can be understood to mean that only the change in orientation during the transformation plays a role. With this choice of the push-forward, norms of structural tensors or (generalized) traces of structural tensors of even degree also remain constant. Thus, the push-forward of the structural tensors, different from the original Lagrangian structural tensors, are not independent of motion.
	However, the push-forward operation is usually not necessary, since the constitutive equations are typically formulated with $\te{C}$, which is related to the reference.}

\paragraph{Invariance of structural tensors}
Let us now consider how the structural tensors behave under orthogonal transformations. Since they describe the underlying material symmetry in the reference configuration $\Set{B}_0$, this can be interpreted as a rotation of $\Set{B}_0$. 

\begin{definition}[Tuple of structural tensors]
	\label{def:structural_tensors}
	Let $\tens{M}{n_\mu}_\mu \in \ts{n_\mu}{}$ with $\mu \in\{1,\ldots,l\}$, $l\in\Setnum{Z}_{\ge 1}$ tensors of order $n_\mu \in \Setnum{Z}_{\ge 1}$ that are  gathered in the tuple $\Set{M}_\Group{G} = \left( \tens{M}{n_1}_1, \ldots,  \tens{M}{n_l}_l \right)$. 
	The tuple $\Set{M}_\Group{G}$ is called tuple of structural tensors of the group $\Group{G}\subseteq\Group{O}(3)$, if it satisfies the condition
	\begin{equation}
		\Set{M}_\Group{G} = \te{Q}\star\Set{M}_\Group{G} \Leftrightarrow\te{Q}\in\Group{G}\subseteq\Group{O}(3) \; .
		\label{eq:struc_ten_set_trafo}
	\end{equation}
\end{definition}

\begin{center}
	\fbox{
		\begin{minipage}{0.9\textwidth}
			Descriptively, the property~\eqref{eq:struc_ten_set_trafo} can be interpreted as follows.
			If the tuple $\Set{M}_\Group{G}$ includes only a single structural tensor $\tens{M}{n}$, i.e., $l=1$, this tensor has to satisfy the condition
			\begin{equation}
				\tens{M}{n} = \te{Q} \star \tens{M}{n}\Leftrightarrow\te{Q}\in\Group{G}\subseteq\Group{O}(3) \commar
				\label{eq:structure_tensor_condition}
			\end{equation}
			thereby completely capturing the anisotropy of a material.\footnotemark\,%
			In the general case, the tuple $\Set{M}_\mathcal{G} = (\tens{M}{n_1}_1, \dots, \tens{M}{n_l}_l)$ may consist of multiple structural tensors.
			Then, Eq.~\eqref{eq:struc_ten_set_trafo} implies
			\begin{equation}
				\tens{M}{n_\mu}_\mu = \te{Q} \star \tens{M}{n_\mu}_\mu\:\forall\:\te{Q}\in\Group{G}\subseteq\Group{O}(3)
				\quad \forall \mu \in \{1, \dots, l\}
				\commar
				\label{eq:structural_tensor_multiple_condition}
			\end{equation}
			i.e., each tensor must (at least) be invariant under all transformations of the symmetry group $\Group{G}\subseteq\Group{O}(3)$.
			In addition, for several $\mu \in \{1, \dots, l\}$, $\tens{M}{n_\mu}_\mu$ can also be invariant with respect to transformations $\te Q \in \Group{O}(3)\setminus\Group{G}$ that are not elements of the symmetry group.
			However, for every transformation $\te Q \in \Group{O}(3)\setminus\Group{G}$ there has to exist (at least) one tensor in the tuple that is not invariant, i.e., $\forall\: \te Q \in \Group{O}(3)\setminus\Group{G}\;\exists\: \mu \in\{1,\ldots,l\}$ such that $\te Q \star \tens{M}{n_\mu}_\mu \ne \tens{M}{n_\mu}_\mu$.
			In other words, the set of transformations that leave the entire tuple $\Set{M}_\Group{G}$ of structural tensors invariant must coincide with the group $\Group{G}$.
		\end{minipage}
	}
\end{center}
\footnotetext{
	Note that Eq.~\eqref{eq:structure_tensor_condition} is equivalent to $\tens{M}{n} = \te{Q} \star \tens{M}{n} \: \forall \te{Q}\in\Group{G}\subseteq\Group{O}(3) \:\land\: \tens{M}{n} \neq \te{Q} \star \tens{M}{n} \: \forall \te Q \in \Group{O}(3)\setminus\Group{G}$.}

\paragraph{Isotropic extension of the elastic potential}
By using the tuple of structural tensors as additional arguments in the Helmholtz free energy, an anisotropic elastic potential $\varPsi(\te C)$ is replaced by a scalar-valued isotropic tensor function 
\begin{align}
	\varPsi:
	\tsposdef \times  \ts{n_1}{} \times \dots \times \ts{n_\mu}{} \times \dots \ts{n_l}{} \to \Setnum{R}_{\geq 0} \; , \; \left(\te{C}, \Set{M}_\Group{G}\right) \mapsto \varPsi\left( \te{C}, \Set{M}_\Group{G} \right)
\end{align}
of multiple arguments, with $n_\mu \in \zset_{\geq 1}$ for all $\mu\in\{1,\ldots,l\}$. Thus, the material symmetry condition~\eqref{eq:material_symmetry} becomes
\begin{equation}
	\varPsi\left( \te{Q} \star \te{C}, \te{Q}\star\Set{M}_\Group{G} \right) = \varPsi\left( \te{C}, \Set{M}_\Group{G} \right) \:\forall\:\te{Q}\in\Group{O}(3) \point
	\label{eq:material_symmetry_isotropic}
\end{equation}
This procedure is called isotropic extension~\citep{Xiao1996, Apel2004} and is also known as Rychlewski's theorem~\citep{zhang_structural_1990, itskov_tensor_2025, man_further_2026}.

\begin{figure}[t]
	\centering
	\includegraphics[clip]{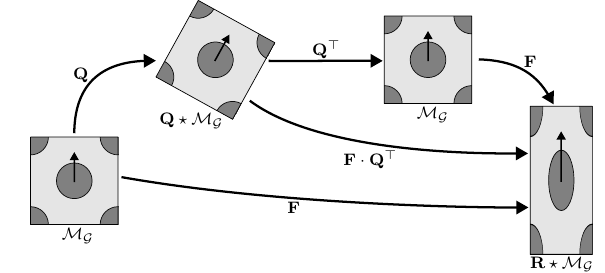}
	\caption{Illustration of the structural tensor concept for obtaining isotropic tensor functions. The material is actively rotated by applying the transformation $\te{Q}\in\Group{O}(3)$ to the structural tensors in $\Set{M}_\Group{G}$. This rotation is then reversed using $\te{Q}^\top$, thereby ensuring that the principle of material symmetry is satisfied $\forall\:\te{Q}\in\Group{O}(3)$. $\te{R} = \te{F}\cdot\te{U}^{-1}$ denotes the rotational component of the deformation gradient.}	
	\label{fig:structure_invariants}
\end{figure}

In order to clearly analyze the connection to the movement of a body, let us now consider the energy density $\varPsi\left( \te{F}, \Set{M}_\Group{G} \right)$ for which we find $\varPsi\left( \te{F}, \Set{M}_\Group{G} \right)=\varPsi\left(\te{F} \cdot \te Q^\top, \te Q \star \Set{M}_\Group{G} \right) \:\forall\: \te{Q}\in\Group{SO}(3)$. Based on this, Fig.~\ref{fig:structure_invariants} provides a visual representation of this relationship: First, the body is rotated by an arbitrary $\te Q\in \Group{SO}(3)$ and thus the structural tensors are actively transformed by $\te{Q}\star\Set{M}_\Group{G}$. Then, the deformation $\te{F} \cdot \te Q^\top$ is applied, where $\te{Q^\top}$ rotates the body in the opposite direction, thereby restoring the original non-rotated configuration for all $\te{Q}\in\Group{SO}(3)$, and afterwards $\te F$ is applied. This is therefore the same as if we directly apply the deformation $\te F$ to the non-rotated configuration with the non-transformed tuple of structural tensors $\Set{M}_\Group{G}$.
As a result, including $\Set{M}_\Group{G}$ ensures that $\varPsi\left( \te{F}, \Set{M}_\Group{G} \right)$ is an isotropic tensor function. The same holds for $\varPsi\left( \te{C}, \Set{M}_\Group{G} \right)$.
\paragraph{Invariants with structural tensors}
To formulate the elastic potential, invariants of $\te{C}$ and $\Set{M}_{\mathcal{G}}$, so-called isotropic invariants, are formed. Thus, the tuple $\Set{H}$ used in Def.~\ref{def:invariant} is specified below by the chosen constitutive variable and the structural tensors.
\begin{definition}[Isotropic invariant]
	\label{def:isotropic_invariant}
	An invariant $I$, which obeys the condition
	\begin{equation}
		I(\Set{H}) = I(\te{Q}\star\Set{H}) \quad \forall\:\te{Q}\in\Group{O}(3) \commar
		\label{eq:invariants_structural}
	\end{equation}
	is called isotropic invariant.
\end{definition}
In hyperelasticity, $\Set{H}=\Set{H}_{\Group{G}}$ consists of a constitutive variable, such as $\te{C}$, and one or multiple structural tensors gathered in the tuple $\Set{M}_{\mathcal{G}}$, i.e., $\Set{H}_{\Group{G}} = \left(\te{C},\tens{M}{n_1}_1,\ldots,\tens{M}{n_l}_l\right)$. To indicate that the tuple now contains the structural tensor(s) belonging to the group $\Group G$, we add $(\cdot)_{\Group G}$ as a subscript.
If either the constitutive variable, here $\te{C}$, or the tuple $\Set{M}_\Group{G}$ of structural tensors is transformed separately by any $\te{Q}\in\Group{G}$, the conditions
\begin{subequations}
	\begin{align}
		I(\te{C},\Set{M}_\Group{G}) &= I(\te{Q}\star\te{C},\Set{M}_\Group{G}) \: \forall\: \te{C}\in\tsposdef \Leftrightarrow \te{Q}\in\Group{G} \subseteq\Group{O}(3) \; \textrm{and}\label{eq:struct_fix}\\
		I(\te{C},\Set{M}_\Group{G}) &= I(\te{C},\te{Q}\star\Set{M}_\Group{G}) \: \forall\: \te{C}\in\tsposdef \Leftrightarrow \te{Q}\in\Group{G}\subseteq\Group{O}(3)
	\end{align}
	\label{eq:structural_invariants_properties}
\end{subequations}
follow, which again restrict the admissible transformations to the corresponding group $\Group{G}$. 

\begin{remark}
	Note that invariants constructed via the structural tensor concept and isotropic extension are equivalent to coordinate-dependent invariants when the structural tensors $\Set{M}_\Group{G}$
	are fixed, cf. Eq.~\eqref{eq:struct_fix}. To be more precise, by evaluating the structural tensors $\Set{M}_\Group{G}$ and $\te C$ in a given reference frame, these invariants reduce to the coordinate-dependent form defined in Eq.~\eqref{eq:smith_needs_rotation}.
	To illustrate this equivalence, we consider the invariant $I=\tr(\te M \cdot \te C)$ defined via a structural tensor $\te M\in \tss{2}$. Let $\te M$ be represented with respect to the standard basis $\{\ve {e}_1, \ve {e}_2, \ve {e}_3\}$ such that its only non-zero component is $M_{11}=1$, while all other components vanish. In this case, the invariant reduces to $I=\tr(\te M \cdot \te C)= M_{ij} C_{ji}=C_{11}$, which corresponds to a coordinate-dependent invariant of $\te C$.
\end{remark}
We also find that it follows
\begin{align}
	\diffp{I(\te X,\Set{Y})}{\te X} \Bigg|_{(\te Q \star \te{C},\Set{M}_\Group{G})} = \te Q \star \diffp{I(\te{X},\Set{Y})} {\te X} \Bigg|_{(\te{C},\Set{M}_\Group{G})} 
	&\Leftrightarrow \te{Q}\in\Group{G}\subseteq\Group{O}(3) \; \textrm{and}\\[3pt]
	\diffp{^2 I(\te X,\Set{Y})}{\te X\partial \te X} \Bigg|_{(\te Q \star \te{C},\Set{M}_\Group{G})} = \te Q \star \diffp{^2 I(\te X,\Set{Y})}{\te X\partial \te X} \Bigg|_{(\te{C},\Set{M}_\Group{G})}
	&\Leftrightarrow \te{Q}\in\Group{G}\subseteq\Group{O}(3) \; .
	\label{eq:structural_invariants_tensors_properties}
\end{align}
This means that tensor-valued derivatives\footnote{For mappings whose domain is the space of symmetric 2nd order tensors $\tss{2}$, we follow the approach proposed by \citet[p.~27]{truesdell_non-linear_1965} and later \citet[Sect.~6]{itskov_tensor_2025} to calculate the Fr\'{e}chet derivative, see also \citet[Appendix A.1]{dammas_phase-field_2026}.} of a scalar invariant
$I(\te{C},\Set{M}_\Group{G})$ are $\Set{G}$-equivariant (Def.~\ref{def:equivariant}) when the group only acts on $\te C$ (with structural tensors $\Set{M}_\Set{G}$ fixed). However, they become $\Group{O}(3)$-equivariant when the group acts on both $\te C$ and $\Set{M}_\Group{G}$.

\begin{center}
	\fbox{
		\begin{minipage}{0.9\textwidth}
			The formulation of the elastic potential in terms of $r\in \Setnum{Z}_{>0}$ structural tensor-based invariants $I_\alpha(\te{C}, \Set{M}_\Group{G})$ with $\alpha\in\{1,\ldots,r\}$ reads
			\begin{align}
				\varPsi_{\Group{G}}:
				\tsposdef \times  \ts{n_1}{} \times \dots \times \ts{n_\mu}{} \times \dots \ts{n_l}{} \to \Setnum{R}_{\geq 0} \; , \; \left(\te{C}, \Set{M}_\Group{G}\right) \mapsto \varPsi_{\Group{G}}\left(I_1(\te{C}, \Set{M}_\Group{G}), \ldots, I_r(\te{C}, \Set{M}_\Group{G}) \right) \; .
				\label{eq:free_energy_strucral}
			\end{align}
			It follows directly from Eq.~\eqref{eq:free_energy_strucral} that the properties~\eqref{eq:invariants_structural} and  \eqref{eq:structural_invariants_properties} are also satisfied by the free energy defined in this way. The potential is thus $\Set{G}$-invariant when the group only acts on $\te C$ (with structural tensors $\Set{M}_\Set{G}$ fixed), and becomes $\Group{O}(3)$-invariant when the group acts on both $\te C$ and $\Set{M}_\Group{G}$.
			Consequently, the properties~\eqref{eq:structural_invariants_tensors_properties} are also fulfilled by the 2nd Piola-Kirchhoff stress tensor and the associated tangent operator both derived from the potential~\eqref{eq:free_energy_strucral}:
			\begin{align}
				\te T_\Group{G}(\te C, \Set{M}_\Group{G})=2\diffp{\psi_\Group{G}(\te C, \Set{M}_\Group{G})}{\te C}
				\quad \text{and} \quad 
				\tens{C}{4}_\Group{G}(\te C, \Set{M}_\Group{G}) = 4 \diffp{{}^2\psi_\Group{G}(\te C, \Set{M}_\Group{G})}{\te C\partial \te C} \; .
			\end{align}
			These quantities are thus $\Set{G}$-equivariant when the group only acts on $\te C$ (with structural tensors $\Set{M}_\Set{G}$ fixed), and become $\Group{O}(3)$-equivariant when the group acts on both $\te C$ and $\Set{M}_\Group{G}$.
	\end{minipage}}
\end{center}

\subsubsection{Isotropic invariants of multiple tensors}
\label{ssec:invariant_polynomials}
Since we consider isotropic scalar-valued tensor functions, results on isotropic invariants from multiple tensors can be used directly. 
One of the most important works in this field is the one by \citet{Boehler1977}, which provides a functional basis of isotropic homogeneous polynomial invariants. Thereby, the original framework is valid for any number of 1st order tensors $\ve{v}_{\alpha}\in\ts{1}{}$, symmetric 2nd order tensors $\te{A}_{\beta}\in\tss{2}$ and skew-symmetric 2nd order tensors $\te{W}_{\gamma}\in\tsa{2}$, cf. Tab.~\ref{tab:boehler_table}.
To maintain consistency with the original work of \citet{Boehler1977}, we have chosen the same notation for the introduced tensors above.

\begin{center}
	\fbox{
		\begin{minipage}{0.9\textwidth}
			Since \citet{Boehler1977} only considered scalar-valued invariants of 1st and 2nd order tensors, the tuple $\Set{M}_\Group{G}$ is naturally restricted to structural tensors of at most 2nd order when using the results from Tab.~\ref{tab:boehler_table}.
		\end{minipage}
	}
\end{center}
Thus, to this point, a functional basis can be obtained from the results of \citet{Boehler1977} for all cases with 1st and 2nd order structural tensors.
\begin{center}
	\fbox{
		\begin{minipage}{0.9\textwidth}
			In general, according to \citet{Pennisi1987}, the functional basis of~\citet{Boehler1977} is minimal.
			However, Boehler’s basis was formulated for non-constant arguments or tensors. Therefore, functional and also polynomial dependencies between invariants arise from the fact that structural tensors are constant~\citep{boehler_simple_1979}. In turn, this implies that a functional basis for a specific anisotropy obtained from Tab.~\ref{tab:boehler_table} is often not minimal~\citep{Apel2004,Ebbing2010} and consists of a large number of invariants.
		\end{minipage}
	}
\end{center}
As a result, numerous works~\citep{Apel2004,Ebbing2010,kalina_neural_2024} are limited to invariants of lower polynomial degree in order to reduce the number of invariants in a material model.

\begin{center}
	\fbox{
		\begin{minipage}{0.9\textwidth}
			Since the invariants of \citet{Boehler1977} are homogeneous polynomials, the question naturally arises whether the functional bases also constitute integrity bases. In general, this is not the case. However, for the class of hyperelasticity problems considered in this work, it turns out that the functional bases are also minimal integrity bases in many cases. This useful observation is discussed in detail in Sect.~\ref{sec:construction_of}.
		\end{minipage}
	}
\end{center}

\section{Reducing the cardinality of functional bases by finding polynomial relations}
\label{sec:construction_of}
In the following, we address three important points that were raised in Sect.~\ref{ssec:invariant_polynomials}:
\begin{enumerate}[label=(\Roman*)]
	\item We want to define functional bases using on structural tensors of order greater than two. 
	\item We would like to reduce the cardinality of functional bases when using structural tensors.
	\item Finally, we want to verify whether the identified reduced functional bases are also minimal integrity bases.
\end{enumerate}
To tackle point~(\Romannum{1}), Sect.~\ref{sec:constructing_xiao}   first introduces the approach by \citet{Xiao1996} to exchange the tuple of structural tensors $\Set{M}_\Group{G}$ with a tuple $\mathcal F_{\mathcal G}$ of so-called structural functions. This reformulation allows the results of \citet{Boehler1977} to be extended to cases involving structural tensors of 3rd, 4th, or higher order.

The cardinality of the functional bases obtained in this way can initially be very large. By searching for functional relationships between the invariants of the functional basis, the number of elements in these functional bases can be reduced. Next, point~(\Romannum{2}) is addressed in Sect.~\ref{sec:elimination}. Thereby, we limit ourselves to the special case of determining whether one invariant of the functional basis can be expressed as a polynomial in the others. If so, that element is not needed as an element of a functional basis.

Finally, point~(\Romannum{3}) is discussed in Sect.~\ref{sec:proof_minimal_integrity_basis}.
To show that the reduced functional bases, which are sets of homogeneous polynomial invariants, are indeed  minimal integrity bases, we introduce two approaches. The most straight-forward way is by referring to well-known results on minimal integrity basis from literature. Since this is not always possible, we further introduce some concepts from invariant theory.

The introduced concepts are subsequently illustrated by means of two specific examples; In Sect.~\ref{sec:example_monoclinic} we discuss the monoclinic system ($\mathcal A_2$) and in Sect.~\ref{sec:example_cubic} we consider a cubic system ($\mathcal A_{10}$).
\subsection{Constructing functional bases via Xiao's approach}
\label{sec:constructing_xiao}
The result of~\citet{Boehler1977} for constructing functional bases, summarized in Tab.~\ref{tab:boehler_table}, is restricted to 1st and 2nd order tensors. However, since not all anisotropies can be represented by structural tensors that are of at most 2nd order~\citep{xiao_2006}, the tuple $\Set{M}_\Group{G}$ of structural tensors was generalized by \citet{Xiao1996} to a tuple $\Set{F}_\Group{G}$ of structural functions
\begin{equation}
	\Set{F}_\Group{G}(\te{C},\Set M_\Group{G}) = \left( \ve{v}_1(\te{C},\Set M_\Group{G}), \dots, \ve{v}_{a}(\te{C},\Set M_\Group{G}), \te{A}_{1}(\te{C},\Set M_\Group{G}), \dots, \te{A}_{b}(\te{C},\Set M_\Group{G}), \te{W}_1(\te{C},\Set M_\Group{G}), \dots, \te{W}_{c}(\te{C},\Set M_\Group{G}) \right) \; , \; 
	\label{eq:structure_functions_set2}
\end{equation}
$a,b,c\in\Setnum{Z}_{\ge 0}$.\footnote{
	Recently, \citet{man_remarks_2018, man_further_2026} proposed a reformulation of the structural tensor concept to enable a formulation of isotropic functions using 1st and 2nd order tensors only, see also~\citet{madadi_representation_2025-1, madadi_representation_2025} or \citet[p.153]{itskov_tensor_2025}. Contrary to Def.~\ref{def:structural_tensors}, in this reformulation, the tuple $\Group{M}_\Group{G}$ is interpreted as a set, which has the consequence that the requirements for the individual structural tensors are relaxed. They no longer need to be invariant under the transformations of a group. Scalar-valued quantities formed using these structural tensors are therefore generally not invariants. Consequently, based on how the structural tensors change under group transformations, these scalar-valued quantities must be combined to construct invariants. This additional step does not arise in the structural tensor concept pursued in this work.}
The tensors $\ve{v}_{\alpha}(\te{C},\Set M_\Group{G})\in\ts{1}{}$, $\te{A}_{\beta}(\te{C},\Set M_\Group{G})\in\tss{2}$ and $\te{W}_{\gamma}(\te{C},\Set M_\Group{G})\in\tsa{2}$ can now be tensor functions depending on $\te{C}$ and a tuple of structural tensors $\Set{M}_\Group{G} = \left( \tens{M}{n_1}_1, \ldots,  \tens{M}{n_l}_l \right)$, where the order of the individual structural tensors $\tens{M}{n_\mu}_\mu \in \ts{n_\mu}{}$, $\mu \in\{1,\ldots,l\}$, $l\in\Setnum{Z}_{\ge 1}$ can be greater than two. Hence, using Tab.~\ref{tab:boehler_table} of \citet{Boehler1977} together with the tuple $\Set{F}_\Group{G}$, a functional basis can be constructed. However, due to the presence of constant structural tensors, this basis may contain numerous invariants that are not functionally or even polynomially independent.
\begin{remark}
	\label{rem:caley_hamiltion}
	In the examined cases, it is sufficient to only consider structural functions up to the order of $\te{C}^2$, cf. \citet{kalina_neural_2024}. For example, structural tensor functions can take the form $\tens{M}{4}:\te{C}$, $\tens{M}{4}:\te{C}^2$, $\tens{M}{4}:\te{C}^3$, where $\tens{M}{4}\in\ts{4}{}$ denotes a constant 4th order structural tensor. By using the Cayley-Hamilton theorem,
	\begin{equation}
		\te{C}^3 = I_1\te{C}^2 - I_2\te{C} + I_3\te{1}
		\label{eq:caley_hamilton}
	\end{equation}
	follows. When Eq.~\eqref{eq:caley_hamilton} is multiplied by $\tens{M}{4}$, it becomes evident that structural functions $\tens{M}{4}:\te{C}^n$ up to at most 2nd order ($n=2$) are sufficient, as higher-order terms ($n>2$) can be expressed in terms of lower-order terms $\tens{M}{4}:\te{C}^n, n\in\left\{0,1,2\right\}$ and the three well-known isotropic or main invariants
	\begin{equation}
		\textnormal
		I_1 = \tr[\te{C}], \quad
		I_2 = \frac{1}{2}\left( \tr^2[\te{C}] - \tr[\te{C}^2] \right) \quad \textrm{and} \quad
		I_3 = \det[\te{C}] \point
		\label{eq:isotropic_invariants}
	\end{equation}
\end{remark}
\subsection{Elimination of polynomially expressible invariants}
\label{sec:elimination}
We are interested in finding polynomial relations between invariants in order to sort out invariants that can be expressed as a polynomial of other invariants.\footnote{
	\citet[Sect. 5]{betten_irreduzible_1992} introduced a combinatorial algorithm to count reducible and irreducible homogeneous polynomial invariants of a 4th order tensor. Finding polynomial relations between invariants is more involved, since also an irreducible invariant, which is not expressible as a product of non-constant invariants of lower degrees, may still be expressible as a polynomial of invariants of lower degrees. Therefore, finding polynomial relations can be seen as an extension of the algorithm of \citet[Sect. 5]{betten_irreduzible_1992}.} 
Therefore, we assume that we have a set $\left\{ \inv{n_1}_1,\inv{n_2}_2,\dots,\inv{n_L}_L \right\}$, $L\in\Setnum{Z}_{\ge 1}$, of homogeneous polynomial invariants, which may not all be polynomially independent. Furthermore, we define $\Set{X}_\Group{G}$ as the tuple of these invariants sorted in ascending order of polynomial degree $n_\nu\in\Setnum{Z}_{>0}$. 
At this point, it is not yet of relevance that the elements of $\Set{X}_\Group{G}$ also form an integrity basis. This has to be verified after the following reduction process was executed.
In order to determine which invariants are polynomially expressible, it is necessary to identify the polynomial relations of the form
\begin{equation}
	\inv{n_\nu}_\nu = f\left( \inv{n_1}_1,\inv{n_2}_2,\dots,\inv{n_{\nu-1}}_{\nu-1},\inv{n_{\nu+1}}_{\nu+1},\dots,\inv{n_L}_{L} \right)
	\commar
	\label{eq:ansatz_syzygies}
\end{equation}%
where $f$ denotes a homogeneous polynomial in a finite number of invariants, which themselves depend on multiple variables. Here, each invariant is a function of the coordinates $C_{11}, C_{22}, C_{33}, C_{23}, C_{13}, C_{12}$ of the symmetric 2nd order tensor $\te{C}\in\tsposdef$ w.r.t. the standard basis $\{\ve e_1, \ve e_2, \ve e_3\}$.
Since identifying polynomial relationships between invariants can become arbitrarily complex, we use a simple framework for determining such relationships using an analytical-numerical approach in the following. We explicitly note that this framework can be inefficient for a high number of invariants in the tuple $\Set{X}_\Group{G}$ and more sophisticated algorithms would then be necessary, cf. \citet{sturmfels_algorithms_2008, derksen_computational_2002, desmorat_computation_2023, lercier_hyperelliptic_2012, lercier_covariant_2017, olive_geometrie_2014, olive_minimal_2017, olive_characterization_2022, taurines_modelisation_2022}. However, for the present cases the following approach has revealed to be sufficient.
\subsubsection{General approach}
\label{sec:analytical_approach}
Initially, the invariant $\inv{n_1}_1$ with $n_1\in\Setnum{Z}_{\geq0}$ is selected as an element ${}^{n_1}\tilde I_1=\inv{n_1}_1$ of the preliminary tuple $\Set{Y}_\Group{G}=\left( {}^{n_1}\tilde I_1 \right)$ of polynomially independent invariants. Subsequently, the next invariant $\inv{n_2}_2$ from the tuple $\Set{X}_\Group{G}$ is considered, and a complete polynomial ansatz $\inv{n_2}_2=f\left( {}^{n_1}\tilde I_1 \right)$ is made to express it as a polynomial in the invariants of $\Set{Y}_\Group{G}$, where each summand of $f\left( {}^{n_1}\tilde I_1 \right)$ has degree $n_2$. If $\inv{n_2}_2$ can be represented as a polynomial in the invariants of $\Set{Y}_\Group{G}$, the invariant is considered trivial and the procedure is repeated for the next invariant $\inv{n_3}_3$ in $\Set{X}_\Group{G}$. Otherwise, ${}^{n_2}\tilde I_2=\inv{n_2}_2$ is added to tuple $\Set{Y}_\Group{G}$, and the polynomial ansatz is extended accordingly for the remaining invariants. By applying this procedure to all remaining invariants, a set $\Set{J}_\Group{G}$ of polynomially independent invariants is obtained, whose elements are the components of the final tuple $\Set Y_\Group{G}$.

\subsubsection{Analytical-numerical approach}
\label{sec:elimination_implementation}
While the overall procedure in Sect.~\ref{sec:analytical_approach} is well understood and also well-known in the literature, cf.~\citet[Algorithm 2.6.1]{derksen_computational_2002}, the primary challenge lies in identifying the polynomial relationships~\eqref{eq:ansatz_syzygies}, or to be more precise, the coefficients of the complete polynomial ansatz. Therefore, we use a simple method for determining the unknown coefficients for each invariant $\inv{n_\mu}_{\mu}$ out of $\Set X_\Group{G}$. For the $\mu$th invariant we start with a numerical step as an initial guess, which is followed by an analytical step using symbolic computations.

\paragraph{Numerical step for the $\mu$th invariant}
Within this paragraph, we do not use the summation convention.
First, for a given group or anisotropy, the tuple $\Set{X}_\Group{G}$ of $L$ invariants is assembled based on the corresponding tuple $\Set{F}_\Group{G}$ of structural functions and by using Tab.~\ref{tab:boehler_table} of \citet{Boehler1977}. Subsequently, the invariants 
\begin{equation}
	\inv{n_\mu}_{\mu,\Lambda}=\inv{n_\mu}_{\mu}\left(\te{C}_\Lambda\right)
	\; , \; \mu\in\{1,2,\ldots,L\}
	\label{eq:invariants_evaluated}
\end{equation}
are evaluated for multiple randomly generated instances $\Lambda$ of the tensor $\te{C}$, where we use $\Lambda\in\{1,2,\ldots,N\}$ for labeling.
Accordingly, the required structural tensors are initialized randomly in accordance with their prescribed construction rules and then kept constant throughout.
The coordinates of these instances are drawn from a normal distribution with a mean of 0 and a standard deviation of 1, where symmetry of $\te{C}$ is ensured but not positive definiteness.\footnote{Positive definiteness does not affect the algorithm, as the invariants of \citet{Boehler1977} are not subject to this constraint either.} Hence, $\inv{n_\mu}_{\mu,\Lambda}\in\rset$ represents a vector ${}^{n_\mu}\V{I}{}_\mu\in\rset^{N\times 1}$. Then, a complete polynomial ansatz for the $\mu$th invariant from $\Set{X}_\Group{G}$ is made:
\begin{equation}
	\inv{n_\mu}_{\mu,\Lambda} = \sum_{\alpha=1}^P {}^{n_\mu}J_{\mu,\Lambda\alpha}\: a_{\mu,\alpha} \quad\textrm{for all states}\:\Lambda \quad \leftrightarrow \quad
	{}^{n_\mu}\V{I}{}_\mu = {}^{n_\mu}\M{J}{}_{\mu} \cdot \V{a}{}_\mu
	\; .
	\label{eq:general_ansatz}
\end{equation}
Thereby, $J_{\mu,\Lambda\alpha}\in\rset$ represent the coefficients of  a matrix ${}^{n_\mu}\M{J}{}_\mu\in\rset^{N\times P}$, with $P$ denoting the number of terms in the ansatz, and $\V{a}{}_{\mu}\in\rset^{P\times1}$ a vector of unknown coefficients $a_{\mu,\alpha}\in\rset$. Each column of the matrix represents an argument in the ansatz, i.e., an invariant
\begin{equation}
	{}^{n_\mu}J_{\mu,\Lambda\alpha} = 
	\prod_{\nu\in\Set{A}_{\mu,\alpha}}
	{}^{n_\nu}\tilde I_{\nu}(\te C_\Lambda)\;
	,\; {}^{n_\nu}\tilde I_{\nu} \in \Set{Y}_\Group{G} \quad\textrm{with}\quad
	\sum_{\nu\in\Set{A}_{\mu,\alpha}}
	n_{\nu}=n_\mu \commar
	\label{eq:ansatz_monomials}
\end{equation}
that is constructed from a (sub)tuple of the polynomially independent invariants ${}^{n_\nu}\tilde I_\nu$ within the current $\Set{Y}_\Group{G}$ for all instances $\Lambda$. 
Thereby, the tuple $\Set{A}_{\mu,\alpha}$ contains the indices $\nu\in\Set{A}_{\mu,\alpha}$ such that the polynomial degree of the product in the polynomially independent invariants ${}^{n_\nu}\tilde I_\nu$ equals the degree $n_\mu$ of the $\mu$th invariant from $\Set X_\Group{G}$. Since multiplication is commutative, it follows from Eq.~\eqref{eq:ansatz_monomials} that of all possible tuples $\Set{A}_{\mu,\alpha}$ only one realization of every allowed permutation is required. Hence, the index $\alpha$ goes from $1$ to the number of different allowed permutations $P$.

To determine the coefficients $\V a{}_\mu$ for the $\mu$th invariant, we formulate the linear optimization problem 
\begin{align}
	\hat{\V a}{}_{\mu} = \underset{\V a{}_{\mu}\in\rset^{P\times 1}}{\arg\min} \left|
	{}^{n_\mu}\V{I}{}_\mu - {}^{n_\mu}\M{J}{}_{\mu} \cdot \V{a}{}_\mu
	\right|^2 
	\label{eq:opt_prob}
\end{align}
that yields the necessary condition $({}^{n_\mu} \M{J}{}_{\mu})^\top \cdot \left[{}^{n_\mu}\V{I}{}_\mu - {}^{n_\mu} \M{J}{}_{\mu}\cdot\V{a}{}_\mu\right] = \V 0 $. We solve the problem given above by using NumPy’s least squares method.
If all entries 
\begin{align}
	\left({}^{n_\mu}J_{\mu,\Lambda1}, {}^{n_\mu}J_{\mu,\Lambda2},\ldots,{}^{n_\mu}J_{\mu,\Lambda\alpha},\ldots,{}^{n_\mu}J_{\mu,\Lambda P}\right)
\end{align}
obtained from Eq.~\eqref{eq:ansatz_monomials} are linear or polynomially independent, it follows that the determined coefficients $\hat a_{\mu,\alpha}$ of the linear optimization task are unique. After solving the problem~\eqref{eq:opt_prob}, the criterion
\begin{equation}
	R_\mu = \underset{\Lambda}{\textrm{max}} \left| \sum_{\alpha=1}^P \hat a_{\mu,\alpha} {}^{n_\mu}J_{\mu,\Lambda\alpha} - \inv{n_\mu}_{\mu,\Lambda} \right| \rightarrow
	\begin{array}{cc}
		R_\mu < \textrm{tol} \rightarrow    & \textrm{polynomially expressible invariant} \\
		R_\mu \geq \textrm{tol} \rightarrow & \textrm{polynomially independent invariant}
	\end{array}
	\label{eq:tolerance_for_invariants}
\end{equation}
is employed to verify whether the given invariant can be accurately represented as a polynomial, which is inspired by the Chebyshev distance. In the cases studied, a number of $\num{1000}$ randomly generated states and a tolerance value $\textrm{tol}=1$ were found to yield reliable results in a couple of seconds. 
If it turns out that the $\mu$th invariant is polynomially expressible and the exact coefficients of the found relation are of interest, the analytical step is carried out. Otherwise, ${}^{n_{\kappa+1}}\tilde I_{\kappa+1} = \inv{n_\mu}_\mu$ is added to the current tuple $\Set{Y}_\Group{G}$ with $\kappa$ elements and the numerical step is directly executed again for the next invariant $\inv{n_{\mu+1}}_{\mu+1}$ in $\Set{X}_{\Set{G}}$.
\paragraph{Analytical step for the $\mu$th invariant}
The polynomial relations found numerically are initially very good approximations, which are then verified symbolically in a subsequent step. For this purpose, we use the SymPy library, which provides the invariants in analytical form. For every found polynomial relation, the verification process begins by examining the solution vector $\hat{\V{a}}{}_\mu$ to determine whether the coefficients $\hat{a}_{\mu,\alpha}$ can be expressed as rational numbers. If this is the case, the identified polynomial relation between the invariants is checked for exactness.
\begin{remark}
	\label{rem:elimation_tips}
	In general, the number of random states must be greater than or equal to the number of terms in the polynomial ansatz. When the ansatz involves high polynomial degrees, numerical errors may accumulate, potentially leading to incorrect identification of invariants as either polynomial independent or expressible. This issue can be mitigated by increasing the precision of floating-point values or by switching to symbolic computation also for the numerical step. However, this comes at the cost of increased computational time. Thus, fully symbolic computation should be avoided if possible and only used to perform the analytical step. Alternatively, one could use more sophisticated algorithms, which are described in \citet{sturmfels_algorithms_2008, derksen_computational_2002, desmorat_computation_2023, lercier_hyperelliptic_2012, lercier_covariant_2017, olive_geometrie_2014, olive_minimal_2017, olive_characterization_2022, taurines_modelisation_2022}, for instance.
\end{remark}
\begin{remark}
	\label{rem:further_usage}
	The presented procedure serves not only to derive a set of polynomially independent invariants, but also to find polynomial relations between the elements of two distinct sets of invariants. Furthermore, this numerical-analytical approach also serves to identify syzygies between polynomial invariants. Syzygies denote algebraic dependencies between invariants in polynomial form, see \citet[Sect. 1.3]{derksen_computational_2002} or \citet[p. 14]{sturmfels_algorithms_2008}. For example, let $I_1, I_2$ and $I_3$ be polynomially independent homogeneous polynomial invariants which share the syzygy $0=I_1^2 - I_2I_3$. Hence, one of the three invariants depends not polynomially but algebraically on the other two invariants. 
	\begin{center}
		\fbox{
			\begin{minipage}{0.9\textwidth}
				We want to note again that algebraic dependence of invariants, i.e., the existence of syzygies, does not always mean that an invariant is expressible as a function of the others. 
				However, this is true in some examples.
				For instance, for the syzygy $0=I_1^2 - I_2I_3$, we may find that
				\begin{equation}
					I_2 = \left\{\begin{array}{ll}
						\dfrac{I_1^2}{I_3} & \textrm{for } I_3 \neq 0 \\[12pt]
						I_1 & \textrm{for } I_3 = 0
					\end{array}\right.
					\label{eq:functional_realtion_example}
				\end{equation}
				holds. Accordingly, $I_2$ is expressible as a function of $I_1$ and $I_3$. Such functional relations between invariants are not in the scope of this paper, but useful to determine minimal functional basis.
			\end{minipage}
		}
	\end{center}
\end{remark}
\subsection{Proof that invariants form a minimal integrity basis}
\label{sec:proof_minimal_integrity_basis}
In the following, we present two approaches to verify that a set of invariants is a minimal integrity basis: one involving a comparison with already established minimal integrity bases of invariants in coordinate-dependent form, and another based on the Hilbert series and a Hironaka decomposition. We would like to point out once again that the reduced invariant sets are not necessarily integrity bases, since we generally obtain functional bases using the approach presented in Section~\ref{sec:constructing_xiao}; see also Remark~\ref{rem:no_complete_set_found}. However, for all groups considered, the specified functional bases are also integrity bases.
\subsubsection{Usage of a known minimal integrity basis}
\label{sec:proof_irreducible}
In order to prove that a set of invariants is also a minimal integrity basis it is sufficient to find the polynomial relations to another known minimal integrity basis like that from \citet{smith_further_1962} or \citet{smith_transversely_1982}, which is constituted by invariants in coordinate-dependent form.
In order to accomplish this, it is required to compute the structural tensors for a fixed orientation in such a way that the orientation coincides with that of the coordinate-dependent invariants.
Additionally, it is essential to verify that the number of invariants in each polynomial degree is equal. The polynomial relations can be obtained in a straightforward manner with the algorithm introduced in Sect.~\ref{sec:elimination}. 
In the following, we present a more formal discussion to support this claim.

\begin{proposition}
	\label{prop:minimal_integrity_basis}
	Let $J_1, \dots, J_n$, $n\in\Setnum{Z}_{\geq0}$, be homogeneous polynomial invariants that are elements of a minimal integrity basis and $\textrm{deg}(J_\alpha) = j_\alpha$, $\alpha\in\{1,\ldots,n\}$ the respective polynomial degrees. 
	Furthermore, let $I_1, \dots, I_n$ with $\textrm{deg}(I_\alpha) = \textrm{deg}(J_\alpha) \:\forall\: \alpha \in\{1,\ldots,n\}$ homogeneous polynomial invariants, and let polynomial relations $I_\alpha=f_\alpha(J_1,\ldots,J_n)$ and $J_\alpha=g_\alpha(I_1,\ldots,I_n)$ be known for all $\alpha\in \{1,\ldots,n\}$.
	Then, $\{I_1, \dots, I_n\}$ is a minimal integrity basis.
\end{proposition}
\begin{proof}
	\label{proof:minimal_integrity_basis}
	First, since all $I_\alpha$ are expressible as polynomials of the form $f_\alpha(J_1,\ldots,J_n)$, all $I_\alpha$ are indeed invariants. 
	Second, since every polynomial invariant $K=h(J_1,\ldots,J_n)$ is expressible as a polynomial function of the form $h(J_1,\ldots,J_n)$, and $J_\alpha=g_\alpha(I_1,\ldots,I_n)$, we can express $K$ as a polynomial function $\hat{h}_\alpha(I_1,\ldots,I_n)$. 
	Thus, $\{I_1, \dots, I_n\}$ is an integrity basis.
	Finally, because of $\textrm{deg}(I_\alpha) = \textrm{deg}(J_\alpha) \:\forall\: \alpha\in\{1,\ldots,n\}$ and the minimality of $\{J_1, \dots, J_n\}$,
	$\{I_1, \dots, I_n\}$ is a minimal integrity basis~\citep[Remark 4.2.]{olive_minimal_2017}. 
\end{proof}
\subsubsection{Hilbert series}
\label{sec:molien_series}
Since, in general, results on minimal integrity bases are not known, Proposition~\ref{prop:minimal_integrity_basis} cannot always be applied. Therefore, we would like to introduce some further concepts from invariant theory within this section that play a central role in many algorithms for determining invariants. 
More specifically, this concerns polynomial rings (Def.~\ref{def:polynomial_ring}), rings of invariant polynomials (Def.~\ref{def:invariant_polynomial}), and the closely related Hilbert series (HS). In what follows, we will elaborate on the concepts for the specific case of hyperelasticity.

We denote by $\PolRing{\Setnum{R}}{x_1,\ldots,x_6}{}$ the polynomial ring in the six variables $x_1,\dots,x_6$ over the real numbers $\Setnum{R}$.%
\footnote{
	Note that in general the number of formal variables and also the field is not restricted to the present case $\PolRing{\Setnum{R}}{x_1,\ldots,x_6}{}$. For example, one could also consider the fields $\Setnum{Q}$ of rational and $\Setnum{C}$ of complex numbers, which both have zero characteristic, and an arbitrary finite number of formal variables.}
For $x_1,\dots,x_6$, we insert the six independent components $C_{11},C_{22},C_{33},C_{23},C_{13},C_{12}$ of the right Cauchy-Green deformation tensor $\te C\in\tsposdef$. It has to be pointed out that there exist infinite possibilities of what we can define as formal variables. In fact, one can also choose $C_{11},C_{22},C_{33},\sqrt{2}C_{23},\sqrt{2}C_{13},\sqrt{2}C_{12}$ instead. Depending on what one wants to show or calculate, one choice is more appealing than another. We use the former to define the ring of polynomials that is invariant under the action of $\Group{G}$ as 
\begin{align}
	\PolRing{\Setnum{R}}{x_1,\ldots,x_6}{\Group{G}}:=\left\{ f\in\PolRing{\Setnum{R}}{x_1,\ldots,x_6}{} \,| \,  f((\te Q \star \te C)_{11},\ldots,(\te Q \star \te C)_{12}) = f(C_{11},\ldots,C_{12}) \, \forall\: \te{Q}\in\Group{G} \right\} \; . 
\end{align}
The HS is an important tool for showing that a given set of invariants $\{I_1,\ldots, I_n\}$, $n\in\Setnum{Z}_{\ge 1}$, generates the ring of invariant polynomials $\PolRing{\Setnum{R}}{x_1,\ldots,x_6}{\Group{G}}$. Descriptively, this means that every invariant polynomial can be expressed as a polynomial in $I_1,\ldots, I_n$.
\begin{center}
	\fbox{
		\begin{minipage}{0.9\textwidth}
			That is, if a set of invariants generates the ring of invariant polynomials, then this set is an integrity basis. Furthermore, a set of polynomially independent generators is a minimal integrity basis. 
		\end{minipage}
	}
\end{center}
In order to introduce the HS properly, we note that the set of all polynomials over a vector space $\Group{V} \cong \mathbb R^6$ is a graded algebra $\Setnum{R}[\Group{V}]$, which is isomorphic to the polynomial ring $\PolRing{R}{x_1,\dots,x_6}{}$~\citep[Sect. 2.1]{derksen_computational_2002}. It can be expressed as the direct sum $\Setnum{R}[\Group{V}] = \oplus_{k=0}^\infty \Setnum{R}[\Group{V}]_k$ of vector spaces $\Setnum{R}[\Group{V}]_k$ of homogeneous polynomials of degree $k$~\citep[Sect. 2.1]{derksen_computational_2002}. 
Since we focus only on group actions on homogeneous polynomials that preserve degree, which are therefore linear, the set of invariant polynomials on a vector space is also a graded algebra $\Setnum{R}[\Group{V}]^{\Group{G}} = \oplus_{k=0}^\infty \Setnum{R}[\Group{V}]_k^{\Group{G}}$~\citep[Sect. 2.1]{derksen_computational_2002}. 
Note that $\Setnum{R}[\Group{V}]^{\Group{G}}_0 = \Setnum{R}[\Group{V}]_0 = \Setnum{R}$. Next, we only consider compact groups $\Group{G}\subseteq\Group{O}(3)$~\citep{zheng_theory_1994}, which are also reductive and therefore linear reductive in zero characteristic~\citep[Sect. 2.2]{derksen_computational_2002}. This leads all rings $\PolRing{R}{x_1,\dots,x_6}{G}$ we consider to be finitely generated~\citep[Theorem 2.2.10]{derksen_computational_2002} and Cohen-Macaulay~\citep[Main Theorem]{hochster_rings_1974}. Since $\Setnum{R}[\Group{V}]_k^{\Group{G}}$ is finite dimensional for all $k\in\Setnum{Z}_{\geq0}$, the HS is defined as
\begin{equation}
	\Hs_\Group{G} := \sum_{k=0}^\infty \textrm{dim}\left(\Setnum{R}[\Group{V}]_k^{\Group{G}}\right)t^k \commar
	\label{eq:hilbert_series}
\end{equation}
where $\textrm{dim}\left(\Setnum{R}[\Group{V}]_k^{\Group{G}}\right)$ denotes the dimension of the respective vector space and $t\in\left(-1,1\right)$ a parameter, cf. \citet[Def. 1.4.1]{derksen_computational_2002}.\footnote{In some works in literature \citep{teranishi_ring_1986,derksen_computational_2002,neusel_invariant_2007}, the HS is also referred to as Poincaré series.} For finite linearly reductive groups, the HS can be computed via Molien's formula
\begin{equation}
	\Hs_\Group{G} = \frac{1}{\left|\Group{G}\right|} \sum_{\te{Q}\in\Group{G}} \frac{1}{\textrm{det}\left( \M{1} - t\M{\varPhi}\left( \te{Q} \right)\right)} \commar
	\label{eq:molien_series}
\end{equation}
cf. \citet[Sect. 2.6, Theorem 3.2.2, Sect. 4.6.1]{derksen_computational_2002}. Within the scope of this work, the quantity $\uuline{1}\in\Setnum{R}^{6\times 6}$ denotes the $6\times 6$ identity matrix and $\M{\varPhi}\in\Setnum{R}^{6\times 6}$ an $6\times 6$ matrix representation of a group element $\te{Q}\in\Group{G}$. 
\begin{center}
	\fbox{
		\begin{minipage}{0.9\textwidth}
			For illustration, instead of writing $\te{Q}\star\te{C}$ for the symmetric 2nd order tensor $\te{C}\in\tsposdef$, we can also write $\M{\varPhi}\cdot\V{C}$, where $\V{C}$ is a vector of the independent coordinates of the tensor $\te{C}$, e.g., a Voigt or Kelvin-Mandel notation. In this case, $\M{\varPhi}$ is a $6\times6$ matrix representation of a group element $\te{Q}\in\Group{G}$ depending on the chosen vector representation $\V{C}$ of the tensor $\te{C}$.
		\end{minipage}
	}
\end{center}
Furthermore, by computing the Taylor series 
\begin{equation}
	\Hs_{\textrm{Tay},\Group{G}} = \sum_{i=1}^{\infty} \frac{1}{i!} \left. \frac{\textrm{d}^{i}\Hs_\Group{G}(t)}{\textrm{d}t^i}\right|_{t=0} = 1 + b_1t + b_2t^2 + b_3t^3 + \textrm{HOT},\quad b_\lambda\in\Setnum{Z}_{\geq 0}
	\label{eq:molien_series_taylor}
\end{equation}
of Eq.~\eqref{eq:molien_series}, we obtain the original form~\eqref{eq:hilbert_series} of the HS once again, where HOT denotes higher order terms. Hence, $\Hs_{\textrm{Tay},\Group{G}} = \Hs_{\Group{G}}$.
\paragraph{Counting invariants}
From the HS in the form of Eq.~\eqref{eq:hilbert_series}, or equivalently Eq.~\eqref{eq:molien_series_taylor}, one can count the dimensions of the individual vector spaces. Hence, we can determine the number of basis elements (linear independent invariants) for each degree without having computed a single invariant. 
This is a very powerful information, as it provides a direction regarding the search for elements of a minimal integrity basis in specific polynomial degrees. The underlying counting process is initiated at the coefficient of the linear term in the parameter $t$, corresponding to the number of linear independent polynomials of polynomial degree one and subsequently progresses to higher degrees. We illustrate that fact using multiple examples in Sects.~\ref{sec:example_monoclinic}~and~\ref{sec:example_cubic} as well as in \ref{app:example_hs_derksen}.
\paragraph{Noether's degree bound}
Given that the HS in Eq.~\eqref{eq:hilbert_series} is an infinite series and the coefficients of the powers of the parameter $t$ allow extracting information about minimal integrity bases, one may ask up to which power of $t$ the HS should be at least evaluated. One possibility for finite groups and characteristic zero is to use Noether's degree bound, which states that an invariant $I\in\PolRing{R}{x_1,\dots,x_6}{\Group{G}}$, that is an element of a minimal integrity basis, can be at most of degree $\textrm{deg}(I)=\left|\Group{G}\right|$.

\paragraph{The Cohen-Macaulay property}
The series representation of HS, together with the Noether degree bound, provides useful information about an invariant ring $\PolRing{R}{x_1,\dots,x_6}{G}$. Nonetheless, a lot more can be said on the structure of the ring because we only consider rings that are Cohen-Macaulay. This property tells us that we can express the polynomial ring of invariant polynomials by
\begin{equation}
	\PolRing{\Setnum{R}}{x_1,\dots,x_6}{G} = \bigoplus\limits_{i=1}^q S_i \, \PolRing{R}{P_1,\dots,P_p}{}_i,\quad p,q\in\Setnum{Z}_{\geq0},\: p\leq6 \commar
	\label{eq:hironaka_decomposition}
\end{equation}
which is known as a Hironaka decomposition~(Def.~\ref{def:hironaka_decomposition}).
\begin{center}
	\fbox{
		\begin{minipage}{0.9\textwidth}
			A Hironaka decomposition is another way to express a ring of invariant polynomials that is Cohen-Macaulay. Such a Hironaka decomposition is not unique. The ring $\PolRing{R}{P_1,\dots,P_{p}}{}\subseteq\PolRing{R}{x_1,\dots,x_6}{G}$ is generated by homogeneous polynomials, so-called primary invariants $P_1,\dots,P_p$. These form a homogeneous system of parameters (h.s.o.p.) (Def.~\ref{def:hsop}) for $\PolRing{R}{x_1,\dots,x_6}{G}$. Primary invariants are algebraically independent and have the property that every polynomial invariant $I\in\PolRing{R}{x_1,\dots,x_6}{G}$ is a solution of a syzygy $0 = I^n + f_{n-1}(P_1,\dots,P_P)I^{n-1} + \dots + f_{0}(P_1,\dots,P_P)$, $n\in\Setnum{Z}_{\geq0}, f_\lambda\in\PolRing{R}{P_1,\dots,P_p}{}$ and $\lambda\in\{0,\dots,n-1\}$. Especially, the secondary invariants $S_i$ in Eq.~\eqref{eq:hironaka_decomposition} are solutions of such syzygies. Thus, according to Eq.~\eqref{eq:hironaka_decomposition}, a set of primary and secondary invariants allow to write every invariant polynomial $g\in\PolRing{R}{x_1,\dots,x_6}{G}$ as a polynomial of the primary and secondary invariants, whereas $g = S_1h_1(P_1,\dots,P_p) + S_2h_2(P_1,\dots,P_p) +\dots + S_qh_q(P_1,\dots,P_p)$ with $q\in\Setnum{Z}_{\geq1}$, $h_\mu(P_1,\dots,P_p)\in\PolRing{R}{P_1,\dots,P_p}{}$ and $\mu\in\{1,\dots,q\}$.\footnotemark \:Therefore, the primary and secondary invariants are also suited as elements of an integrity basis. Such an integrity basis is generally not minimal, since the elements of the joint sets of primary or secondary invariants can be polynomially dependent. These invariants may be necessary for a Hironaka decomposition, but not for an integrity basis. In conclusion, a minimal integrity basis is given by the polynomially independent primary and secondary invariants.\footnotemark
		\end{minipage}
	}
\end{center}
\addtocounter{footnote}{-1}
\footnotetext{
	By definition, one secondary invariant is always the number $1$. Consider, for example, an invariant $g\in\PolRing{R}{x_1,\dots,x_6}{G}$ which is expressible by $g=h(P_1,\dots,P_p) = Sh(P_1,\dots,P_p)$ with $h(P_1,\dots,P_p)\in\PolRing{R}{P_1,\dots,P_p}{}$. Thus, $S=1$.}
\stepcounter{footnote}
\footnotetext{
	The number $1$, which is always a secondary invariant, is also not required for a minimal integrity basis. Furthermore, please note that for a given minimal integrity basis, it may not be always possible to choose a set of primary and secondary invariants directly from the elements of that basis. For instance, in order to specify a set of primary invariants, it can be required to build a homogeneous polynomial of the elements of the integrity basis.} 

For rings that are Cohen-Macaulay, the HS can be expressed in the special form
\begin{equation}
	\Hs_\Group{G} = \frac{1 + a_1t + a_2t^2 + \dots +a_{s-1}t^{s-1} + a_st^s}{\left(1-t\right)^{p_{1}} \left(1-t^2\right)^{p_{2}} \dots  \left(1-t^{n-1}\right)^{p_{n-1}} \left(1-t^n\right)^{p_n}  } \quad\textrm{with}\quad a_\lambda,p_\mu,s\in\Setnum{Z}_{\geq0} \quad\textrm{and}\quad \lambda,\mu,n \in\Setnum{Z}_{>0} \; ,
	\label{eq:molien_series_prim_sek}
\end{equation}
which relates to a Hironaka decomposition. The factors of the denominator provide information about the number $p_\mu$ and polynomial degree $\mu$ of the primary invariants, while the coefficients $a_\lambda$ of the numerator indicate how many secondary invariants exist in the corresponding polynomial degree $\lambda$~\citep{sturmfels_algorithms_2008}. 
\begin{center}
	\fbox{
		\begin{minipage}{0.9\textwidth}
			It is very important to note that it is possible to find a different admissible rational form of the HS~\eqref{eq:molien_series_prim_sek} that does not belong to any Hironaka decomposition. With admissible we mean that the numerator is a polynomial $1+a_1t+\dots+a_u$ with coefficients $a_\lambda\in\Setnum{Z}_{\geq0}$ and $\lambda,u\in\Setnum{Z}_{\geq0}$, while the denominator is given by $\left(1-t\right)^{p_{1}} \dots \left(1-t^n\right)^{p_n}$. Hence, no information about primary and secondary variants can be derived from it~\citep{dixmier_serie_1982}. To illustrate this, we examine one prominent example in \ref{app:example_hs_derksen} where one finds an admissible rational form of the HS~\eqref{eq:molien_series_prim_sek} without a corresponding Hironaka decomposition. 
			Nevertheless, in many algorithms for calculating invariants, a rational form of the HS is computed first, and then an attempt is made to determine a set of primary invariants. If this search is successful, then the rational form of the HS belongs to an Hironaka decomposition and the calculation of secondary invariants follows. Otherwise, another valid rational form of the HS is determined, and the search for primary invariants is repeated.
		\end{minipage}
	}
\end{center}
\begin{remark}
	\label{rem:primary_invars_is_number_of_coords}
	Since the rational form of the HS is not unique and may not belong to any Hironaka decomposition, we find it useful, under the restrictions made, to keep the following in mind when searching for an admissible rational form:
	\begin{enumerate}
		\item The sum $\sum_{i=1}^n\textrm{deg}(P_i)$ of the degrees of the primary invariants should be as low as possible~\citep[Sect. 3.3.1]{derksen_computational_2002}.
		\item For a finite group that is a subgroup of the general linear group, the number $n$ of primary invariants $P_1,\dots,P_n$ is equal to the number of independent tensor coordinates, cf.~\citep[Proposition 2.1.1]{sturmfels_algorithms_2008}.
		\item The number $s$ of secondary invariants is determined by $s=\Pi_{i=1}^n \textrm{deg}(P_i) / |\Group{G}|$ for a finite group $\Group{G}$~\citep[Proposition 2.3.6]{sturmfels_algorithms_2008}.
		\item For the case of a finite group, the maximal degree of any secondary invariant is at most $\sum_{i=1}^n\textrm{deg}(P_i)-n$~\citep[Proposition 3.9.5]{derksen_computational_2002}.
	\end{enumerate}
\end{remark}
\paragraph{Calculation of the HS in hyperelasticity}
In hyperelasticity, we focus on scalar invariants of a symmetric 2nd order tensor, which has six independent coordinates. Thus, for identifying the matrix $\M{\varPhi}\in\Setnum{R}^{6\times6}$, we begin by referring to Eq.~\eqref{eq:C_transformation}, where the symmetric tensor $\te{C}\in\tsposdef$ is transformed by an element $\te{Q}\in\Group{O}(3)$. However, since we are now interested in material symmetry, we will only consider $\te{Q}\in\Group{G}\subseteq\Group{O}(3)$.
Instead of performing this transformation directly in $\tss{2}$, it can alternatively be carried out by transforming the tensor in vector notation $\V{C}\in\Setnum{R}^{6\times1}$ with a transformation matrix $\M{\varPhi}$. 
For this purpose, the relationships
\begin{equation}
	C_\alpha = T^\alpha_{ij}C_{ij} \quad \textrm{and} \quad C_{ij} = \tilde{T}_\alpha^{ij}C_\alpha
	\label{eq:conditions_vector_tensor}
\end{equation}
should hold between the tensor coordinates $C_{ij}$ of $\te C$ for the standard basis $\{\ve e_1,\ve e_2, \ve e_3\}$ and the vector coordinates $C_\alpha$, where $T^\alpha_{ij}$ and $\tilde{T}_\alpha^{ij}$ are suitable transformation coefficients. We note that 
\begin{equation}
	T^\alpha_{ij} \tilde{T}_\beta^{ij} = \delta_{\alpha\beta} \quad \textrm{and} \quad T^\alpha_{ij} \tilde{T}_\alpha^{kl} = \frac{1}{2}\left( \delta_{ik}\delta_{jl} + \delta_{il}\delta_{jk} \right)
	\label{eq:conditions_on_trafo_coeffs}
\end{equation}
have to be true. By imposing $T^\alpha_{ij} = \tilde{T}_\alpha^{ij}$, Eq.~\eqref{eq:conditions_vector_tensor} becomes
\begin{equation}
	C_\alpha = T^\alpha_{ij}C_{ij} \quad \textrm{and} \quad C_{ij} = T^\alpha_{ij}C_\alpha
	\label{eq:conditions_vector_tensor2}
\end{equation}
and the product
\begin{equation}
	\te{C}:\te{C} = \V{C} \cdot \V{C} \quad \leftrightarrow \quad C_{ij}C_{ji} = C_\alpha C_\alpha
	\label{eq:norm_spaces}
\end{equation}
is equal in both spaces. The transformation coefficients
\begin{equation}
	T^{\alpha}_{ij} = \left\{ \begin{array}{lll}
		1 & \alpha=i=j \\
		\frac{1}{\sqrt{2}} & (\alpha,i,j) = \left\{ (4,2,3), (4,3,2), (5,1,3), (5,3,1), (6,1,2), (6,2,1) \right\} \\
		0 & \textrm{else}
	\end{array}\right.
	\label{eq:vector_notation}
\end{equation}
fulfill the relations in Eq.~\eqref{eq:conditions_vector_tensor2}. Hence, the resulting vector $\V{C}$ corresponds to the Kelvin-Mandel notation of the tensor $\te{C}$~\citep{mandel_generalisation_1965, brannon_rotation_2018}. In this representation, $\te{Q}\star\te{C}$ is represented by $\M{\varPhi}\cdot\V{C}$, where $\M{\varPhi}$ is the $6 \times 6$ rotation matrix in Kelvin-Mandel notation, whose coefficients $\Phi_{\alpha\beta}$ can be computed using
\begin{equation}
	\Phi_{\alpha\beta} = T^{\alpha}_{ik} T^{\beta}_{jl} Q_{ij}Q_{kl}\quad , \quad \te{Q}\in\Group{G} \point
	\label{eq:six-dimensional-rotation}
\end{equation}
Thus, the full matrix $\M{\varPhi}$ is denoted by
\begin{equation}
	\left[ \Phi_{\alpha\beta} \right]\!=\! \begin{pmatrix}
		Q^2_{11} & Q^2_{12} & Q^2_{13} & \sqrt{2}Q_{12}Q_{13} & \sqrt{2}Q_{11}Q_{13} & \sqrt{2}Q_{11}Q_{12} \\
		Q^2_{21} & Q^2_{22} & Q^2_{23} & \sqrt{2}Q_{22}Q_{23} & \sqrt{2}Q_{21}Q_{23} & \sqrt{2}Q_{21}Q_{22} \\
		Q^2_{31} & Q^2_{32} & Q^2_{33} & \sqrt{2}Q_{32}Q_{33} & \sqrt{2}Q_{31}Q_{33} & \sqrt{2}Q_{31}Q_{32} \\
		\sqrt{2}Q_{21}Q_{31} & \sqrt{2}Q_{22}Q_{32} & \sqrt{2}Q_{23}Q_{33} & Q_{22}Q_{33}+Q_{23}Q_{32} & Q_{21}Q_{33}+Q_{23}Q_{31} & Q_{21}Q_{32}+Q_{22}Q_{31} \\
		\sqrt{2}Q_{11}Q_{31} & \sqrt{2}Q_{12}Q_{32} & \sqrt{2}Q_{13}Q_{33} & Q_{12}Q_{33}+Q_{13}Q_{32} & Q_{11}Q_{33}+Q_{13}Q_{31} & Q_{11}Q_{32}+Q_{12}Q_{31} \\
		\sqrt{2}Q_{11}Q_{21} & \sqrt{2}Q_{12}Q_{22} & \sqrt{2}Q_{13}Q_{23} & Q_{12}Q_{23}+Q_{13}Q_{22} & Q_{11}Q_{23}+Q_{13}Q_{21} & Q_{11}Q_{22}+Q_{12}Q_{21}
	\end{pmatrix} \point
	\label{eq:6d_rot_mat_2nd_sym_tensor}
\end{equation}
From Eq.~\eqref{eq:six-dimensional-rotation} and the transformation coefficients $T^\alpha_{ij}$, it follows directly that the matrix $\M{\varPhi}$ is diagonal if the transformation $\te{Q}\in\Group{G}$ lies in the principle-axis state. Furthermore, for any $\te{Q}=\pm\rot{a}{\varphi}\in\Group{G}\subseteq\Group{O}(3)$, the eigenvalues of $\pm\rot{a}{\varphi}$ are denoted by $\left\{ \pm 1, \pm e^{i\varphi}, \pm e^{-i\varphi}\right\}$ for an arbitrary rotation angle $\varphi\in[0,2\pi)$. Finally, this allows us to express the determinant
\begin{equation}
	\textrm{det}\left( \M{1} - t \M{\varPhi}(\te{Q}) \right) = f(\varphi) = 
	(1-t)^2 (1-2t\cos\left(\varphi\right) + t^2) (1-2t\cos\left(2\varphi\right) + t^2)
	\label{eq:molien_series_determinant}
\end{equation}
in Eq.~\eqref{eq:molien_series} as a function of the rotation angle of a group element.\footnote{For any even-order tensor, which is not a two-point tensor (cf. Footnote~\ref{foot:two-point}), transformations with $+\rot{a}{\varphi}$ and $-\rot{a}{\varphi}$ are identical, see Remark~\ref{rem:inversion_does_not_lead}. This equivalence is also reflected in the fact that the expressions of the determinant for both $+\rot{a}{\varphi}$ and $-\rot{a}{\varphi}$ yield the same result. In contrast, for odd-order tensors, the transformation behavior differs, resulting in two distinct formula for $+\rot{a}{\varphi}$ and $-\rot{a}{\varphi}$.} Hence, Eq.~\eqref{eq:molien_series_determinant} allows to easily calculate the HS.
\subsection{Example: Monoclinic crystal system}
\label{sec:example_monoclinic}
The monoclinic crystal system comprises the point groups $\Group{C}_2$, $\Group{C}_{1h}$ and $\Group{C}_{2h}$ leading to the same anisotropy $\aniso{2}$ and equivalent integrity basis in hyperelasticity. We show for this anisotropy the construction and reduction of the corresponding functional basis using the above introduced framework. Furthermore, we show that the obtained reduced functional basis is also a minimal integrity basis.
\subsubsection{Structural tensors and functional basis}
\label{ssec:monoclinic_functional_basis}
Since the functional bases are equivalent for the tensor $\te{C}\in\tsposdef$ and $\Group{C}_2$, $\Group{C}_{1h}$ and $\Group{C}_{2h}$, we only consider the group $\Group{C}_{2h}$ here. The corresponding structural tensors $\te{P}_2\in\tss{2}$ and $\te{G}_1\in\tsa{2}$, introduced by \citet{Zheng1993}, are denoted by
\begin{equation}
	\te{P}_2 = \ve{a}_1\ve{a}_1 - \ve{a}_2\ve{a}_2 \quad \textrm{and} \quad \te{G}_1 = \tens{E}{3}\cdot\ve{a}_3 = \ve{a}_1\ve{a}_2 - \ve{a}_2\ve{a}_1 \commar
	\label{eq:example_monoclinic_structural_tensor}
\end{equation}
where $\tens{E}{3}\in\tsa{3}$ is the permutation tensor ($\tens{E}{3}$-Pseudotensor or Levi-Civita symbol) and $\ve{a}_1$, $\ve{a}_2$ and $\ve{a}_3\in\ts{1}{}$ orthonormal 1st order tensors. First, we determine a functional basis using the table of \citet{Boehler1977}, where we define the tuple of structural functions by
\begin{equation}
	\Set{F}_{\Group{C}_{2h}}=\Set{M}_{\Group{C}_{2h}}=\left( \te{P}_2, \te{G}_1 \right)
	\label{eq:example_monoclinic_structural_functions}
\end{equation}
and the tuple of argument tensors by
\begin{equation}
	\Set{H}_{\Group{C}_{2h}}=\left( \te{C},\te{P}_2, \te{G}_1 \right) \point
	\label{eq:example_monoclinic_isotropic_extension_functions}
\end{equation}
Thus, a functional basis consisting of 14 invariants
\begin{equation}
	\begin{aligned}
		&\tr[\te{C}],\: \tr[\te{C}^2],\: \tr[\te{C}^3],\:
		\tr[\te{C}\cdot\te{P}_2],\: \tr[\te{C}^2\cdot\te{P}_2],\:\tr[\te{C}\cdot\te{P}_2^2],\:\tr[\te{C}^2\cdot\te{P}_2^2],\: \tr[\te{C}\cdot\te{G}_1^2],\: \tr[\te{C}^2\cdot\te{G}_1^2], \\
		& \tr[\te{C}^2\cdot\te{G}_1^2\cdot\te{C}\cdot\te{G}_1],\: \tr[\te{C}\cdot\te{P}_2\cdot\te{G}_1],\: \tr[\te{C}^2\cdot\te{P}_2\cdot\te{G}_1],\: \tr[\te{C}\cdot\te{P}_2^2\cdot\te{G}_1],\: \tr[\te{C}\cdot\te{G}_1^2\cdot\te{P}_2\cdot\te{G}_1] 
	\end{aligned}
	\label{eq:example_monoclinic_complete_invariants}
\end{equation}
is obtained from Tab.~\ref{tab:boehler_table} of \citet{Boehler1977}. In the functional basis~\eqref{eq:example_monoclinic_complete_invariants}, we have already sorted out invariants that consist solely of structural tensors, since these are merely constants. The next step is to identify polynomial dependencies between the invariants of Eq.~\eqref{eq:example_monoclinic_complete_invariants} in order to reduce the cardinality of the function basis.
\subsubsection{Minimal integrity basis}
\label{ssec:monoclinic_integrity_basis}
By applying the procedure outlined in Sect.~\ref{sec:elimination} for eliminating polynomially expressible invariants, we obtain the reduced functional basis in Tab.~\ref{tab:example_monoclinic_integrity_basis} for group $\Group{C}_{2h}$ from the initial functional basis~\eqref{eq:example_monoclinic_complete_invariants}. 
As a proof that the functional basis in Tab.~\ref{tab:example_monoclinic_integrity_basis} is also a minimal integrity basis, we utilize Proposition~\ref{prop:minimal_integrity_basis} by calculating the polynomial relations between the invariants from Tab.~\ref{tab:example_monoclinic_integrity_basis} and those proposed by \citet{smith_further_1962}. These are determined using again the procedure described in Sect.~\ref{sec:elimination}. To achieve this, the orientation of the structural tensors must be aligned with that of Smith’s coordinate-dependent invariants. Therefore, $[\ve{a}_1]=\left(0,1,0\right)^\top$, $[\ve{a}_2]=\left(0,0,1\right)^\top$ and $[\ve{a}_3] = \left(1,0,0\right)^\top$ are set accordingly, where we choose the cartesian basis $\{\ve{e}_1, \ve{e}_2, \ve{e}_3\}$ as reference frame. The polynomial relationships between the two invariant sets can be found in the supplementary material.
\begin{table}[ht]
	\centering
	\caption{Minimal integrity bases for $\Group{C}_{2h}$ or $\Group{A}_{2}$ and a symmetric tensor $\te{C}\in\tsposdef$ with the structural tensors $\te{P}_2\in\tss{2}$ and $\te{G}_1\in\tsa{2}$ of Eq.~\eqref{eq:example_monoclinic_structural_tensor}.}
	\label{tab:example_monoclinic_integrity_basis}
	\renewcommand{\arraystretch}{1.}
	\begin{small}
		\begin{tabularx}{0.94\textwidth}{p{15.cm}}
			\toprule 
			$I_1 = \tr[\te{C}] \hspace{1.5em}
			I_2 = \tr[\te{C}\cdot\te{G}_1^2] \hspace{1.5em}
			I_3 = \tr[\te{C}\cdot\te{P}_2] \hspace{1.5em}
			I_4 = \tr[\te{C}\cdot\te{P}_2\cdot\te{G}_1] \hspace{1.5em}
			I_5 = \tr[\te{C}^2] \hspace{1.5em}
			I_6 = \tr[\te{C}^2\cdot\te{P}_2] \vspace{3pt} \hspace{1.5em}
			I_7 = \tr[\te{C}^2\cdot\te{P}_2\cdot\te{G}_1]$ \\
			\bottomrule
		\end{tabularx}
	\end{small}
\end{table}
\subsubsection{Hilbert series}
\label{ssec:monoclinic_hs}
Since we have already proven with the help of Proposition~\ref{prop:minimal_integrity_basis} that the invariants in Tab.~\ref{tab:example_monoclinic_integrity_basis} form a minimal integrity basis, the HS for this example merely represents a consistency check. On the other hand, if no minimal integrity basis were known from the literature, we would have to use the HS anyway to check whether the invariants actually form a minimal integrity basis.

The groups $\Group{C}_2$, $\Group{C}_{1h}$ and $\Group{C}_{2h}$ lead to the same HS in the present case, which is why the HS only has to be computed for one of the three. We show the calculation using $\Group{C}_2$. This group is generated by $\rot{z}{\pi}$ and contains two group elements that hold one angle of $\pi$ and one angle of $0$, respectively. Hence, with Eqs.~\eqref{eq:molien_series}~and~\eqref{eq:molien_series_determinant}, the HS
\begin{equation}
	\Hs_{\Group{C}_2} = \Hs_{\Group{C}_{1h}} = \Hs_{\Group{C}_{2h}} = \frac{1}{2}\left[ \frac{1}{\left(1-t\right)^6} + \frac{1}{\left( 1-t \right)^2 \left( 1-t^2\right)^2} \right]
	\label{eq:example_monoclinic_calc_molien_series}
\end{equation}
can be computed. After building the common denominator of Eq.~\eqref{eq:example_monoclinic_calc_molien_series},
\begin{equation}
	\Hs_{\Group{C}_2} = \Hs_{\Group{C}_{1h}} = \Hs_{\Group{C}_{2h}} = \frac{1 + t^2}{\left(1-t\right)^4\left(1-t^2\right)^2} \point
	\label{eq:example_monoclinic_molien_series}
\end{equation}
follows, which is already in the desired form of Eq.~\eqref{eq:molien_series_prim_sek}. The corresponding Taylor series expansion, see Eq.~\eqref{eq:molien_series_taylor}, is denoted by
\begin{equation}
	\Hs_{\textrm{Tay},\Group{C}_2} = \Hs_{\textrm{Tay},\Group{C}_{1h}} = \Hs_{\textrm{Tay},\Group{C}_{2h}} = 1 + 4t + 13t^2 + 32t^3 + 70t^4 + 136^5 + 246t^6 + 416t^7 + \textrm{HOT} \point
	\label{eq:example_monoclinic_molien_series_taylor}
\end{equation}
\paragraph{Information of Eq.~\eqref{eq:example_monoclinic_molien_series_taylor}}
The HS~\eqref{eq:example_monoclinic_molien_series_taylor} can be analyzed as follows. 
First, the constant term $1$ indicates, trivially, that any number is an invariant. Then term $4t$ points out that four linear independent invariants exist at polynomial degree one, which are the invariants $I_1$, $I_2$, $I_3$ and $I_4$ of Tab.~\ref{tab:example_monoclinic_integrity_basis}. With these linear invariants, ten linear or polynomially independent products, $I_1^2$, $I_2^2$, $I_3^2$, $I_4^2$, $I_1I_2$, $I_1I_3$, $I_1I_4$, $I_2I_3$, $I_2I_4$ and $I_3I_4$, can be formed at quadratic degree; thus, these invariants are polynomially expressible in the invariants of lower degrees and even reducible, cf. Def.~\ref{def:irreducibility}. The term $13t^2$ then indicates that 13 independent invariants exist of quadratic degree, but since only ten of them can be build in terms of fundamental invariants of lower degrees, it follows that there must be three fundamental invariants of degree two, namely $I_5$, $I_6$ and $I_7$.
From the fundamental invariants $I_1,\dots,I_7$ of degrees one and two, 32 linear independent invariants can be formed with cubic degree. Considering the next term, $32t^3$, we conclude that no fundamental invariants exists of cubic degree. At degree four, 71 products or invariants can be formed from the seven fundamental invariants of lower degrees. However, the term $70t^4$ indicates that only 70 linear independent invariants exist at this degree. From this, we conclude that there exists at least one syzygy or polynomial relation between the 71 products: $\Set{I} = \left\{ I_1^4, I_1^3I_2, I_1^3I_3, \dots \right\}$. Indeed, when formulated with the invariants of Tab.~\ref{tab:example_monoclinic_integrity_basis}, we have the relation
\begin{equation}
	\begin{aligned}
		0 = & - 4I_1^4 - 4I_5^2 + 16I_7^2 + 16I_6^2 + 4I_4^2I_5 - I_4^4 + 4I_3^2I_5 - 2I_3^2I_4^2 - I_3^4 + 32I_2I_4I_7 + 32I_2I_3I_6 + 12I_2^2I_5 + 10I_2^2I_4^2 + 10I_2^2I_3^2 \\
		&- 9I_2^4 + 16I_1I_2I_5 - 8I_1I_2I_4^2 - 8I_1I_2I_3^2 - 24I_1I_2^3 + 8I_1^2I_5 - 4I_1^2I_4^2 - 4I_1^2I_3^2 - 28I_1^2I_2^2 - 16I_1^3I_2
	\end{aligned} \point
	\label{eq:example_monoclinic_degree_four_syzygy}
\end{equation}
By using this polynomial relation, 70 linear independent products remain: $\Set{I}\setminus\left\{ I_1^4 \right\}$, for example. This is exactly what the HS tells, too. In the higher polynomial degrees, this counting process can be continued and would postulate that only syzygies between fundamental invariants exist, but no new fundamental invariants. However, it should be noted that, for example, four of the syzygies at degree five can be obtained by multiplying the degree-four syzygy in Eq.~\eqref{eq:example_monoclinic_degree_four_syzygy} by the fundamental invariants $I_1$, $I_2$, $I_3$ and $I_4$ of degree one, respectively.\footnote{As stated by Noether's degree bound, we only found homogeneous polynomial invariants that are elements of a minimal integrity that have polynomial degree less or equal to the group order $\left|\Group{C}_2\right|=2$. Therefore, we could have also stopped the counting process in polynomial degree two, if we are not interested in possible syzygies between the invariants of Tab.~\ref{tab:example_monoclinic_integrity_basis}.}
\paragraph{Information of Eq.~\eqref{eq:example_monoclinic_molien_series}}
\citet{smith_further_1962} has specified a suitable set of primary and secondary invariants, so we already know that the form of the HS in Eq.~\eqref{eq:example_monoclinic_molien_series} corresponds to a Hironaka decomposition. 
From the denominator in Eq.~\eqref{eq:example_monoclinic_molien_series} it can be observed that there exist the following primary invariants: four of degree one and two of degree two. The numerator indicates that the following secondary invariants are additionally required: the number $1$ and one of degree two. Since only one secondary invariant, in addition to the secondary invariant 1, is required, this secondary invariant must be irreducible.
\paragraph{Hironaka decomposition}
In the next step, we check whether we can find a Hironaka decomposition that matches the proposed rational form~\eqref{eq:example_monoclinic_molien_series} of the HS without relying on the result of \citet{smith_further_1962}. 
\begin{lemma}
	\label{lemma:example_monoclinic_hironaka_lemma}
	Let $I_1,\dots,I_n$, $n\in\Setnum{Z}_{>1}$, be homogeneous polynomials. Furthermore, let there exist a polynomial relation in the form of $I_1^2 = I_1A + B$, where $A=A(I_2,\dots,I_n)$ and $B=B(I_2,\dots,I_n)$ are homogeneous polynomials.
	Then, for every $m\in\Setnum{Z}_{\geq2}$, $I_1^m$ is expressible as a polynomial that is linear in $I_1$.
\end{lemma}
\begin{proof}
	\label{proof:lemma:example_monoclinic_hironaka_lemma}
	We proof this by induction. \\[3pt]
	\begin{tabular}{cp{12cm}}
		Induction start ($m=2$): & Since $I_1^2 = I_1A + B$, $I_1^2$ is expressible as a polynomial linear in $I_1$. \\[3pt]
		Induction step ($m+1$): & 
		Admit the statement for $I_1^m$, then $I_1^m = I_1A + B$. After multiplication with $I_1$, it follows: $I_1^{m+1}=I_1^2A + I_1B$. By induction start, we finally substitute $I_1^2$ by a polynomial which is linear in $I_1$.
	\end{tabular} \\
\end{proof}
\begin{proposition}
	\label{prop:example_monoclinic_hironaka}
	The ring of invariant polynomials $\PolRingE{R}{C_{11},C_{22},C_{33},C_{23},C_{13},C_{12}}{\Group{C}_{2h}}=\PolRingE{R}{I_1,\dots,I_7}{}$, generated by the homogeneous polynomial invariants $I_1,\dots,I_7$ of Tab.~\ref{tab:example_monoclinic_integrity_basis} has the Hironaka decomposition
	\begin{equation}
		\PolRingE{R}{C_{11},C_{22},C_{33},C_{23},C_{13},C_{12}}{\Group{C}_{2h}} = \PolRingE{R}{I_1,I_2,I_3,I_4,I_5,I_6}{} \oplus I_7\PolRingE{R}{I_1,I_2,I_3,I_4,I_5,I_6}{} \point
		\label{eq:hsop_monoclinic}
	\end{equation}
\end{proposition}
\begin{proof}
	\label{proof:example_monoclinic_hironaka}
	Since $\PolRingE{R}{C_{11},C_{22},C_{33},C_{23},C_{13},C_{12}}{\Group{C}_{2h}}=\PolRingE{R}{I_1,\dots,I_7}{}$, we can write an arbitrary invariant polynomial $p\in\PolRingE{R}{C_{11},C_{22},C_{33},C_{23},C_{13},C_{12}}{\Group{C}_{2h}}$ as 
	\begin{equation}
		p = \sum_{a=0}^{r} \sum_{b=0}^{s} \sum_{c=0}^{t} \sum_{d=0}^{u} \sum_{e=0}^{v} \sum_{f=0}^{w} \sum_{g=0}^{x} A_{abcdefg}I_1^a I_2^b I_3^c I_4^d I_5^e I_6^f I_7^g
		\label{eq:monoclinic_proof_arbitrary_polyomial_long}
	\end{equation}
	where $r,s,t,u,v,w,x\in\Setnum{Z}_{\geq0}$ and $A_{abcdef}\in\Setnum{R}$. We will further use summation convention and just write
	\begin{equation}
		p = A_{abcdefg}I_1^a I_2^b I_3^c I_4^d I_5^e I_6^f I_7^g \point
		\label{eq:monoclinic_proof_arbitrary_polyomial}
	\end{equation}
	Next, we consider the rational form of the HS~\eqref{eq:example_monoclinic_molien_series} and try to find corresponding primary and secondary invariants. By doing so, we choose $I_1,\dots,I_6$ as candidates for primary invariants and $1, I_7$ as secondary invariants. We rewrite the syzygy~\eqref{eq:example_monoclinic_degree_four_syzygy} as 
	\begin{equation}
		I_7^2 =  I_7J + K\commar
		\label{eq:monoclinic_proof_syz}
	\end{equation}
	where $J=J(I_1,\dots,I_6)$ and $K=(I_1,\dots,I_6)$ are homogeneous polynomials. Next, we split the polynomial~\eqref{eq:monoclinic_proof_arbitrary_polyomial} into two parts
	\begin{equation}
		p = B_{abcdef(2\alpha)}I_1^a I_2^b I_3^c I_4^d I_5^e I_6^f I_7^{2\alpha} + C_{abcdef(2\alpha+1)}I_1^a I_2^b I_3^c I_4^d I_5^e I_6^f I_7^{2\alpha}I_7, \quad \alpha\in\Setnum{Z}_{\geq0} ,\:
		B_{abcdef(2\alpha)}, C_{abcdef(2\alpha+1)} \in\Setnum{R}
		\commar
		\label{eq:monoclinic_proof_split}
	\end{equation}
	containing even and odd exponents of $I_7$, respectively. Because of Eq.~\eqref{eq:monoclinic_proof_syz} and Lemma~\ref{lemma:example_monoclinic_hironaka_lemma} it directly follows that we can express the polynomial~\eqref{eq:monoclinic_proof_split} by
	\begin{equation}
		p = \tilde{B}_{abcdef}I_1^a I_2^b I_3^c I_4^d I_5^e I_6^f + I_7\tilde{C}_{abcdef}I_1^a I_2^b I_3^c I_4^d I_5^e I_6^f ,\:
		\tilde{B}_{abcdef}, \tilde{C}_{abcdef} \in\Setnum{R} \point
		\label{eq:monoclinic_proof_poylnom_in_hironaka}
	\end{equation}
	This is a Hironaka decomposition that relates to the rational form of the HS~\eqref{eq:example_monoclinic_calc_molien_series}. Hence, we can express the ring of invariant polynomials by $\PolRingE{R}{C_{11},C_{22},C_{33},C_{23},C_{13},C_{12}}{\Group{C}_{2h}} = \PolRingE{R}{I_1,I_2,I_3,I_4,I_5,I_6}{} \oplus I_7\PolRingE{R}{I_1,I_2,I_3,I_4,I_5,I_6}{}$.
\end{proof}
\subsubsection{Quadratic energy function}
\label{ssec:example_monoclinic_quadratic_energy}
In order to illustrate one possible use of the determined minimal integrity basis, we consider a monoclinic $(\Group{A}_{2})$ Saint-Venant-Kirchhoff model, given by the homogeneous quadratic Helmholtz free energy
\begin{equation}
	\varPsi^{\Group{C}_{2h}} = \alpha_1I_1^2 + \alpha_2I_1I_2 + \alpha_3I_1I_3 + \alpha_4I_1I_4 + \alpha_5I_2^2 + \alpha_6I_2I_3 + \alpha_7I_2I_4 + \alpha_8I_3^2 + \alpha_9I_3I_4 + \alpha_{10}I_4^2 + \alpha_{11}I_5 + \alpha_{12}I_6 + \alpha_{13}I_7 
	\label{eq:saint_ven_energy_monoclinic}
\end{equation}
with some material constants $\alpha_\lambda\in\Setnum{R}$, $\lambda\in\{1,\ldots,13\}$.%
\footnote{
	\label{foot:params_pos_def}
	In order to derive a positive definite stiffness from the Helmholtz free energy, the coefficients can not be chosen arbitrarily. } 
The invariants in Eq.~\eqref{eq:saint_ven_energy_monoclinic} are taken from Tab.~\ref{tab:example_monoclinic_integrity_basis} and are formulated in the components of the Green-Lagrange strain tensor $\te{E}\in\tss{2}$ instead of $\te{C}\in\tsposdef$. This is possible because both $\te{E}$ and $\te{C}$ are symmetric 2nd order tensors, cf. Footnote~\ref{foot:same_invariants}.
Please note that all ten reducible invariants or products $I_1^2, I_1I_2,\dots,I_4^2$ and the three fundamental invariants $I_5,I_6$ and $I_7$ are polynomially independent from each another. Furthermore, the Helmholtz free energy.~\eqref{eq:saint_ven_energy_monoclinic} consists of 13 terms, which is no coincidence, since this is the number determined by $13t^2$ of the HS~\eqref{eq:example_monoclinic_molien_series_taylor}, i.e., the maximum number of polynomially independent homogeneous polynomial invariants with polynomial degree two. 
Next, the three vectors specifying the structural tensors are chosen as $[\ve{a}_1]=\left(0,1,0\right)^\top$, $[\ve{a}_2]=\left(0,0,1\right)^\top$ and $[\ve{a}_3] = \left(1,0,0\right)^\top$ with respect to the cartesian basis $\{\ve{e}_1, \ve{e}_2, \ve{e}_3\}$. By taking the second derivative
\begin{equation}
	C_{ijkl}^{\Group{C}_{2h}} = \frac{\partial \varPsi^{\Group{C}_{2h}}}{\partial E_{ij} \partial E_{kl}}
	\label{eq:elast_tensor_monoclinic_example}
\end{equation}
one can compute the coordinates $C_{ijkl}^{\Group{C}_{2h}}$ of the constant elasticity tensor $\tens{C}{4}^{\Group{C}_{2h}}\in\ts{4}{}$. The 4th order elasticity tensor in Kelvin-Mandel notation is
\begin{equation}
	\small{
		\left[C_{\alpha\beta}^{\Group{C}_{2h}}\right] = \begin{pmatrix}
			C_{11} & C_{12} & C_{13} & \sqrt{2}C_{14} & 0 & 0 \\
			C_{12} & C_{22} & C_{23} & \sqrt{2}C_{24} & 0 & 0 \\
			C_{13} & C_{23} & C_{33} & \sqrt{2}C_{34} & 0 & 0 \\
			\sqrt{2}C_{14} & \sqrt{2}C_{24} & \sqrt{2}C_{34} & 2C_{44} & 0 & 0 \\
			0 & 0 & 0 & 0 & 2C_{55} & 2C_{56} \\
			0 & 0 & 0 & 0 & 2C_{56} & 2C_{66} \\
		\end{pmatrix}
		\quad\textrm{with}\:\:\:
		\begin{array}{ll}
			C_{11} = 2(\alpha_1 + \alpha_{11}), \quad C_{12} = 2\alpha_1 - \alpha_2 + \alpha_3,\\
			C_{13} = 2\alpha_1 - \alpha_2 - \alpha_3, \quad C_{14} = \alpha_4, \\
			C_{22} = 2(\alpha_1 + \alpha_{11} + \alpha_{12} - \alpha_2 + \alpha_3 + \alpha_5 - \alpha_6 + \alpha_8) \\
			C_{23} = 2(\alpha_1 - \alpha_2 + \alpha_5 - \alpha_8),\\
			C_{24} = \alpha_{13} + \alpha_4 - \alpha_7 + \alpha_9, \\
			C_{33} = 2(\alpha_1 + \alpha_{11} - \alpha_{12} - \alpha_2 - \alpha_3 + \alpha_5 + \alpha_6 + \alpha_8), \\
			C_{34} = \alpha_{13} + \alpha_4 - \alpha_7 - \alpha_9, \quad C_{44} = 2\alpha_{10} + \alpha_{11}, \\
			C_{55} = \alpha_{11} - \frac{\alpha_{12}}{2}, \quad C_{56} = \frac{\alpha_{13}}{2},\\
			C_{66} = \alpha_{11} + \frac{\alpha_{12}}{2}
		\end{array}
	}
	\label{eq:example_monoclinic_stiffness}
\end{equation}
where $C_{\alpha\beta}^{\Group{C}_{2h}} = T^\alpha_{ij}T^\beta_{kl} C_{ijkl}^{\Group{C}_{2h}}$ with the transformation coefficients $T^\alpha_{ij}$ defined by Eq.~\eqref{eq:vector_notation}. Given the chosen orientation of the vectors $\ve{a}_1, \ve{a}_2$ and $\ve{a}_3$, the elasticity matrix~\eqref{eq:example_monoclinic_stiffness} reveals the characteristic occupancy for this anisotropy, see \citet[Appendix A.2]{altenbach_kontinuumsmechanik_2018} or \citet[Sect. 4.6.2.2.]{Apel2004} for example.
\subsection{Example: Cubic crystal system}
\label{sec:example_cubic}
The cubic crystal system comprises the groups $\Group{T}$, $\Group{T}_h$, $\Group{O}$, $\Group{T}_d$ and $\Group{O}_h$ leading only to two different anisotropies $\aniso{10}$ and $\aniso{11}$ in hyperelasticity. We show for the anisotropy $\aniso{10}$ the construction and reduction of the corresponding functional basis using the introduced framework. Furthermore, we show that the obtained reduced functional basis is also a minimal integrity basis.
\subsubsection{Structural tensor and functional basis}
\label{ssec:example_cubic_functional_basis}
Since the functional bases are equivalent for the tensor $\te{C}\in\tsposdef$ and $\Group{T}$ as well as $\Group{T}_h$, it is sufficient to only consider $\Group{T}_h$. The corresponding structural tensor, introduced by \citet{Zheng1993}, is denoted as 
\begin{equation}
	\tens{\Lambda}{4} = \ve{a}_1\ve{a}_1\ve{a}_2\ve{a}_2 - \ve{a}_2\ve{a}_2\ve{a}_1\ve{a}_1 + \ve{a}_3\ve{a}_3\ve{a}_1\ve{a}_1 - \ve{a}_1\ve{a}_1\ve{a}_3\ve{a}_3 + \ve{a}_2\ve{a}_2\ve{a}_3\ve{a}_3 - \ve{a}_3\ve{a}_3\ve{a}_2\ve{a}_2 \in\Set{L}^4 \commar
	\label{eq:example_cubic_structural_tensor}
\end{equation}
where $\ve{a}_1$, $\ve{a}_2$, $\ve{a}_3\in\Set{L}^1$ are orthonormal. With respect to $\Group{T}_h$, these 1st order tensors can be interpreted as the three two-fold axes of this group~\citep{Xiao1996}. To be able to determine a functional basis with the table of \citet{Boehler1977}, we define the tuple of structural functions by
\begin{equation}
	\Set{F}_{\Group{T}_{h}}=\left( \tens{\Lambda}{4}:\te{C}, \tens{\Lambda}{4}:\te{C}^2 \right)
	\label{eq:example_cubic_structural_functions}
\end{equation}
and the tuple of argument tensors by
\begin{equation}
	\Set{H}_{\Group{T}_{h}}=\left( \te{C}, \tens{\Lambda}{4}:\te{C}, \tens{\Lambda}{4}:\te{C}^2 \right) \commar
	\label{eq:example_cubic_isotropic_extension_functions}
\end{equation}
which are all symmetric 2nd order tensors.
We thus obtain a functional basis containing the 22 invariants
\begin{equation}
	\begin{aligned}
		&\tr[\te{C}],\: \tr[\te{C}^2],\: \tr[\te{C}^3],\:
		\tr[\tens{\Lambda}{4}:\te{C}],\: \tr[(\tens{\Lambda}{4}:\te{C})^2],\: \tr[(\tens{\Lambda}{4}:\te{C})^3],\:
		\tr[(\tens{\Lambda}{4}:\te{C}^2)],\: \tr[(\tens{\Lambda}{4}:\te{C}^2)^2],\: \tr[(\tens{\Lambda}{4}:\te{C}^2)^3], \\
		& \tr[\te{C}\cdot(\tens{\Lambda}{4}:\te{C})],\: \tr[\te{C}^2\cdot(\tens{\Lambda}{4}:\te{C})],\:	\tr[\te{C}\cdot(\tens{\Lambda}{4}:\te{C})^2],\: \tr[\te{C}^2\cdot(\tens{\Lambda}{4}:\te{C})^2],\: \tr[\te{C}\cdot(\tens{\Lambda}{4}:\te{C}^2)],\: \tr[\te{C}^2\cdot(\tens{\Lambda}{4}:\te{C}^2)],\: \\
		& \tr[\te{C}\cdot(\tens{\Lambda}{4}:\te{C}^2)^2],\: \tr[\te{C}^2\cdot(\tens{\Lambda}{4}:\te{C}^2)^2],\: \tr[(\tens{\Lambda}{4}:\te{C})\cdot(\tens{\Lambda}{4}:\te{C}^2)],\: \tr[(\tens{\Lambda}{4}:\te{C})^2\cdot(\tens{\Lambda}{4}:\te{C}^2)],\: \\
		& \tr[(\tens{\Lambda}{4}:\te{C})\cdot(\tens{\Lambda}{4}:\te{C}^2)^2],\: \tr[(\tens{\Lambda}{4}:\te{C})^2\cdot(\tens{\Lambda}{4}:\te{C}^2)^2],\: \tr[ \te{C}\cdot(\tens{\Lambda}{4}:\te{C})\cdot(\tens{\Lambda}{4}:\te{C}^2)] \; .
	\end{aligned}
	\label{eq:example_cubic_complete_invariants}
\end{equation}
\subsubsection{Minimal integrity basis}
\label{ssec:example_cubic_integrity_basis}
By applying the procedure outlined in Sect.~\ref{sec:elimination} for eliminating polynomially expressible invariants, we obtain the reduced functional basis in Tab.~\ref{tab:example_cubic_integrity_basis} for group $\Group{T}_{h}$ from the initial functional basis Eq.~\eqref{eq:example_cubic_complete_invariants}. 
As a proof that the functional basis in Tab.~\ref{tab:example_cubic_integrity_basis} is also a minimal integrity basis, we utilize Proposition~\ref{prop:minimal_integrity_basis} by calculating the polynomial relations between the invariants from Tab.~\ref{tab:example_cubic_integrity_basis} and those proposed by \citet{smith_further_1962}.
To achieve this, the orientation of the structural tensor must be aligned with that of Smith’s invariants, which is why $[\ve{a}_1]=\left(1,0,0\right)^\top$, $[\ve{a}_2]=\left(0,1,0\right)^\top$ and $[\ve{a}_3] = \left(0, 0, 1\right)^\top$ are set accordingly with respect to the cartesian basis $\{\ve{e}_1, \ve{e}_2, \ve{e}_3\}$. The polynomial relationships between the two invariant sets can be found in the supplementary material.
\begin{table}[ht]
	\centering
	\caption{
		Minimal integrity basis for $\Group{A}_{10}$ using the symmetric 2nd order tensor $\te{C}\in\tsposdef$ and the structural tensor $\tens{\Lambda}{4}\in\ts{4}{}$ of group $\Group{T}_h$, which is defined in Eq.~\eqref{eq:example_cubic_structural_tensor}.}
	\label{tab:example_cubic_integrity_basis}
	\renewcommand{\arraystretch}{1.}
	\begin{small}
		\begin{tabularx}{0.88\textwidth}{p{15.cm}}
			\toprule 
			$I_1 = \tr[\te{C}] \hspace{1.5em}
			I_2 = \tr[\te{C}^2] \hspace{1.5em}
			I_3 = \tr[(\tens{\Lambda}{4}:\te{C})^2] \hspace{1.5em}
			I_4 = \tr[\te{C}^3] \hspace{1.5em}
			I_5 = \tr[(\tens{\Lambda}{4}:\te{C})^3] \hspace{1.5em}
			I_6 = \tr[\te{C}^2\cdot(\tens{\Lambda}{4}:\te{C})] \vspace{3pt}\newline
			I_7 = \tr[\te{C}\cdot(\tens{\Lambda}{4}:\te{C})^2] \hspace{1.5em}
			I_8 = \tr[(\tens{\Lambda}{4}:\te{C})\cdot(\tens{\Lambda}{4}:\te{C}^2)] \hspace{1.5em}
			I_9 = \tr[(\tens{\Lambda}{4}:\te{C}^2)^2] \hspace{1.5em}
			I_{10} = \tr[\te{C}^2\cdot(\tens{\Lambda}{4}:\te{C})^2] \vspace{3pt}\newline
			I_{11} = \tr[(\tens{\Lambda}{4}:\te{C})^2\cdot(\tens{\Lambda}{4}:\te{C}^2)] \hspace{1.5em}
			I_{12} = \tr[\te{C}\cdot(\tens{\Lambda}{4}:\te{C}^2)^2] \hspace{1.5em}
			I_{13} = \tr[(\tens{\Lambda}{4}:\te{C})\cdot(\tens{\Lambda}{4}:\te{C}^2)^2] \hspace{1.5em}
			I_{14} = \tr[(\tens{\Lambda}{4}:\te{C}^2)^3]$ \\
			\bottomrule
		\end{tabularx}
	\end{small}
\end{table}
\subsubsection{Hilbert series}
\label{ssec:example_cubic_hilbert_series}
Since we have already proven with the help of Proposition~\ref{prop:minimal_integrity_basis} that the invariants in Tab.~\ref{tab:example_cubic_integrity_basis} form a minimal integrity basis, the HS for this example merely represents a consistency check. On the other hand, if no minimal integrity basis were known from the literature, we would have to use the HS anyway to check whether the invariants actually form a minimal integrity basis.
Both $\Group{T}$ and $\Group{T}_h$ lead to the same HS, which is why the HS only has to be computed for $\Group{T}$. This group is generated by $\rot{z}{\pi}$ and $\rot{k}{\frac{2\pi}{3}}$ and contains twelve group elements that hold 4-times a rotational angle of $\frac{2\pi}{3}$ and $\frac{4\pi}{3}$, respectively, 3-times an angle of $\pi$ and one angle of $0$. Hence, with Eqs.~\eqref{eq:molien_series}~and~\eqref{eq:molien_series_determinant}, the HS
\begin{equation}
	\Hs_{\Group{T}} = \Hs_{\Group{T}_h} = \frac{1}{12}\left[ \frac{1}{\left(1-t\right)^6} + \frac{8}{\left(1-t^3\right)^2} + \frac{3}{\left( 1-t \right)^2 \left( 1-t^2\right)^2} \right]
	\label{eq:example_cubic_calc_molien_series}
\end{equation}
can be computed. To bring the HS in the desired form~\eqref{eq:molien_series_prim_sek}, a multiplication with $(1+t)(1+t^2)/((1+t)(1+t^2))$ is necessary resulting in
\begin{equation}
	\Hs_{\Group{T}} = \Hs_{\Group{T}_h} = \frac{1 + 3t^3 + 2t^4 + 2t^5 + 3t^6 + t^9}{\left(1-t\right)\left(1-t^2\right)^2\left(1-t^3\right)^2\left(1-t^4\right)} \commar
	\label{eq:example_cubic_molien_series}
\end{equation}
for which we already know that this form corresponds to a Hironaka decomposition~\citep{smith_further_1962}.
The corresponding Taylor series expansion is denoted by
\begin{equation}
	\Hs_{\textrm{Tay},\Group{T}} = \Hs_{\textrm{Tay},\Group{T}_h} = 1 + t + 3t^2 + 8t^3 + 14t^4 + 26t^5 + 48t^6 + 76t^7 + \textrm{HOT} \point
	\label{eq:example_cubic_molien_series_taylor}
\end{equation}
\paragraph{Information of Eq.~\eqref{eq:example_cubic_molien_series_taylor}}
The Taylor series~\eqref{eq:example_cubic_molien_series_taylor} can be analyzed as follows: First, the constant term $1$ indicates, trivially, that any number is an invariant. Then the term $t$ points out that one polynomially independent invariant exists at polynomial degree one, which is $I_1$. With this linear invariant, the product $I_1^2$ can be formed at quadratic degree. Thus, this invariant is polynomially expressible in $I_1$ and also reducible. The term $3t^2$ then indicates that three polynomially independent invariants exist of quadratic degree, but since $I_1^2$ is polynomially expressible, it follows that there must be two fundamental invariants of degree two, namely $I_2$ and $I_3$. From the fundamental invariants $I_1$, $I_2$ and $I_3$ of degrees one and two, three polynomially independent products $I_1^3$, $I_1I_2$ and $I_1I_3$ can be formed with cubic degree. Considering the next term, $8t^3$, eight polynomially independent invariants exist in degree three. Subtracting three because of the three polynomially independent products $I_1^3$, $I_1I_2$ and $I_1I_3$, we conclude that five fundamental invariants, $I_4$ to $I_8$, exist of cubic degree. This counting process can be continued up to degree six, where it becomes evident that we need the fundamental invariants $I_9$ to $I_{13}$.

At degree six, 51 products can be formed from the fundamental invariants of lower degrees: $\Set{I} = \left\{ I_1^6, I_1^4I_2, I_1^4I_3, \dots \right\}$. However, the term $48t^6$ indicates that only 48 polynomially independent invariants exist at this degree. From this, one conclude that at least three of those 51 products are not polynomially independent and at least three syzygies must exist in polynomial degree six. In fact, exactly four syzygies 
\begin{subequations}
	\begin{align}
		0 &= I_8^2 + 3I_6^2 - I_3I_9 \\
		0 &= 3I_7I_8 + 3I_5I_6 - 3I_3I_{10} + I_2I_3^2 - I_1I_3I_8 \\
		0 &= 3I_6I_7 - I_5I_8 + I_3I_{11} - I_1I_3I_6 \\
		0 &= - 18I_7^2 - 6I_5^2 + I_3^3 + 12I_1I_3I_7 - 2I_1^2I_3^2
	\end{align}
	\label{eq:example_cubic_degree_six_syzygies}%
\end{subequations}
exist at this polynomial degree, written in the invariants of Tab.~\ref{tab:example_cubic_integrity_basis}. Therefore, four of the 51 products are not independent, reducing the total number of products from 51 to 47 linear independent products: $\Set{I}\setminus\left\{ I_3I_9, I_1I_3I_8, I_1I_3I_6, I_1^2I_3^2 \right\}$, for example. Then, since the Taylor series~\eqref{eq:example_cubic_molien_series_taylor} at this polynomial degree postulates 48 linear independent invariants, there must be one additional fundamental invariant, which is $I_{14}$. For higher polynomial degrees only new syzygies follow but no new fundamental invariants.
\paragraph{Information of Eq.~\eqref{eq:example_cubic_molien_series}}
\citet{smith_further_1962} has specified a suitable set of primary and secondary invariants, so we already know that the form of the HS in Eq.~\eqref{eq:example_cubic_molien_series} corresponds to a Hironaka decomposition.
From the denominator in Eq.~\eqref{eq:example_cubic_molien_series} it can be observed that there exist the following primary invariants: one of degree one, two of degree two, two of degree three and one of degree four. The numerator tells that the following secondary invariants are additionally necessary: the secondary invariant 1 as well as three secondary invariants of degree three, two of degree four, two of degree five, three of degree six and one of degree nine. Furthermore, the secondary invariant of polynomial degree nine and two secondary invariants of polynomial degree six are reducible~\citep{smith_further_1962}. These are necessary for a corresponding Hironaka decomposition, but not for a minimal integrity basis.
\paragraph{Hironaka decomposition}
Since we are primarily interested in minimal integrity bases in this work, we do not determine a Hironaka decomposition for the current example. However, as in \ref{app:example_hs_derksen}, we would like to illustrate that it is also possible in this example to determine a rational representation of the HS that does not correspond to a Hironaka decomposition. To do this, we multiply Eq.~\eqref{eq:example_cubic_molien_series_taylor} by $(1-t^3)/(1-t^3)$, resulting in
\begin{equation}
	\Hs_{\Group{T}} = \Hs_{\Group{T}_h} = \frac{1 + 2t^3 + 3t^4 + 2t^5 + t^8}{\left(1-t\right)\left(1-t^2\right)^2\left(1-t^3\right)^3} \point
	\label{eq:example_cubic_molien_series_wrong}
\end{equation}
\begin{proposition}
	\label{prop:cubic_no_hironaka_decomposition}
	Let $\PolRingE{R}{C_{11},C_{22},C_{33},C_{23},C_{13},C_{12}}{\Group{T}_{h}}=\PolRing{R}{I_1,\dots,I_{14}}{}$, where $I_1,\dots,I_{14}$ are the invariants of Tab.~\ref{tab:example_cubic_integrity_basis}. Then the rational form of the HS~\eqref{eq:example_cubic_molien_series_wrong} does not correspond to any Hironaka decomposition of the invariant ring $\PolRingE{R}{C_{11},C_{22},C_{33},C_{23},C_{13},C_{12}}{\Group{T}_{h}}$.
\end{proposition}
\begin{proof}
	\label{proof:cubic_no_hironaka_decomposition}
	Let us, for a contradiction, assume that there exists a Hironaka decomposition according to the HS~\eqref{eq:example_cubic_molien_series_wrong} consisting of primary invariants $\invar{1}{P_1}, \invar{2}{P_2}, \invar{2}{P_3}, \invar{3}{P_4}, \invar{3}{P_5}, \invar{3}{P_6}$ and secondary invariants $\invar{1}{S_1}=1, \invar{3}{S_2}, \invar{3}{S_3}, \invar{4}{S_4}, \invar{4}{S_5}, \invar{4}{S_6}, \invar{5}{S_7}, \invar{5}{S_8}$, $\invar{8}{S_9}$, where the index on the top left denotes the polynomial degree. Accordingly, with these it is possible to determine any polynomial invariant as a polynomial in the primary and secondary invariants. 
	Now we consider the HS~\eqref{eq:example_cubic_molien_series_taylor} in its series representation, where we know that their exists one invariant in polynomial degree six that can not be expressed as a polynomial of the invariants of lower degrees, in our case $I_{14}$. Therefore, $\invar{1}{P_1}, \invar{2}{P_2}, \invar{2}{P_3}, \invar{3}{P_4}, \invar{3}{P_5}, \invar{3}{P_6}, \invar{1}{S_1}, \invar{3}{S_2}, \invar{3}{S_3}$, $\invar{4}{S_4}, \invar{4}{S_5}, \invar{4}{S_6}, \invar{5}{S_7}, \invar{5}{S_8}$ are also insufficient to express $I_{14}$ as a polynomial.
	Furthermore, since there exists no further primary or secondary invariant of polynomial degree six, $I_{14}$ can not be expressed as a polynomial. This is a contradiction. 
	Hence, it exists no Hironaka decomposition according to the rational form of the HS~\eqref{eq:example_cubic_molien_series_wrong}.
\end{proof}
\begin{remark}
	\label{rem:no_complete_set_found}
	\citet{Xiao1996} used an additional structural tensor $\tens{T}{4}_h^a\in\ts{4}{}$ for $\Group{T}_h$, which is anti-symmetric in the first two indices and symmetric in the last two indices. The corresponding tuple of structural functions is defined by \mbox{$\Set{F}_{\Group{T}_{h}}=\left( \tens{\Lambda}{4}:\te{C}, \tens{T}{4}_h^a:\te{C} \right)$}. However, this does not lead to an integrity basis when used in conjunction with the table provided by \citet{Boehler1977}, as it results in only one invariant of polynomial degree five, whereas the HS~\eqref{eq:example_cubic_molien_series_taylor} tells that two invariants of degree five are required for an integrity basis. This is because Boehler’s basis does not necessarily constitute an integrity basis. However, for most cases considered in this work, the structural functions provided by \citet{Xiao1996} lead to an integrity basis with the invariants defined by \citet{Boehler1977}, but this is not the case for $\Group{T}_h$, $\Group{T}_d$ and $\Group{C}_{3i}$. To overcome this, we simply choose different structural functions in these cases. However, we emphasize that this will not work in general.
\end{remark}

\subsubsection{Quadratic energy function}
\label{ssec:example_cubic_quadratic_energy_function}
As for the monoclinic example in Sect.~\ref{sec:example_monoclinic}, we define a cubic $(\Group{A}_{10})$ Saint-Venant-Kirchhoff model in terms of the homogeneous quadratic Helmholtz free energy
\begin{equation}
	\varPsi^{\Group{T}_{h}} = \alpha_1I_1^2 + \alpha_2I_2 + \alpha_3I_3
	\label{eq:saint_ven_energy_cubic}
\end{equation}
with some material constants $\alpha_\lambda\in\Setnum{R}$, $\lambda \in\{1,2,3\}$. This energy potential is formulated by using the invariants of Tab.~\ref{tab:example_cubic_integrity_basis}, where we replace $\te{C}\in\tsposdef$ with the Green-Lagrange strain tensor $\te{E}\in\tss{2}$. Again, please note that $I_1^2,I_2$ and $I_3$ are polynomially independent, which is the maximum number of polynomially independent invariants with polynomial degree two, cf. $3t^2$ of the HS~\eqref{eq:example_cubic_molien_series_taylor}.
We further specify: $[\ve{a}_1]=\left(1,0,0\right)^\top$, $[\ve{a}_2]=\left(0,1,0\right)^\top$ and $[\ve{a}_3] = \left(0,0,1\right)^\top$, thus fixing the structural tensor with respect to the cartesian basis $\{\ve{e}_1, \ve{e}_2, \ve{e}_3\}$. By taking the second derivative
\begin{equation}
	C_{ijkl}^{\Group{T}_{h}} = \frac{\partial \varPsi^{\Group{T}_{h}}}{\partial E_{ij} \partial E_{kl}} \; ,
	\label{eq:elast_tensor_cubic_example}
\end{equation}
we compute the coordinates $C_{ijkl}^{\Group{T}_{h}}$ of the constant elasticity tensor $\tens{C}{4}^{\Group{T}_{h}}\in\ts{4}{}$. The 4th order elasticity tensor in Kelvin-Mandel notation is
\begin{equation}
	\small{
		\left[C_{\alpha\beta}^{\Group{T}_{h}}\right] = \begin{pmatrix}
			C_{11} & C_{12} & C_{13} & 0 & 0 & 0 \\
			C_{12} & C_{22} & C_{23} & 0 & 0 & 0 \\
			C_{13} & C_{23} & C_{33} & 0 & 0 & 0 \\
			0 & 0 & 0 & 2C_{44} & 0 & 0 \\
			0 & 0 & 0 & 0 & 2C_{55} & 0 \\
			0 & 0 & 0 & 0 & 0 & 2C_{66} \\
		\end{pmatrix}
		\quad\textrm{with}\quad
		\begin{array}{ll}
			C_{11} = C_{22} = C_{33} = 2(\alpha_1 + \alpha_2 + 2\alpha_3),\\
			C_{12} = C_{13} = C_{23} = 2(\alpha_1 - \alpha_3),\\
			C_{44} = C_{55} = C_{66} = \alpha_2,
		\end{array}
	}
	\label{eq:example_cubic_stiffness}
\end{equation}
where $C_{\alpha\beta}^{\Group{T}_{h}} = T^\alpha_{ij}T^\beta_{kl} C_{ijkl}^{\Group{T}_{h}}$ with the transformation coefficients $T^\alpha_{ij}$ defined by Eq.~\eqref{eq:vector_notation}. Given the chosen orientation of the vectors $\ve{a}_1, \ve{a}_2$ and $\ve{a}_3$, the elasticity matrix~\eqref{eq:example_cubic_stiffness} reveals the characteristic occupancy for this anisotropy, see \citet[Appendix A.5]{altenbach_kontinuumsmechanik_2018} or \citet[Sect. 4.6.10.22]{Apel2004} for example.
\section{Minimal integrity bases for all common anisotropy classes in finite strain hyperelasticity}
\label{sec:complete_and_irr_sets}
In this section, the Hilbert series (HS), structural tensors, structural functions and minimal integrity bases are either calculated or provided for all relevant anisotropies $\aniso1$--$\aniso{14}$ in hyperelasticity. As an additional example, the anisotropy $\aniso{15}$, arising from the icosahedral non-crystal system, is also considered. A detailed example for the monoclinic anisotropy $\aniso2$ and cubic anisotropy $\aniso{10}$ is presented in Sects.~\ref{sec:example_monoclinic}~and~\ref{sec:example_cubic}, respectively.
\subsection{Hilbert series}
\label{sec:molien_series_anisotropies}
First, we would like to point out again that the representation of the HS in its rational form~\eqref{eq:molien_series_prim_sek} is not unique and may not correspond to any Hironaka decomposition, see Sects.~\ref{sec:molien_series}~and~\ref{sec:example_cubic}. However, \citet{smith_further_1962} has already provided a Hironaka decomposition for the anisotropies $\Group{A}_1$--$\Group{A}_{11}$, so that the rational form of the HS presented in the following can be validated directly. Similarly, we specify corresponding Hironaka decompositions for the anisotropies $\Group{A}_{12}$ and $\Group{A}_{13}$ in Sect.~\ref{sec:irreducible_integrity bases}.
Next, the HS in rational form for anisotropy $\Group{A}_{14}$ aligns with the well-know structure of the invariant ring in the isotropic case. For the final anisotropy $\Group{A}_{15}$, we postulate that the rational form of the HS actually belongs to a Hironaka decomposition, which is justified in Sect.~\ref{sec:irreducible_integrity bases}.

For all finite groups, the HS can be computed analogously to the examples outlined in Sects.~\ref{sec:example_monoclinic}~and~\ref{sec:example_cubic}. In contrast, the calculation for the infinite groups is more complicated and requires transforming Eq.~\eqref{eq:molien_series} into an integral, which is shown below. The results of the calculated series are summarized in Tab.~\ref{tab:molien_series}.
\begin{table}[!ht]
	\centering
	\caption{HS and the coefficients of their respective Taylor series up to order seven of a symmetrical 2nd order tensor and all considered groups. $(P,S,F)$ denotes the number of primary-, secondary- and fundamental invariants, where the number $1$ is also counted as a secondary invariant.}
	\label{tab:molien_series}
	\renewcommand{\arraystretch}{1.}
	\begin{small}
		\begin{tabularx}{0.90\textwidth}{p{3.1cm}p{5.4cm}p{4.6cm}p{1cm}}
				\toprule
				Group/Anisotropy & Hilbert series & Coefficients of series expansion & $(P,S,F)$ \\
				\midrule
				$\Group{C}_1$, $\Group{C}_i$ / $\aniso{1}$ & 
				$\dfrac{1}{\left( 1-t \right)^6}$ & 
				$\left\{ 1,6,21,56,126,252,462,792  \right\}$ &
				$(6,1,6)$ \\[12pt]
				$\Group{C}_2$, $\Group{C}_{1h}$, $\Group{C}_{2h}$ / $\aniso{2}$ & 
				$\dfrac{1 + t^2}{\left(1-t\right)^4\left(1-t^2\right)^2}$ & 
				$\left\{ 1,4,13,32,70,136,246,416  \right\}$ &
				$(6,2,7)$ \\[12pt]
				$\Group{D}_2$, $\Group{C}_{2v}$, $\Group{D}_{2h}$ / $\aniso{3}$ & 
				$\dfrac{1 + t^3}{\left(1-t\right)^3\left(1-t^2\right)^3}$ & 
				$\left\{ 1,3,9,20,42,78,138,228  \right\}$ &
				$(6,2,7)$ \\[12pt]
				$\Group{C}_4$, $\Group{C}_{2i}$, $\Group{C}_{4h}$ / $\aniso{4}$ & 
				$\dfrac{1 + t^2 + 4t^3 + t^4 + t^6}{\left(1-t\right)^2\left(1-t^2\right)^3\left(1-t^4\right)}$ & 
				$\left\{ 1,2,7,16,36,68,124,208 \right\}$ &
				$(6,8,12)$ \\[12pt]
				$\Group{D}_4$, $\Group{C}_{4v}$, $\Group{D}_{2d}$, $\Group{D}_{4h}$ / $\aniso{5}$ & 
				$\dfrac{1 + 2t^3 + t^6}{\left(1-t\right)^2\left(1-t^2\right)^3\left(1-t^4\right)}$ & 
				$\left\{ 1,2,6,12,25,44,77,124  \right\}$ &
				$(6,4,8)$ \\[12pt]
				$\Group{C}_3$, $\Group{C}_{3i}$ / $\aniso{6}$ & 
				$\dfrac{1 + 2t^2 + 6t^3 + 2t^4 + t^6}{\left(1-t\right)^2\left(1-t^2\right)^2\left(1-t^3\right)^2}$ & 
				$\left\{ 1,2,7,20,42,84,156,264 \right\}$ &
				$(6,12,14)$ \\[12pt]
				$\Group{D}_3$, $\Group{C}_{3v}$, $\Group{D}_{3d}$ / $\aniso{7}$ & 
				$\dfrac{1 + t^2 + 2t^3 + t^4 + t^6}{\left(1-t\right)^2\left(1-t^2\right)^2\left(1-t^3\right)^2}$ & 
				$\left\{ 1,2,6,14,28,52,93,152  \right\}$ &
				$(6,6,9)$ \\[12pt]
				$\Group{C}_6$, $\Group{C}_{3h}$, $\Group{C}_{6h}$ / $\aniso{8}$ & 
				$\dfrac{1 + 3t^3 + 2t^4 + 2t^5 + 3t^6 + t^9}{\left(1-t\right)^2\left(1-t^2\right)^2\left(1-t^3\right)\left(1-t^6\right)}$ & 
				$\left\{ 1,2,5,12,24,46,84,140  \right\}$ &
				$(6,12,14)$ \\[12pt]
				$\Group{D}_6$, $\Group{C}_{6v}$, $\Group{D}_{3h}$, $\Group{D}_{6h}$ / $\aniso{9}$ & 
				$\dfrac{1 + t^3 + t^4 + t^5 + t^6 + t^9}{\left(1-t\right)^2\left(1-t^2\right)^2\left(1-t^3\right)\left(1-t^6\right)}$ & 
				$\left\{ 1,2,5,10,19,33,57,90  \right\}$ &
				$(6,6,9)$ \\[12pt]
				$\Group{T}$, $\Group{T}_{h}$ / $\aniso{10}$ & 
				$\dfrac{1 + 3t^3 + 2t^4 + 2t^5 + 3t^6 + t^9}{\left(1-t\right)\left(1-t^2\right)^2\left(1-t^3\right)^2\left(1-t^4\right)}$ & 
				$\left\{ 1,1,3,8,14,26,48,76  \right\}$ &
				$(6,12,14)$ \\[12pt]
				$\Group{O}$, $\Group{T}_{d}$, $\Group{O}_{h}$ / $\aniso{11}$ & 
				$\dfrac{1 + t^3 + t^4 + t^5 + t^6 + t^9}{\left(1-t\right)\left(1-t^2\right)^2\left(1-t^3\right)^2\left(1-t^4\right)}$ & 
				$\left\{ 1,1,3,6,11,18,32,48  \right\}$ &
				$(6,6,9)$ \\[12pt]
				$\Group{C}_{\infty}$, $\Group{C}_{\infty h}$ / $\aniso{12}$ & 
				$\dfrac{1+t^3}{\left(1-t\right)^2\left(1-t^2\right)^2\left(1-t^3\right)}$ & 
				$\left\{ 1,2,5,10,18,30,48,72  \right\}$ &
				$(5,2,6)$ \\[12pt]
				$\Group{D}_{\infty}$, $\Group{C}_{\infty v}$, $\Group{D}_{\infty h}$ / $\aniso{13}$ & 
				$\dfrac{1}{\left(1-t\right)^2\left(1-t^2\right)^2\left(1-t^3\right)}$ & 
				$\left\{ 1,2,5,9,16,25,39,56  \right\}$ &
				$(5,1,5)$ \\[12pt]
				$\Group{K}$, $\Group{K}_h$ / $\aniso{14}$ & 
				$\dfrac{1}{\left(1-t\right)\left(1-t^2\right)\left(1-t^3\right)}$ & 
				$\left\{ 1,1,2,3,4,5,7,8  \right\}$ &
				$(3,1,3)$ \\[12pt]
				$\Group{I}$, $\Group{I}_h$ / $\aniso{15}$ & 
				$\dfrac{1 + t^5 + 2t^6 + t^7 + t^{12}}{\left(1-t\right)\left(1-t^2\right) \left(1-t^3\right)^2 \left(1-t^4\right) \left(1-t^5\right)}$ & 
				$\left\{ 1,1,2,4,6,10,17,24  \right\}$ &
				$(6,6,10)$ \\
				\bottomrule
			\end{tabularx}
		\end{small}
	\end{table}
	
	\subsubsection{Cylindrical systems}
	The groups $\Group{C}_\infty$ and $\Group{C}_{\infty h}$ lead to the same HS for the symmetric 2nd order tensor $\te{C}\in\tss{2}$. In order to minimize the calculation effort for the HS, group $\Group{C}_\infty$ is used, which contains all pure rotations $\rot{z}{\varphi}$ around an axis. To perform the calculation of the HS, Eq.~\eqref{eq:molien_series} will be written in such a way that the definition of a Riemann integral is obtained. To do so, $\varphi$ is substituted by $k\Delta\varphi$, where $k\in\Setnum{Z}_{>0}$ labels all $n$ group elements of a given finite group ${\Group{C}}_n \approx \Group{C}_\infty$. To ensure that $\Group C_n$ equals $\Group C_\infty$, the limit is taken, i.e., $n$ goes to infinity and $\Delta\varphi$ approaches zero. Hence, with the relationship $\Delta\varphi=\frac{2\pi}{n}$, Eq.~\eqref{eq:molien_series} becomes
	\begin{equation}
		\Hs_{\Group{C}_\infty} \coloneqq
		\lim_{\substack{n \to \infty \\ \Delta\varphi \to 0}} \frac{1}{n}\sum_{k=1}^{n} \frac{1}{f(k\Delta\varphi)} 
		= \lim_{\substack{n \to \infty \\ \Delta\varphi \to 0}} \frac{1}{2\pi}\sum_{k=1}^{n} \Delta\varphi \frac{1}{f(k\Delta\varphi)}
		= \frac{1}{2\pi} \integg{0}{2\pi}{\varphi} \frac{1}{f(\varphi)}.
		\label{eq:cylindriacal_molien_Cinfty_part1}
	\end{equation}
	Inserting the definition of $f(\varphi)$ from Eq.~\eqref{eq:molien_series_determinant}, the integral
	\begin{equation}
		\Hs_{\Group{C}_\infty} = 
		\frac{1}{2\pi} \integg{0}{2\pi}{\varphi} \frac{1}{(1-t)^2 (1-2t\cos\left(\varphi\right) + t^2) (1+2t-4t\cos^2\left(\varphi\right) + t^2)}
		\label{eq:eq:cylindriacal_molien_Cinfty_part2}
	\end{equation}
	is obtained. This integral is solved with the computer algebra program Maxima (\url{https://maxima.sourceforge.io/}) and the final HS is presented in Tab.~\ref{tab:molien_series}.
	
	The point groups $\Group{D}_\infty$, $\Group{C}_{\infty v}$ and $\Group{D}_{\infty h}$ lead to the same HS for the symmetrical 2nd order tensor $\te{C}\in\tss{2}$. In order to minimize the calculation effort for the HS, group $\Group{D}_\infty$ is used, which contains all pure rotations $\rot{z}{\varphi}$ around an axis as well as all rotations $\rot{z}{\varphi}\cdot\rot{x}{\pi}$. Since $\rot{x}{\pi}$ is diagonal, the product $\rot{z}{\varphi}\cdot\rot{x}{\pi}$ is commutative. With that information it follows directly that for every $\rot{z}{\varphi}$ there exists exactly one product $\rot{z}{\varphi}\cdot\rot{x}{\pi}$ in $\Group{D}_\infty$. Furthermore, we note that the resulting rotation angle $\alpha(\varphi)$ of every possible product $\rot{z}{\varphi}\cdot\rot{x}{\pi}$ is always $\pi$. Hence, we distinguish between $\rot{z}{\varphi}$ and $\rot{z}{\varphi}\cdot\rot{x}{\pi}$ in $\Group{D}_\infty$ by splitting the HS into two sums and take the limit analogously to Eq.~\eqref{eq:cylindriacal_molien_Cinfty_part1}:
	\begin{equation}
		\Hs_{\Group{D}_\infty} \coloneqq
		\lim_{\substack{n \to \infty \\ \Delta\varphi \to 0}} \frac{1}{2n}\sum_{k=1}^{n} \left(\frac{1}{f(k\Delta\varphi)} +
		\frac{1}{f(\alpha(k\Delta\varphi))} \right) = \frac{1}{2}\Hs_{C_\infty} + 
		\frac{1}{4\pi} \integg{0}{2\pi}{\varphi} \frac{1}{\left(1-t\right)^2 \left(1-t^2\right)^2} \point
		\label{eq:cylindrical_molien_Dinfty_part1}
	\end{equation}
	The final result for the HS of $\Group{D}_\infty$ is presented in Tab.~\ref{tab:molien_series}.
	\subsubsection{Spherical system}
	For the isotropic case, i.e., $\Group{SO}(3)$ and $\Group{O}(3)$, we first define the rotation angle $\varphi$ in Eq.~\eqref{eq:molien_series_determinant} as a function of the three Euler angles $\alpha$, $\beta$ and $\gamma$ implicitly by
	\begin{equation}
		\cos[\varphi(\alpha, \beta, \gamma)] = \frac{1}{2}\left( \cos(\alpha+\gamma) + \cos(\beta) + \cos(\alpha+\gamma)\cos(\beta) - 1 \right) \commar
		\label{eq:rot_angles_with_euler_angles}
	\end{equation}
	thereby obtaining the expression
	\begin{equation}
		g\left(t,\varphi(\alpha,\beta,\gamma)\right)= \left(1-t\right)^2 \left(1-2t\cos{[\varphi(\alpha, \beta, \gamma)]} + t^2\right) \left(1 +2t -4t\cos^2{[\varphi(\alpha, \beta, \gamma)]} + t^2\right) \point
		\label{eq:long_integrand_isotropic}
	\end{equation}
	Then, the HS can be calculated by the integral 
	\begin{equation}
		\Hs_{\Group{O}(3)} = \Hs_{\Group{SO}(3)} = \frac{1}{8\pi^2}\integg{0}{2\pi}{\alpha}\integg{0}{\pi}{\beta}\integg{0}{2\pi}{\gamma}  \frac{\sin(\beta)}{g\left(t,\varphi(\alpha,\beta,\gamma)\right)} \commar
		\label{eq:integral_isotropic}
	\end{equation}
	where $\sin(\beta) / 8\pi^2$ denotes the Haar probability measure~\citep{ecker_haar_2025}.%
	\footnote{
		To compute the Haar probability measure for $\Group{SO}(3)$, we first consider an element $R_{kl}(\alpha,\beta,\gamma)\ve{e}_k\ve{e}_l=\rot{z}{\alpha}\cdot\rot{x}{\beta}\cdot\rot{z}{\gamma} \in\Group{SO}(3)$. By determining the coefficients $g_{\lambda\mu} = \diffp{R_{kl}}{\lambda} \diffp{R_{kl}}{\mu}$, $\lambda,\mu\in\{ \alpha,\beta,\gamma \}$, of the metric tensor $\te{g}\in\tss{2}$, we can then calculate the determinant $g=\det(\te{g})=8\sin^2(\beta)$. The Haar probability measure is given by $\frac{\sqrt{g}}{n}=\frac{\sin(\beta)}{8\pi^2}$, where $n=\integg{0}{2\pi}{\alpha}\integg{0}{\pi}{\beta}\integg{0}{2\pi}{\gamma} \sqrt{g} = 8\pi^2\sqrt{8}$ is a normalization.}
	Since calculating the integral~\eqref{eq:integral_isotropic} is quiet challenging, the integrand is first expanded into a Taylor series
	\begin{equation}
		g(t,\varphi(\alpha,\beta,\gamma) = \sum_{i=1}^{\infty} \frac{1}{i!} \left.\frac{\textrm{d}^ig}{dt^i}\right|_{t=0}
	\end{equation}
	up to a finite order, which is then integrated term by term. This yields the HS
	\begin{equation}
		\Hs_{\Group{O}(3),\textrm{Tay}} = \Hs_{\Group{SO}(3),\textrm{Tay}} =
		1 + 1t + 2t^2 + 3t^3 + 4t^4 + 5t^5 + 7t^6 + 8t^7 + \textrm{HOT} \point
		\label{eq:molien_series_isotropic_sym_tensor}
	\end{equation}
	The series~\eqref{eq:molien_series_isotropic_sym_tensor} can subsequently be expressed in rational form 
	\begin{equation}
		\Hs_{\Group{O}(3)} = \Ms_{\Group{SO}(3)} = \frac{1}{\left(1-t\right) \left(1-t^2\right) \left(1-t^3\right)} \commar
		\label{eq:molien_series_isotropic_rational_form}
	\end{equation}
	which indicates the existence of exactly three primary invariants in degrees one, two, and three. This is fully consistent with the well-known invariants in the isotropic case. Furthermore, Eq.~\eqref{eq:molien_series_isotropic_rational_form} is also validated by \citet{shariff_smallest_2023}, who showed that the number of primary invariants is three less than the number of independent tensor coordinates in the isotropic case.
	\begin{remark}
		\label{rem:molien_isotropic}
		The maximum order to which the Taylor series expansion in Eq.~\eqref{eq:molien_series_isotropic_sym_tensor} must be carried out can be limited by the result of \citet{spencer_invariants_1961}, which states that the highest degree of a fundamental invariant for any number of non-constant 2nd order symmetric tensors is at most six.
		Therefore, forming the Taylor series beyond the 6th order is only meaningful if syzygies exist among the invariants of interest. However, in the isotropic case and only one symmetric 2nd order tensor, only primary invariants are present and thus no syzygies.
	\end{remark}
	\paragraph{Information regarding linear elasticity}
	In contrast to the nonlinear theory, fewer anisotropies can be distinguished for linear elasticity, i.e., $\ve \sigma = \tens{C}{4}:\ve \varepsilon$ with $\ve \varepsilon\in\tss{2}$ being the linearized or technical strain tensor. This is since quadratic energy densities have to be considered in that case and thus
	only invariants with polynomial degree two or less are relevant. Which anisotropies coincide can be inferred directly from the coefficients of the Taylor series in Tab.~\ref{tab:molien_series}, as well as from the knowledge of which groups are subgroups of other groups. 
	This is because if an invariant is invariant with respect to all transformations of a group $\Group{G}$, then it is also invariant with respect to all transformations of a subgroup $\Group{G}_{\textrm{sub}}\subseteq\Group{G}$. If the number of fundamental invariants up to a given degree then matches for both groups, it is no longer possible to distinguish whether $\Group{G}_{\textrm{sub}}\subseteq\Group{G}$ or $\Group{G}$ is considered from those invariants.
	
	Consider the groups $\Group{T}$ and $\Group{O}$ as an example or equivalently speaking the two cubic anisotropies $\aniso{10}$ and $\aniso{11}$. We know that $\Group{T}\subseteq\Group{O}$. Thus, every $\Group{O}$-invariant is also a $\Group{T}$-invariant. Furthermore, the first three coefficients of the respective Taylor series in Tab.~\ref{tab:molien_series} read $1$, $1$ and $3$ in both cases. From this it is obvious that the number of fundamental invariants are also equal up to polynomial degree two. But since the number of fundamental invariants are equal and every $\Group{O}$-invariant is also a $\Group{T}$-invariant, the fundamental invariants have to be identical up to polynomial degree two or can be chosen identical. Hence, the two cubic anisotropies $\aniso{10}$ and $\aniso{11}$ can not be distinguished when only considering invariants up to degree two.
	It should be noted that we deliberately did not choose group $\Group{T}_h$ and group $\Group{O}$ as examples, since $\Group{T}_h$ is not a subgroup of $\Group{O}$. However, as also noted in Remark~\ref{rem:inversion_does_not_lead}, the invariants for a 2nd order tensor and groups $\Group{T}$ and $\Group{T}_h$ are identical. Therefore, the discussion would fall back to groups $\Group{T}$ and $\Group{O}$ anyway.
	
	By doing this analogously for all anisotropies examined, it first follows that the four anisotropies $\aniso{8}$, $\aniso{9}$, $\aniso{12}$ and $\aniso{13}$ are indistinguishable in linear elasticity. Analogously $\aniso{10}$ and $\aniso{11}$ as well as $\aniso{14}$ and $\aniso{15}$ can not be distinguished when only fundamental invariants up to polynomial degree 2 are considered, respectively. Hence, the 15 considered anisotropies in this work reduce to 10 anisotropies in linear elasticity~\citep{Apel2004,otto_data-driven_2025}. However, regarding the elasticity tensor $\tens{C}{4}$, $\aniso{4}$ can not be distinguished from $\aniso{5}$ and $\aniso{6}$ not from $\aniso{7}$, respectively. This is because, by rotating the elasticity tensor appropriately, the characteristic symmetries of the tensor can no longer be distinguished in these cases. Thus, only 8 anisotropies are identifiable in linear elasticity by looking at the elasticity tensor~\citep{Moakher2006}.
	\subsection{Structural tensors and structural functions}
	\label{sec:structural_tensors}
	Some point groups result in the same anisotropies in hyperelasticity, see Remark~\ref{rem:inversion_does_not_lead}. Nevertheless, in order to provide an alternative but equivalent set of invariants in the next section, all groups studied by \citet{Xiao1996,xiao_unified_1997,Xiao_icosahedral} and relevant to this work are taken into account. The definitions of the required structural tensors are listed in Tab.~\ref{tab:structrual_tensors}, while Tab.~\ref{tab:isotropic_extension} provides an overview of the structural tensors and structural functions used for all considered groups. Furthermore, in Tabs.~\ref{tab:single_to_multiple}~and~\ref{tab:single_to_multiple_icosahedral} the conversions of single structural tensors~\citep{Zheng1993} to the multiple structural tensors~\citep{Xiao1996} are presented. We emphasize that, for the trigonal and hexagonal crystal systems, the formulas presented in Tab.~\ref{tab:single_to_multiple} refer to a specific relative orientation between the vectors associated with the respective structural tensors. 
	\begin{table}[ht]
		\centering
		\caption{Definitions of structural tensors $\tens{P}{n}_n$, $\tens{P}{n}_n^{'}$, $\tens{\Lambda}{4}$, $\tens{\Sigma}{3}$ and $\tens{\Theta}{4}$ from \citet{Zheng1993} as well as $\tens{D}{4}_{4h}$, $\tens{D}{4}_{3d}$, $\tens{D}{3}_{3h}$, $\tens{D}{6}_{6h}$, $\tens{I}{6}_{h}$, $\tens{I}{8}_h$ and $\tens{I}{10}_h$ from \citet{Xiao1996,xiao_unified_1997}. $\ve{a}_\sigma\cdot\ve{a}_\tau=\delta_{\sigma\tau}$, $\ve{a}_3\cdot\ve{l}_\sigma=0$, $\ve{l}_\sigma\cdot\ve{l}_\tau\overset{\sigma\neq\tau}{=}-1/2$ and $\ve{l}_\sigma\cdot\ve{l}_\sigma=\ve{n}_\sigma\cdot\ve{n}_\sigma=1$, where $\ve{n}_\sigma$ denotes the six five-fold axes of $\Group{I}_h$.}
		\label{tab:structrual_tensors}
		\renewcommand{\arraystretch}{1.}
		\begin{small}
			\begin{tabularx}{0.94\textwidth}{p{2.8cm}p{2.6cm}p{3.0cm}p{3.5cm}p{4.8cm}}
				\toprule
				$\te{P}_2^{'} = \ve{a}_1\ve{a}_2 + \ve{a}_2\ve{a}_1$ &
				$\te{P}_2 = \ve{a}_1\ve{a}_1 - \ve{a}_2\ve{a}_2$ &
				$\tens{P}{3}_3^{'} = \ve{a}_1\te{P}_2^{'} + \ve{a}_2\te{P}_2$ &
				$\tens{P}{4}_4^{'} = \te{P}_2\te{P}_2^{'} + \te{P}_2^{'}\te{P}_2$ &
				$\tens{P}{4}_4 = \te{P}_2\te{P}_2 - \te{P}_2^{'}\te{P}_2^{'}$ \\
				$\tens{P}{6}_6 = \te{P}_2\tens{P}{4}_4 - \te{P}_2^{'}\tens{P}{4}_4^{'}$ &
				$\tens{D}{4}_{4h} = \sum_{\sigma=1}^2 \overset{4}{\otimes}\ve{a}_\sigma$ &
				$\tens{D}{4}_{3d} = \sum_{\sigma=1}^3 \tens{E}{3}\cdot\overset{3}{\otimes}\ve{l}_\sigma$ &
				$\tens{D}{3}_{3h} = \sum_{\sigma=1}^3 \overset{3}{\otimes}\ve{l}_\sigma$ &
				$\tens{D}{6}_{6h} = \sum_{\sigma=1}^3 \overset{6}{\otimes}\ve{l}_\sigma$ \\
				$\tens{I}{6}_h = \sum_{\sigma=1}^6 \overset{6}{\otimes}\ve{n}_\sigma$ &
				$\tens{I}{8}_{h} = \sum_{\sigma=1}^6 \overset{8}{\otimes}\ve{n}_\sigma$ &
				$\tens{I}{10}_{h} = \sum_{\sigma=1}^6 \overset{10}{\otimes}\ve{n}_\sigma$ &
				$\tens{C}{4}_{3i} = \sum_{\sigma=1}^3 \ve{a}_3\ve{l}_\sigma\ve{l}_\sigma\ve{l}_\sigma$ &
				$\tens{\Theta}{4} = \sum_{\sigma=1}^3 \overset{4}{\otimes}\ve{a}_\sigma$ \\[4pt]
				\multicolumn{3}{l}{$\tens{\Sigma}{3} = \ve{a}_1\ve{a}_2\ve{a}_3 + \ve{a}_3\ve{a}_1\ve{a}_2 + \ve{a}_2\ve{a}_3\ve{a}_1 + \ve{a}_3\ve{a}_2\ve{a}_1 + \ve{a}_1\ve{a}_3\ve{a}_2 + \ve{a}_2\ve{a}_1\ve{a}_3$} &
				\multicolumn{2}{l}{$\tens{C}{4}_{3i}^{\textrm{s}} = \frac{1}{2}\sum_{\sigma=1}^3 \left(\ve{a}_3\ve{l}_\sigma\ve{l}_\sigma\ve{l}_\sigma + \ve{l}_\sigma\ve{a}_3\ve{l}_\sigma\ve{l}_\sigma\right)$} \\[4pt]
				\multicolumn{5}{l}{$\tens{\Lambda}{4} = \ve{a}_1\ve{a}_1\ve{a}_2\ve{a}_2 - \ve{a}_2\ve{a}_2\ve{a}_1\ve{a}_1 + \ve{a}_2\ve{a}_2\ve{a}_3\ve{a}_3 - \ve{a}_3\ve{a}_3\ve{a}_2\ve{a}_2 + \ve{a}_3\ve{a}_3\ve{a}_1\ve{a}_1 - \ve{a}_1\ve{a}_1\ve{a}_3\ve{a}_3$} \\
				\bottomrule
			\end{tabularx}
		\end{small}
	\end{table}
	\begin{table}[p]
		\centering
		\caption{Structural tensors and isotropic extension functions of a single tensor $\te{C}\in\tss{2}$ of all common anisotropies in hyperelasticity as well as the icosahedral anisotropy, see \citet{Zheng1993,Xiao1996,xiao_unified_1997}. Definitions of the structural tensors are provided in Tab.~\ref{tab:structrual_tensors}, while Tabs.~\ref{tab:single_to_multiple}~and~\ref{tab:single_to_multiple_icosahedral} illustrate how the listed multiple structural tensors can be derived from the single structural tensor.}
		\label{tab:isotropic_extension}
		\renewcommand{\arraystretch}{1.5}
		\begin{small}
			\begin{tabularx}{0.92\textwidth}{p{2.5cm}p{2.9cm}p{3.4cm}p{5.9cm}}
				\toprule
				Group/Anisotropy & Single structural tensor & Multiple structural tensors & Structural functions \\
				\midrule
				$\Group{C}_i$ / $\Group{G}_{2}$ / $\aniso{1}$
				& $\te{P}_2^{'} + \tens{E}{3}\cdot\ve{a}_1$ 
				& $\tens{E}{3}\cdot\ve{a}_1, \tens{E}{3}\cdot\ve{a}_2$ 
				& $\tens{E}{3}\cdot\ve{a}_1, \tens{E}{3}\cdot\ve{a}_2$ \\
				\midrule
				$\Group{C}_{1h}$ / $\Group{G}_{3}$ / $\aniso{2}$
				& $\ve{a}_3\ve{a}_3\ve{a}_2 + \tens{E}{3}\cdot\ve{a}_3\ve{a}_2$
				& $\ve{a}_1, \ve{a}_2$ 
				& $\ve{a}_1, \ve{a}_2$ \\
				$\Group{C}_{2h}$ / $\Group{G}_{5}$ / $\aniso{2}$ 
				& $\te{P}_2 + \tens{E}{3}\cdot\ve{a}_3$
				& $\tens{E}{3}\cdot\ve{a}_3, \te{P}_2$ 
				& $\tens{E}{3}\cdot\ve{a}_3, \te{P}_2$ \\
				\midrule
				$\Group{C}_{2v}$ / $\Group{G}_{6}$ / $\aniso{3}$ 
				& $\ve{a}_3\te{P}_2 + \overset{3}{\otimes}\ve{a}_3$
				& $\te{P}_{2}, \ve{a}_3$
				& $\te{P}_{2}, \ve{a}_3$ \\
				$\Group{D}_{2h}$ / $\Group{G}_{8}$ / $\aniso{3}$ 
				& $\te{P}_{2}$
				& -
				& $\te{P}_{2}$ \\
				\midrule
				$\Group{C}_{2i}$ / $\Group{G}_{9}$ / $\aniso{4}$ 
				& $\ve{a}_3\te{P}_2 + \tens{E}{3}\cdot\te{P}_2$ 
				& $\tens{D}{4}_{4h}, \tens{\Sigma}{3}, \tens{E}{3}\cdot\ve{a}_3$
				& $\tens{D}{4}_{4h}:\te{C}, \tens{\Sigma}{3}:\te{C}, \tens{E}{3}\cdot\ve{a}_3$ \\
				$\Group{C}_{4h}$ / $\Group{G}_{11}$ / $\aniso{4}$ 
				& $\tens{P}{4}_4 + \tens{E}{3}\cdot\overset{3}{\otimes}\ve{a}_3$ 
				& $\tens{D}{4}_{4h},\tens{E}{3}\cdot\ve{a}_3$ 
				& $\tens{D}{4}_{4h}:\te{C}, \tens{D}{4}_{4h}:\te{C}^2, \tens{E}{3}\cdot\ve{a}_3$ \\
				\midrule
				$\Group{C}_{4v}$ / $\Group{G}_{13}$ / $\aniso{5}$ 
				& $\ve{a}_3\tens{P}{4}_4 + \overset{5}{\otimes}\ve{a}_3$
				& $\tens{D}{4}_{4h}, \ve{a}_3$
				& $\tens{D}{4}_{4h}:\te{C}, \tens{D}{4}_{4h}:\te{C}^2, \ve{a}_3$ \\
				$\Group{D}_{2d}$ / $\Group{G}_{12}$ / $\aniso{5}$ 
				& $\ve{a}_3\te{P}_2^{'}$
				& $\tens{D}{4}_{4h}, \tens{\Sigma}{3}, \ve{a}_3\ve{a}_3$
				& $\tens{D}{4}_{4h}:\te{C}, \tens{\Sigma}{3}:\te{C}, \ve{a}_3\ve{a}_3$ \\
				$\Group{D}_{4h}$ / $\Group{G}_{15}$ / $\aniso{5}$ 
				& $\tens{P}{4}_4$ 
				& $\tens{D}{4}_{4h}, \ve{a}_3\ve{a}_3$
				& $\tens{D}{4}_{4h}:\te{C}, \tens{D}{4}_{4h}:\te{C}^2, \ve{a}_3\ve{a}_3$ \\
				\midrule
				$\Group{C}_{3i}$ / $\Group{G}_{17}$ / $\aniso{6}$ 
				& $\ve{a}_3\tens{P}{3}_3^{'} + \tens{E}{3}\cdot\overset{3}{\otimes}\ve{a}_3$ 
				& $\tens{C}{4}_{3i}, \tens{E}{3}\cdot\ve{a}_3$
				& $\tens{C}{4}_{3i}^{\textrm{s}}:\te{C}, \tens{C}{4}_{3i}^{\textrm{s}}:\te{C}^2, \tens{E}{3}\cdot\ve{a}_3$ \\
				\midrule
				$\Group{C}_{3v}$ / $\Group{G}_{18}$ / $\aniso{7}$ 
				& $\tens{P}{3}_3^{'} + \overset{3}{\otimes}\ve{a}_3$ 
				& $\tens{D}{4}_{3d}, \ve{a}_3$
				& $\tens{D}{4}_{3d}:\te{C}, \tens{D}{4}_{3d}:\te{C}^2, \ve{a}_3$ \\
				$\Group{D}_{3d}$ / $\Group{G}_{20}$ / $\aniso{7}$ 
				& $\ve{a}_3\tens{P}{3}_3^{'}$ 
				& $\tens{D}{4}_{3d}, \ve{a}_3\ve{a}_3$
				& $\tens{D}{4}_{3d}:\te{C}, \tens{D}{4}_{3d}:\te{C}^2, \ve{a}_3\ve{a}_3$ \\
				\midrule
				$\Group{C}_{3h}$ / $\Group{G}_{21}$ / $\aniso{8}$ 
				& $\ve{a}_3\ve{a}_3\tens{P}{3}_3^{'} + \tens{E}{3}\cdot\ve{a}_3\tens{P}{3}_3^{'}$ 
				& $\tens{D}{3}_{3h}, \tens{E}{3}\cdot\ve{a}_3$
				& $\tens{D}{3}_{3h}:\te{C}, \tens{D}{3}_{3h}:\te{C}^2, \tens{E}{3}\cdot\ve{a}_3$ \\
				$\Group{C}_{6h}$ / $\Group{G}_{23}$ / $\aniso{8}$ 
				& $\tens{P}{6}_6 + \tens{E}{3}\cdot\overset{5}{\otimes}\ve{a}_3$ 
				& $\tens{D}{6}_{6h}, \tens{E}{3}\cdot\ve{a}_3$
				& $\tens{D}{6}_{6h}::\left(\te{C}\te{C}\right), \tens{D}{6}_{6h}::\left(\te{C}^2\te{C}^2\right), \tens{E}{3}\cdot\ve{a}_3$ \\
				\midrule
				$\Group{C}_{6v}$ / $\Group{G}_{25}$ / $\aniso{9}$ 
				& $\te{a}_3\tens{P}{6}_6 + \overset{7}{\otimes}\ve{a}_3$
				& $\tens{D}{6}_{6h}, \ve{a}_3$
				& $\tens{D}{6}_{6h}::\left(\te{C}\te{C}\right), \tens{D}{6}_{6h}::\left(\te{C}^2\te{C}^2\right), \ve{a}_3$ \\
				$\Group{D}_{3h}$ / $\Group{G}_{24}$ / $\aniso{9}$ 
				& $\tens{P}{3}_3^{'}$ 
				& $\tens{D}{3}_{3h}, \ve{a}_3\ve{a}_3$
				& $\tens{D}{3}_{3h}:\te{C}, \tens{D}{3}_{3h}:\te{C}^2, \ve{a}_3\ve{a}_3$ \\
				$\Group{D}_{6h}$ / $\Group{G}_{27}$ / $\aniso{9}$ 
				& $\tens{P}{6}_6$
				& $\tens{D}{6}_{6h}, \ve{a}_3\ve{a}_3$
				& $\tens{D}{6}_{6h}::\left(\te{C}\te{C}\right), \tens{D}{6}_{6h}::\left(\te{C}^2\te{C}^2\right), \ve{a}_3\ve{a}_3$ \\
				\midrule
				$\Group{T}_{h}$ / $\Group{G}_{29}$ / $\aniso{10}$ 
				& $\tens{\Lambda}{4}$ 
				& - 
				& $\tens{\Lambda}{4}:\te{C}, \tens{\Lambda}{4}:\te{C}^2$ \\
				\midrule
				$\Group{T}_{d}$ / $\Group{G}_{30}$ / $\aniso{11}$ 
				& $\tens{\Sigma}{3}$ 
				& - 
				& $\tens{\Sigma}{3}:\te{C}, \tens{\Sigma}{3}:\te{C}^2$ \\
				$\Group{O}_{h}$ / $\Group{G}_{32}$ / $\aniso{11}$ 
				& $\tens{\Theta}{4}$
				& -
				& $\tens{\Theta}{4}:\te{C}, \tens{\Theta}{4}:\te{C}^2$ \\
				\midrule
				$\Group{C}_\infty$ / $\Group{T}_{1}$ / $\aniso{12}$ 
				& $\overset{3}{\otimes}\ve{a}_3+\tens{E}{3}\cdot\ve{a}_3\ve{a}_3$ 
				& $\ve{a}_3, \tens{E}{3}\cdot\ve{a}_3$ 
				& $\ve{a}_3, \tens{E}{3}\cdot\ve{a}_3$ \\
				$\Group{C}_{\infty h}$ / $\Group{T}_{3}$ / $\aniso{12}$ 
				& $\tens{E}{3}\cdot\ve{a}_3$ 
				& - 
				& $\tens{E}{3}\cdot\ve{a}_3$ \\
				\midrule
				$\Group{C}_{\infty v}$ / $\Group{T}_{2}$ / $\aniso{13}$ & $\ve{a}_3$        & - & $\ve{a_3}$ \\
				$\Group{D}_{\infty h}$ / $\Group{T}_{4}$ / $\aniso{13}$ & $\ve{a}_3\ve{a}_3$  & - & $\ve{a}_3\ve{a}_3$ \\
				\midrule
				$\Group{K}_h$ / $\aniso{14}$ & $\te{1}$      & - & $\te{1}$ \\
				\midrule
				$\Group{I}_h$ / $\aniso{15}$ & $\tens{I}{6}_h$ 
				& $\tens{I}{6}_h, \tens{I}{8}_h, \tens{I}{10}_h$
				& $\tens{I}{6}_h::(\te{C}\te{C}), \tens{I}{6}_h::(\te{C}\te{C}^2)$ \\
				\bottomrule
			\end{tabularx}
		\end{small}
	\end{table}
	\begin{table}[p]
		\centering
		\caption{Calculation of multiple structural tensors from single structural tensors for all groups considered in Tab.~\ref{tab:isotropic_extension} except $\Group{C}_\infty$ and $\Group{I}_h$. Relations between structural tensors of the latter two are presented in Tab.~\ref{tab:single_to_multiple_icosahedral}. Definitions of all structural tensors are provided in Tab.~\ref{tab:structrual_tensors}. 
			\newline
			\textcolor{red}{$\blacktriangle$} $\ve{l}_1=\ve{a}_1$, $\ve{l}_2=-\frac{1}{2}\ve{a}_1+\frac{\sqrt{3}}{2}\ve{a}_2$ and $\ve{l}_3=-\frac{1}{2}\ve{a}_1-\frac{\sqrt{3}}{2}\ve{a}_2$  $\quad\mid\quad$  \textcolor{blue}{$\blacktriangledown$} $\ve{l}_1=\ve{a}_2$, $\ve{l}_2=\frac{\sqrt{3}}{2}\ve{a}_1-\frac{1}{2}\ve{a}_2$ and $\ve{l}_3=-\frac{\sqrt{3}}{2}\ve{a}_1-\frac{1}{2}\ve{a}_2$.}
		\label{tab:single_to_multiple}
		\renewcommand{\arraystretch}{1.}
		\begin{small}
			\begin{tabularx}{\textwidth}{p{2.2cm}p{3.4cm}p{10cm}}
				\toprule
				Group/Anisotropy & Single structural tensor $\tens{M}{n}$ & Multiple structural tensors \\
				\midrule
				$\Group{C}_i$ / $\Group{G}_{2}$ / $\aniso{1}$
				& $\te{M} = \te{P}_2^{'} + \tens{E}{3}\cdot\ve{a}_1$ 
				& $\tens{E}{3}\cdot\ve{a}_1 = \frac{1}{2}\left(M_{ij}-M_{ji}\right)\ve{e}_i\ve{e}_j$ \quad 
				$\tens{E}{3}\cdot\ve{a}_2 = \frac{1}{2}\left( M_{ki}M_{jk} - M_{ik}M_{kj}\right)\ve{e}_i\ve{e}_j$ \\
				\midrule
				$\Group{C}_{1h}$ / $\Group{G}_{3}$ / $\aniso{2}$
				& $\tens{M}{3} = \ve{a}_3\ve{a}_3\ve{a}_2 + \tens{E}{3}\cdot\ve{a}_3\ve{a}_2$
				& $\ve{a}_1 = M_{ikk}\ve{e}_i$ \quad $\ve{a}_2 = M_{kki}\ve{e}_i$ \\
				$\Group{C}_{2h}$ / $\Group{G}_{5}$ / $\aniso{2}$ 
				& $\te{M} = \te{P}_2 + \tens{E}{3}\cdot\ve{a}_3$
				& $\tens{E}{3}\cdot\ve{a}_3 = \frac{1}{2}\left(M_{ij}-M_{ji}\right)\ve{e}_i\ve{e}_j$ \quad
				$\te{P}_2 = \frac{1}{2}\left(M_{ij}+M_{ji}\right)\ve{e}_i\ve{e}_j$ \\
				\midrule
				$\Group{C}_{2v}$ / $\Group{G}_{6}$ / $\aniso{3}$ 
				& $\tens{M}{3} = \ve{a}_3\te{P}_2 + \overset{3}{\otimes}\ve{a}_3$
				& $\te{P}_{2} = \left(M_{kkm}M_{mij} - M_{kki}M_{mmj}\right)\ve{e}_i\ve{e}_j$ \quad
				$\ve{a}_3 = M_{kki}\ve{e}_i$ \\
				\midrule
				$\Group{C}_{2i}$ / $\Group{G}_{9}$ / $\aniso{4}$ 
				& $\tens{M}{3} = \ve{a}_3\te{P}_2 + \tens{E}{3}\cdot\te{P}_2$ 
				& $\tens{D}{4}_{4h} = \frac{1}{200}\left[ 4A_{im}A_{mj}A_{kn}A_{nl} + 25\left(A_{ij}-B_{ij}\right) \left(A_{kl}-B_{kl}\right)  \right]\ve{e}_i\ve{e}_j\ve{e}_k\ve{e}_l$ \newline
				with \quad $A_{ij} = E_{ikl}M_{klj}$ and $B_{ij} = E_{ikl}\left( M_{jkl} + M_{kjl} \right)$ \newline
				$\tens{\Sigma}{3} = \left( A_{ijk} + A_{jik} + A_{ikj} \right)\ve{e}_i\ve{e}_j\ve{e}_k$ \newline
				with \quad $A_{ijk} = M_{ijk} - \frac{1}{3}\left( M_{imn}M_{lmn} - \frac{1}{2}M_{min}M_{mln} \right) M_{ljk}$\newline
				$\tens{E}{3}\cdot\ve{a}_3 = \left( M_{kli} - B_{kli} \right) B_{klj} \ve{e}_i\ve{e}_j$ \newline
				with \quad $B_{kli} = \frac{1}{3}\left( M_{kmn}M_{omn} - \frac{1}{2}M_{mkn}M_{mon} \right) \left( M_{oli} + M_{loi} \right)$ \\
				$\Group{C}_{4h}$ / $\Group{G}_{11}$ / $\aniso{4}$ 
				& $\tens{M}{4} = \tens{P}{4}_4 + \tens{E}{3}\cdot\overset{3}{\otimes}\ve{a}_3$ 
				& $\tens{D}{4}_{4h} = \frac{1}{32}\left[ 4\left(M_{ijkl}+M_{jikl}+M_{ijmn}M_{nmkl}\right) + M_{imno}M_{onmj}M_{kpqr}M_{rqpl} \right]\ve{e}_i\ve{e}_j\ve{e}_k\ve{e}_l$ \newline
				$\tens{E}{3}\cdot\ve{a}_3 = M_{ijkk}\ve{e}_i\ve{e}_j$ \\
				\midrule
				$\Group{C}_{4v}$ / $\Group{G}_{13}$ / $\aniso{5}$ 
				& $\tens{M}{5} = \ve{a}_3\tens{P}{4}_4 + \overset{5}{\otimes}\ve{a}_3$
				& $\tens{D}{4}_{4h} = \frac{1}{32}\left[ 8P_{ijkl} + 4P_{ijmn}P_{nmkl} + P_{imno}P_{onmj}P_{kpqr}P_{rqpl} \right]\ve{e}_i\ve{e}_j\ve{e}_k\ve{e}_l$ \newline
				with \quad $P_{ijkl} = M_{mmnno}M_{oijkl} - M_{mmnni}M_{ooppj}M_{qqrrk}M_{ssttl}$ \newline
				$\ve{a}_3 = M_{jjkki}\ve{e}_i$ \\
				$\Group{D}_{2d}$ / $\Group{G}_{12}$ / $\aniso{5}$ 
				& $\tens{M}{3} = \ve{a}_3\te{P}_2^{'}$
				& $\tens{D}{4}_{4h} = \frac{1}{2}\left[ M_{min}M_{mnj}M_{okp}M_{opl} + M_{min}M_{mnk}M_{ojp}M_{opl} - M_{mkj}M_{mil} \right] \ve{e}_i\ve{e}_j\ve{e}_k\ve{e}_l$ \newline
				$\tens{\Sigma}{3} = \left( M_{ijk} + M_{jik} + M_{kji} \right)\ve{e}_i\ve{e}_j\ve{e}_k$ \newline
				$\ve{a}_3\ve{a}_3 = \frac{1}{2}M_{ikl}M_{jkl}\ve{e}_i\ve{e}_j$ \\
				$\Group{D}_{4h}$ / $\Group{G}_{15}$ / $\aniso{5}$ 
				& $\tens{M}{4} = \tens{P}{4}_4$ 
				& $\tens{D}{4}_{4h} = \frac{1}{32}\left[ 8M_{ijkl} + 4M_{ijmn}M_{nmkl} + M_{imno}M_{onmj}M_{kpqr}M_{rqpl} \right]\ve{e}_i\ve{e}_j\ve{e}_k\ve{e}_l$ \newline
				$\ve{a}_3\ve{a}_3 = E_{ilk}E_{jmn}M_{mopq}M_{qpoj}M_{krst}M_{tsrn}\ve{e}_i\ve{e}_j$ \\
				\midrule
				$\Group{C}_{3i}$ / $\Group{G}_{17}$ / $\aniso{6}$ \textcolor{blue}{$\blacktriangledown$}
				& $\tens{M}{4} = \ve{a}_3\tens{P}{3}_3^{'} + \tens{E}{3}\cdot\overset{3}{\otimes}\ve{a}_3$ 
				& $\tens{C}{4}_{3i} = -\frac{3}{16}\left( 4M_{ijkl} - M_{ijmm}E_{kno}M_{nopp}E_{lqr}M_{qrss} \right)   \ve{e}_i\ve{e}_j\ve{e}_k\ve{e}_l$ \:\:
				$\tens{E}{3}\cdot\ve{a}_3 = M_{ijkk}\ve{e}_i\ve{e}_j$ \\
				\midrule
				$\Group{C}_{3v}$ / $\Group{G}_{18}$ / $\aniso{7}$ \textcolor{red}{$\blacktriangle$}
				& $\tens{M}{3} = \tens{P}{3}_3^{'} + \overset{3}{\otimes}\ve{a}_3$ 
				& $\tens{D}{4}_{3d} = \frac{3}{4}\left[ M_{jll}M_{ikm} - M_{ill}M_{jkm} + \left( M_{ill}M_{jnn} - M_{jll}M_{inn} \right) M_{koo}M_{mpp} \right] \ve{e}_i\ve{e}_j\ve{e}_k\ve{e}_m$ \newline
				$\ve{a}_3 = M_{ikk}\ve{e}_i$ \\
				$\Group{D}_{3d}$ / $\Group{G}_{20}$ / $\aniso{7}$ \textcolor{red}{$\blacktriangle$}
				& $\tens{M}{4} = \ve{a}_3\tens{P}{3}_3^{'}$ 
				& $\tens{D}{4}_{3d} = \frac{3}{4}\left( M_{jikl} - M_{ijkl} \right)\ve{e}_i\ve{e}_j\ve{e}_k\ve{e}_l$ \quad
				$\ve{a}_3\ve{a}_3 = \frac{1}{4}M_{iklm}M_{jklm}\ve{e}_i\ve{e}_j$ \\
				\midrule
				$\Group{C}_{3h}$ / $\Group{G}_{21}$ / $\aniso{8}$ \textcolor{blue}{$\blacktriangledown$}
				& $\tens{M}{5} = \ve{a}_3\ve{a}_3\tens{P}{3}_3^{'} + \tens{E}{3}\cdot\ve{a}_3\tens{P}{3}_3^{'}$ 
				& $\tens{D}{3}_{3h} = -\frac{3}{4}M_{mmijk}\ve{e}_i\ve{e}_j\ve{e}_k$ \quad
				$\tens{E}{3}\cdot\ve{a}_3 = \frac{1}{8}\left( M_{ijklm}M_{nnklm} - M_{jiklm}M_{nnklm} \right)\ve{e}_i\ve{e}_j$ \\
				$\Group{C}_{6h}$ / $\Group{G}_{23}$ / $\aniso{8}$ \textcolor{blue}{$\blacktriangledown$}
				& $\tens{M}{6} = \tens{P}{6}_6 + \tens{E}{3}\cdot\overset{5}{\otimes}\ve{a}_3$ 
				& $\tens{D}{6}_{6h} = \frac{3}{64}\left(2I_{ijklmn} - M_{ijklmn} - M_{jiklmn}\right) \ve{e}_i\ve{e}_j\ve{e}_k\ve{e}_l\ve{e}_m\ve{e}_n$ \newline 
				with \quad $I_{ijklmn} = \frac{2}{3}\left(I_{ij}I_{klmn} + I_{ik}I_{jlmn} + I_{il}I_{kjmn} + I_{im}I_{kljn} + I_{in}I_{klmj}\right)$, \newline
				$I_{klmn} = I_{kl}I_{mn} + I_{km}I_{ln} + I_{kn}I_{ml}$ \quad and \quad $I_{ij} = \frac{1}{17}M_{iklmno}M_{jklmno}$\newline
				$\tens{E}{3}\cdot\ve{a}_3 = M_{ijkkll}\ve{e}_i\ve{e}_j$ \\
				\midrule
				$\Group{C}_{6v}$ / $\Group{G}_{25}$ / $\aniso{9}$ \textcolor{blue}{$\blacktriangledown$}
				& $\tens{M}{7} = \te{a}_3\tens{P}{6}_6 + \overset{7}{\otimes}\ve{a}_3$
				& $\tens{D}{6}_{6h} = \frac{3}{32}\left(I_{ijklmn} -P_{ijklmn}\right)  \ve{e}_i\ve{e}_j\ve{e}_k\ve{e}_l\ve{e}_m\ve{e}_n$ \newline 
				with \quad $I_{ijklmn} = \frac{2}{3}\left(I_{ij}I_{klmn} + I_{ik}I_{jlmn} + I_{il}I_{kjmn} + I_{im}I_{kljn} + I_{in}I_{klmj}\right)$, \newline
				$I_{klmn} = I_{kl}I_{mn} + I_{km}I_{ln} + I_{kn}I_{ml}$, \quad
				$I_{ij} = \frac{1}{16}\left(M_{piklmno}M_{pjklmno} - A_iA_j\right)$,\newline
				$P_{ijklmn} = A_rM_{rijklmn} - A_iA_jA_kA_lA_mA_n$ \quad and \quad $A_i = M_{jjkklli}$ \newline
				$\ve{a}_3 = M_{jjkklli}\ve{e}_i$ \\
				$\Group{D}_{3h}$ / $\Group{G}_{24}$ / $\aniso{9}$ \textcolor{blue}{$\blacktriangledown$}
				& $\tens{M}{3} = \tens{P}{3}_3^{'}$ 
				& $\tens{D}{3}_{3h} = -\frac{3}{4}M_{ijk}\ve{e}_i\ve{e}_j\ve{e}_k$ \quad
				$\ve{a}_3\ve{a}_3 = \frac{1}{4}E_{ikl}E_{jnm}M_{kmo}M_{oln} \ve{e}_i\ve{e}_j$ \\
				$\Group{D}_{6h}$ / $\Group{G}_{27}$ / $\aniso{9}$ \textcolor{blue}{$\blacktriangledown$}
				& $\tens{M}{6} = \tens{P}{6}_6$
				& $\tens{D}{6}_{6h} = \frac{3}{32}\left(I_{ijklmn} - M_{ijklmn}\right) \ve{e}_i\ve{e}_j\ve{e}_k\ve{e}_l\ve{e}_m\ve{e}_n$ \newline 
				with \quad $I_{ijklmn} = \frac{2}{3}\left(I_{ij}I_{klmn} + I_{ik}I_{jlmn} + I_{il}I_{kjmn} + I_{im}I_{kljn} + I_{in}I_{klmj}\right)$, \newline
				$I_{klmn} = I_{kl}I_{mn} + I_{km}I_{ln} + I_{kn}I_{ml}$ \quad and \quad $I_{ij} = \frac{1}{16}M_{iklmno}M_{jklmno}$\newline
				$\ve{a}_3\ve{a}_3 = \left(\delta_{ij} - \frac{1}{16}M_{iklmno}M_{jklmno} \right) \ve{e}_i\ve{e}_j$ \\
				\bottomrule
			\end{tabularx}
		\end{small}
	\end{table}
	\begin{table}[ht]
		\centering
		\caption{Calculation of multiple structural tensors from single structural tensors for the groups $\Group{C}_\infty$ and $\Group{I}_h$ considered in Tab.~\ref{tab:isotropic_extension}. The definitions of all structural tensors are listed in Tab.~\ref{tab:structrual_tensors}.}
		\label{tab:single_to_multiple_icosahedral}
		\renewcommand{\arraystretch}{1.}
		\begin{small}
			\begin{tabularx}{0.88\textwidth}{p{2.4cm}p{3.4cm}p{10cm}}
				\toprule
				Group/Anisotropy & Single structural tensor $\tens{M}{n}$ & Multiple structural tensors \\
				\midrule
				$\Group{C}_\infty$ / $\Group{T}_{1}$ / $\aniso{12}$ 
				& $\tens{M}{3} = \overset{3}{\otimes}\ve{a}_3+\tens{E}{3}\cdot\ve{a}_3\ve{a}_3$ 
				& $\ve{a}_3 = M_{jji}\ve{e}_i$ \quad
				$\tens{E}{3}\cdot\ve{a}_3 = \frac{1}{2}M_{llk}\left( M_{ijk} - M_{jik} \right)\ve{e}_i\ve{e}_j$ \\
				\midrule
				$\Group{I}_h$ / $\aniso{15}$ & $\tens{M}{6} = \tens{I}{6}_h$ 
				& $\tens{I}{8}_h = \frac{1}{4}\left[ 5M_{abcdef}M_{ghijef} - M_{abcdee}M_{ghijff} \right] \ve{e}_a\ve{e}_b\ve{e}_c\ve{e}_d \ve{e}_g\ve{e}_h\ve{e}_i\ve{e}_j$ \newline
				$\tens{I}{10}_h = \frac{1}{16}\left[ 25M_{abcdef}M_{efghij}M_{ijklmn} - 4M_{abcdee}M_{ghklmn} \right.$ \newline
				${}\hspace{1.22cm} \left.- 5M_{abcdef}M_{ghiief}M_{jjklmn} \right] \ve{e}_a\ve{e}_b\ve{e}_c\ve{e}_d \ve{e}_g\ve{e}_h\ve{e}_k\ve{e}_l \ve{e}_m\ve{e}_n$\\
				\bottomrule
			\end{tabularx}
		\end{small}
	\end{table}
	
	It should be noted that, in contrast to \citet{Xiao1996}, certain modifications were made to the selected structural tensors and structural functions for groups $\Group{T}_h$, $\Group{T}_d$ and $\Group{C}_{3i}$ to obtain an integrity basis using the table of \citet{Boehler1977}, see also Remark~\ref{rem:no_complete_set_found}. Regarding the icosahedral group $\Group{I}_h$, in contrast to \citet{xiao_unified_1997, Xiao_icosahedral}, the isotropic extension functions are formulated solely based on the 6th order structural tensor. 
	\subsection{Minimal integrity bases}
	\label{sec:irreducible_integrity bases}
	With the procedure described in Sect.~\ref{sec:elimination}, all minimal integrity bases in Tabs.~\ref{tab:integrtiy_bases_1}~and~\ref{tab:integrtiy_bases_2} are computed using the structural functions listed in Tab.~\ref{tab:isotropic_extension}. The minimal integrity basis for the isotropic case or anisotropy $\Group{A}_{14}$ is well-known in the literature and does not require a further proof. To demonstrate that the results of Tabs.~\ref{tab:integrtiy_bases_1}~and~\ref{tab:integrtiy_bases_2} for the anisotropies $\Group{A}_1$--$\Group{A}_{11}$ are minimal integrity bases, we determine the polynomial relationships between the proposed invariants and the invariants by \citet{smith_further_1962}. The determined relations are provided in the supplementary material, along with a comparison of the number of invariants at each polynomial degree. According to Proposition~\ref{prop:minimal_integrity_basis}, this proves that the proposed invariants for the anisotropies $\Group{A}_1$--$\Group{A}_{11}$ form minimal integrity bases.
	\begin{table}[ht]
		\centering
		\caption{Minimal integrity bases. For anisotropies $\Group{A}_2$, $\Group{A}_4$, and $\Group{A}_5$, two equivalent minimal integrity bases are specified using different structural tensors, respectively. $\te{G}_1 = \tens{E}{3}\cdot\ve{a}_3$.}
		\label{tab:integrtiy_bases_1}
		\renewcommand{\arraystretch}{1.}
		\begin{small}
			\begin{tabularx}{1.0\textwidth}{p{2.2cm}p{13.8cm}}
				\toprule
				Group/Anisotropy & Minimal integrity basis \\
				\midrule
				$\Group{C}_i$ / $\Group{A}_1$ &                   
				$I_1 = \tr[\te{C}] \hspace{1.5em}
				I_2 = \tr[\te{C}\cdot(\tens{E}{3}\cdot\ve{a}_1)^2] \hspace{1.5em}
				I_3 = \tr[\te{C}\cdot(\tens{E}{3}\cdot\ve{a}_2)^2] \hspace{1.5em}
				I_4 = \tr[\te{C}\cdot\tens{E}{3}\cdot\ve{a}_1\cdot\tens{E}{3}\cdot\ve{a}_2] \vspace{3pt}\newline
				I_5 = \tr[\te{C}\cdot(\tens{E}{3}\cdot\ve{a}_1)^2\cdot\tens{E}{3}\cdot\ve{a}_2] \hspace{1.5em}
				I_6 = \tr[\te{C}\cdot\tens{E}{3}\cdot\ve{a}_1\cdot(\tens{E}{3}\cdot\ve{a}_2)^2]$ \\
				\midrule
				$\Group{C}_{1h}$ / $\Group{A}_2$ &                      
				$I_1 = \tr[\te{C}] \hspace{1.5em}
				I_2 = \ve{a}_1\cdot\te{C}\cdot\ve{a}_1 \hspace{1.5em}
				I_3 = \ve{a}_2\cdot\te{C}\cdot\ve{a}_2 \hspace{1.5em}
				I_4 = \ve{a}_1\cdot\te{C}\cdot\ve{a}_2 \hspace{1.5em}
				I_5 = \tr[\te{C}^2] \hspace{1.5em}
				I_6 = \ve{a}_1\cdot\te{C}^2\cdot\ve{a}_2 \vspace{3pt}\newline
				I_7 = \ve{a}_2\cdot\te{C}^2\cdot\ve{a}_2$ \\[22pt]
				$\Group{C}_{2h}$ / $\Group{A}_2$ &   $I_1 = \tr[\te{C}] \hspace{1.5em}
				I_2 = \tr[\te{C}\cdot\te{G}_1^2] \hspace{1.5em}
				I_3 = \tr[\te{C}\cdot\te{P}_2] \hspace{1.5em}
				I_4 = \tr[\te{C}\cdot\te{P}_2\cdot\te{G}_1] \hspace{1.5em}
				I_5 = \tr[\te{C}^2] \hspace{1.5em}
				I_6 = \tr[\te{C}^2\cdot\te{P}_2] \vspace{3pt}\newline
				I_7 = \tr[\te{C}^2\cdot\te{P}_2\cdot\te{G}_1]$ \\
				\midrule
				$\Group{C}_{2v}$, $\Group{D}_{2h}$ / $\Group{A}_3$ &
				$I_1 = \tr[\te{C}] \hspace{1.5em}
				I_2 = \tr[\te{C}\cdot\te{P}_2] \hspace{1.5em}
				I_3 = \tr[\te{C}\cdot\te{P}_2^2] \hspace{1.5em}
				I_4 = \tr[\te{C}^2] \hspace{1.5em}
				I_5 = \tr[\te{C}^2\cdot\te{P}_2] \hspace{1.5em}
				I_6 = \tr[\te{C}^2\cdot\te{P}_2^2] \vspace{3pt}\newline
				I_7 = \tr[\te{C}^3]$ \\
				\midrule
				$\Group{C}_{2i}$ / $\Group{A}_4$ &  
				$I_1 = \tr[\te{C}] \hspace{1.5em}
				I_2 = \tr[\tens{D}{4}_{4h}:\te{C}] \hspace{1.5em}
				I_3 = \tr[\te{C}^2] \hspace{1.5em}
				I_4 = (\tens{\Sigma}{3}:\te{C})\cdot(\tens{\Sigma}{3}:\te{C}) \hspace{1.5em}
				I_5 = \tr[\te{C}^2\cdot\te{G}_1^2] \vspace{3pt}\newline
				I_6 = \tr[\te{C}\cdot(\tens{D}{4}_{4h}:\te{C})\cdot\te{G}_1] \hspace{1.5em}
				I_7 = \tr[\te{C}^3] \hspace{1.5em}
				I_8 = \tr[\te{C}^2\cdot\te{G}_1^2\cdot\te{C}\cdot\te{G}_1] \hspace{1.5em}
				I_9 = (\tens{\Sigma}{3}:\te{C})\cdot\te{C}\cdot(\tens{\Sigma}{3}:\te{C}) \vspace{3pt}\newline
				I_{10} = (\te{C}\cdot(\tens{\Sigma}{3}:\te{C}))\cdot\te{G}_1\cdot(\tens{\Sigma}{3}:\te{C}) \hspace{1.5em}
				I_{11} = (\tens{\Sigma}{3}:\te{C})\cdot\te{C}^2\cdot(\tens{\Sigma}{3}:\te{C}) \hspace{1.5em}
				I_{12} = (\te{C}^2\cdot(\tens{\Sigma}{3}:\te{C}))\cdot\te{G}_1\cdot(\tens{\Sigma}{3}:\te{C})$ \\[36pt]
				$\Group{C}_{4h}$ / $\Group{A}_4$ & 
				$I_1 = \tr[\te{C}] \hspace{1.5em}
				I_2 = \tr[\tens{D}{4}_{4h}:\te{C}] \hspace{1.5em}
				I_3 = \tr[\te{C}^2] \hspace{1.5em}
				I_4 = \tr[(\tens{D}{4}_{4h}:\te{C})^2] \hspace{1.5em}
				I_5 = \tr[\tens{D}{4}_{4h}:\te{C}^2] \vspace{3pt}\newline
				I_6 = \tr[\te{C}\cdot(\tens{D}{4}_{4h}:\te{C})\cdot\te{G}_1] \hspace{1.5em}
				I_7 = \tr[\te{C}^3] \hspace{1.5em} \hspace{1.5em}
				I_8 = \tr[\te{C}^2\cdot\te{G}_1^2\cdot\te{C}\cdot\te{G}_1] \hspace{1.5em}
				I_9 = \tr[\te{C}^2\cdot(\tens{D}{4}_{4h}:\te{C})] \vspace{3pt}\newline
				I_{10} = \tr[\te{C}^2\cdot(\tens{D}{4}_{4h}:\te{C})\cdot\te{G}_1] \hspace{1.5em}
				I_{11} = \tr[(\tens{D}{4}_{4h}:\te{C}^2)^2] \hspace{1.5em}
				I_{12} = \tr[\te{C}^2\cdot(\tens{D}{4}_{4h}:\te{C}^2)\cdot\te{G}_1]$ \\
				\midrule
				$\Group{C}_{4v}$, $\Group{D}_{4h}$ / $\Group{A}_5$ &  
				$I_1 = \tr[\te{C}] \hspace{1.5em}
				I_2 = \tr[\tens{D}{4}_{4h}:\te{C}] \hspace{1.5em}
				I_3 = \tr[\te{C}^2] \hspace{1.5em}
				I_4 = \tr[(\tens{D}{4}_{4h}:\te{C})^2] \hspace{1.5em}
				I_5 = \tr[\tens{D}{4}_{4h}:\te{C}^2] \hspace{1.5em}
				I_6 = \tr[\te{C}^3] \vspace{3pt}\newline
				I_7 = \tr[\te{C}^2\cdot(\tens{D}{4}_{4h}:\te{C})] \hspace{1.5em}
				I_8 = \tr[(\tens{D}{4}_{4h}:\te{C}^2)^2]$ \\[22pt]
				$\Group{D}_{2d}$ / $\Group{A}_5$ & 
				$I_1 = \tr[\te{C}] \hspace{1.5em}
				I_2 = \tr[\tens{D}{4}_{4h}:\te{C}] \hspace{1.5em}
				I_3 = \tr[\te{C}^2] \hspace{1.5em}
				I_4 = (\tens{\Sigma}{3}:\te{C})\cdot(\tens{\Sigma}{3}:\te{C}) \hspace{1.5em}
				I_5 = \tr[\te{C}^2\cdot\ve{a}_3\ve{a}_3] \vspace{3pt}\newline
				I_6 = \tr[\te{C}^3] \hspace{1.5em}
				I_7 = (\tens{\Sigma}{3}:\te{C})\cdot\te{C}\cdot(\tens{\Sigma}{3}:\te{C}) \hspace{1.5em}
				I_8 = (\tens{\Sigma}{3}:\te{C})\cdot\te{C}^2\cdot(\tens{\Sigma}{3}:\te{C})$ \\
				\bottomrule
			\end{tabularx}
		\end{small}
	\end{table}
	\begin{table}[p]
		\centering
		\caption{Minimal integrity bases. For anisotropies $\Group{A}_7$, $\Group{A}_8$, $\Group{A}_9$ and $\Group{A}_{11}$, two equivalent minimal integrity bases are specified using different structural tensors, respectively. $\te{G}_1 = \tens{E}{3}\cdot\ve{a}_3$.}
		\label{tab:integrtiy_bases_2}
		\renewcommand{\arraystretch}{1.}
		\begin{small}
			\begin{tabularx}{\textwidth}{p{2.2cm}p{13.8cm}}
				\toprule
				Group/Anisotropy & Minimal integrity basis \\
				\midrule
				$\Group{C}_{3i}$ / $\Group{A}_6$ &
				$I_1 = \tr[\te{C}] \hspace{1.5em}
				I_2 = \tr[\te{C}\cdot\te{G}_1^2] \hspace{1.5em}
				I_3 = \tr[\te{C}^2] \hspace{1.5em}
				I_4 = \tr[(\tens{C}{4}_{3i}^{\textrm{s}}:\te{C})^2] \hspace{1.5em}
				I_5 = \tr[\te{C}\cdot(\tens{C}{4}_{3i}^{\textrm{s}}:\te{C})] \vspace{3pt}\newline
				I_6 = \tr[\te{C}\cdot(\tens{C}{4}_{3i}^{\textrm{s}}:\te{C})\cdot\te{G}_1] \hspace{1.5em}
				I_7 = \tr[\te{C}^3] \hspace{1.5em}
				I_8 = \tr[\te{C}^2\cdot\te{G}_1^2\cdot\te{C}\cdot\te{G}_1] \hspace{1.5em}
				I_9 = \tr[\te{C}^2\cdot(\tens{C}{4}_{3i}^{\textrm{s}}:\te{C})] \vspace{3pt}\newline
				I_{10} = \tr[\te{C}\cdot(\tens{C}{4}_{3i}^{\textrm{s}}:\te{C})^2] \hspace{1.5em}
				I_{11} = \tr[\te{C}\cdot(\tens{C}{4}_{3i}^{\textrm{s}}:\te{C}^2)] \hspace{1.5em}
				I_{12} = \tr[\te{C}^2\cdot(\tens{C}{4}_{3i}^{\textrm{s}}:\te{C})\cdot\te{G}_1] \vspace{3pt}\newline
				I_{13} = \tr[\te{C}\cdot(\tens{C}{4}_{3i}^{\textrm{s}}:\te{C})^2\cdot\te{G}_1] \hspace{1.5em}
				I_{14} = \tr[\te{C}\cdot(\tens{C}{4}_{3i}^{\textrm{s}}:\te{C}^2)\cdot\te{G}_1]$ \\
				\midrule
				$\Group{C}_{3v}$ / $\Group{A}_7$ &  
				$I_1 = \tr[\te{C}] \hspace{1.5em}
				I_2 = \ve{a}_3\cdot\te{C}\cdot\ve{a}_3 \hspace{1.5em}
				I_3 = \tr[\te{C}^2] \hspace{1.5em}
				I_4 = \tr[(\tens{D}{4}_{3d}:\te{C})^2] \hspace{1.5em}
				I_5 = (\te{C}\cdot\ve{a}_3)\cdot(\tens{D}{4}_{3d}:\te{C})\cdot\ve{a}_3 \vspace{3pt}\newline
				I_6 = \tr[\te{C}^3] \hspace{1.5em}
				I_7 = \tr[\te{C}\cdot(\tens{D}{4}_{3d}:\te{C})^2] \hspace{1.5em}
				I_8 = (\te{C}^2\cdot\ve{a}_3)\cdot(\tens{D}{4}_{3d}:\te{C})\cdot\ve{a}_3 \hspace{1.5em}
				I_9 = (\te{C}\cdot\ve{a}_3)\cdot(\tens{D}{4}_{3d}:\te{C}^2)\cdot\ve{a}_3$ \\[22pt]
				$\Group{D}_{3d}$ / $\Group{A}_7$ & 
				$I_1 = \tr[\te{C}] \hspace{1.5em}
				I_2 = \tr[\te{C}\cdot\ve{a}_3\ve{a}_3] \hspace{1.5em}
				I_3 = \tr[\te{C}^2] \hspace{1.5em}
				I_4 = \tr[(\tens{D}{4}_{3d}:\te{C})^2] \hspace{1.5em}
				I_5 = \tr[\te{C}\cdot\ve{a}_3\ve{a}_3\cdot(\tens{D}{4}_{3d}:\te{C})] \vspace{3pt}\newline
				I_6 = \tr[\te{C}^3] \hspace{1.5em}
				I_7 = \tr[\te{C}\cdot(\tens{D}{4}_{3d}:\te{C})^2] \hspace{1.5em}
				I_8 = \tr[\te{C}^2\cdot\ve{a}_3\ve{a}_3\cdot(\tens{D}{4}_{3d}:\te{C})] \hspace{1.5em}
				I_9 = \tr[\te{C}\cdot\ve{a}_3\ve{a}_3\cdot(\tens{D}{4}_{3d}:\te{C}^2)]$ \\
				\midrule
				$\Group{C}_{3h}$ / $\Group{A}_8$ & 
				$I_1 = \tr[\te{C}] \hspace{1.5em}
				I_2 = \tr[\te{C}\cdot\te{G}_1^2] \hspace{1.5em}
				I_3 = \tr[\te{C}^2] \hspace{1.5em}
				I_4 = (\tens{D}{3}_{3h}:\te{C})\cdot(\tens{D}{3}_{3h}:\te{C}) \hspace{1.5em}
				I_5 = \tr[\te{C}^3] \vspace{3pt}\newline
				I_6 = \tr[\te{C}^2\cdot\te{G}_1^2\cdot\te{C}\cdot\te{G}_1] \hspace{1.5em}
				I_7 = (\tens{D}{3}_{3h}:\te{C})\cdot\te{C}\cdot(\tens{D}{3}_{3h}:\te{C}) \hspace{1.5em}
				I_8 = (\te{C}\cdot(\tens{D}{3}_{3h}:\te{C}))\cdot\te{G}_1\cdot(\tens{D}{3}_{3h}:\te{C}) \vspace{3pt}\newline
				I_9 = (\tens{D}{3}_{3h}:\te{C})\cdot\te{C}^2\cdot(\tens{D}{3}_{3h}:\te{C}) \hspace{1.5em}
				I_{10} = (\te{C}^2\cdot(\tens{D}{3}_{3h}:\te{C}))\cdot\te{G}_1\cdot(\tens{D}{3}_{3h}:\te{C}) \vspace{3pt}\newline
				I_{11} = (\tens{D}{3}_{3h}:\te{C}^2)\cdot\te{C}\cdot(\tens{D}{3}_{3h}:\te{C}^2) \hspace{1.5em}
				I_{12} = (\te{C}\cdot(\tens{D}{3}_{3h}:\te{C}^2))\cdot\te{G}_1\cdot(\tens{D}{3}_{3h}:\te{C}^2) \vspace{3pt}\newline
				I_{13} = (\tens{D}{3}_{3h}:\te{C}^2)\cdot\te{C}^2\cdot(\tens{D}{3}_{3h}:\te{C}^2) \hspace{1.5em}
				I_{14} = (\te{C}^2\cdot(\tens{D}{3}_{3h}:\te{C}^2))\cdot\te{G}_1\cdot(\tens{D}{3}_{3h}:\te{C}^2)$ \\[50pt]
				$\Group{C}_{6h}$ / $\Group{A}_8$ & 
				$I_1 = \tr[\te{C}] \hspace{1.5em}
				I_2 = \tr[\te{C}\cdot\te{G}_1^2] \hspace{1.5em}
				I_3 = \tr[\te{C}^2] \hspace{1.5em}
				I_4 = \tr[\tens{D}{6}_{6h}::(\te{C}\te{C})] \hspace{1.5em}
				I_5 = \tr[\te{C}^3] \vspace{3pt}\newline
				I_6 = \tr[\te{C}^2\cdot\te{G}_1^2\cdot\te{C}\cdot\te{G}_1] \hspace{1.5em}
				I_7 = \tr[\te{C}\cdot(\tens{D}{6}_{6h}::(\te{C}\te{C}))] \hspace{1.5em}
				I_8 = \tr[\te{C}\cdot(\tens{D}{6}_{6h}::(\te{C}\te{C}))\cdot\te{G}_1] \vspace{3pt}\newline
				I_9 = \tr[\te{C}^2\cdot(\tens{D}{6}_{6h}::(\te{C}\te{C}))] \hspace{1.5em}
				I_{10} = \tr[\te{C}^2\cdot(\tens{D}{6}_{6h}::(\te{C}\te{C}))\cdot\te{G}_1] \hspace{1.5em}
				I_{11} = \tr[\te{C}\cdot(\tens{D}{6}_{6h}::(\te{C}^2\te{C}^2))] \vspace{3pt}\newline
				I_{12} = \tr[\te{C}\cdot(\tens{D}{6}_{6h}::(\te{C}^2\te{C}^2))\cdot\te{G}_1] \hspace{1.5em}
				I_{13} = \tr[\te{C}^2\cdot(\tens{D}{6}_{6h}::(\te{C}^2\te{C}^2))] \hspace{1.5em}
				I_{14} = \tr[\te{C}^2\cdot(\tens{D}{6}_{6h}::(\te{C}^2\te{C}^2))\cdot\te{G}_1]$ \\
				\midrule
				$\Group{C}_{6v}$,  $\Group{D}_{6h}$ / $\Group{A}_9$ & 
				$I_1 = \tr[\te{C}] \hspace{1.5em}
				I_2 = \tr[\te{C}\cdot\ve{a}_3\ve{a}_3] \hspace{1.5em}
				I_3 = \tr[\te{C}^2] \hspace{1.5em}
				I_4 = \tr[\tens{D}{6}_{6h}::(\te{C}\te{C})] \hspace{1.5em}
				I_5 = \tr[\te{C}^3] \vspace{3pt}\newline
				I_6 = \tr[\te{C}\cdot(\tens{D}{6}_{6h}::(\te{C}\te{C}))] \hspace{1.5em}
				I_7 = \tr[\te{C}^2\cdot(\tens{D}{6}_{6h}::(\te{C}\te{C}))] \hspace{1.5em}
				I_8 = \tr[\te{C}\cdot(\tens{D}{6}_{6h}::(\te{C}^2\te{C}^2))] \vspace{3pt}\newline
				I_9 = \tr[\te{C}^2\cdot(\tens{D}{6}_{6h}::(\te{C}^2\te{C}^2))]$ \\[22pt]
				$\Group{D}_{3h}$ / $\Group{A}_9$ & 
				$I_1 = \tr[\te{C}] \hspace{1.5em}
				I_2 = \tr[\te{C}\cdot\ve{a}_3\ve{a}_3] \hspace{1.5em}
				I_3 = \tr[\te{C}^2] \hspace{1.5em}
				I_4 = (\tens{D}{3}_{3h}:\te{C})\cdot(\tens{D}{3}_{3h}:\te{C}) \hspace{1.5em}
				I_5 = \tr[\te{C}^3] \vspace{3pt}\newline
				I_6 = (\tens{D}{3}_{3h}:\te{C})\cdot\te{C}\cdot(\tens{D}{3}_{3h}:\te{C}) \hspace{1.5em}
				I_7 = (\tens{D}{3}_{3h}:\te{C})\cdot\te{C}^2\cdot(\tens{D}{3}_{3h}:\te{C}) \hspace{1.5em}
				I_8 = (\tens{D}{3}_{3h}:\te{C}^2)\cdot\te{C}\cdot(\tens{D}{3}_{3h}:\te{C}^2) \vspace{3pt}\newline
				I_9 = (\tens{D}{3}_{3h}:\te{C}^2)\cdot\te{C}^2\cdot(\tens{D}{3}_{3h}:\te{C}^2)$ \\
				\midrule
				$\Group{T}_{h}$ / $\Group{A}_{10}$ &  
				$I_1 = \tr[\te{C}] \hspace{1.5em}
				I_2 = \tr[\te{C}^2] \hspace{1.5em}
				I_3 = \tr[(\tens{\Lambda}{4}:\te{C})^2] \hspace{1.5em}
				I_4 = \tr[\te{C}^3] \hspace{1.5em}
				I_5 = \tr[(\tens{\Lambda}{4}:\te{C})^3] \hspace{1.5em}
				I_6 = \tr[\te{C}^2\cdot(\tens{\Lambda}{4}:\te{C})] \vspace{3pt}\newline
				I_7 = \tr[\te{C}\cdot(\tens{\Lambda}{4}:\te{C})^2] \hspace{1.5em}
				I_8 = \tr[(\tens{\Lambda}{4}:\te{C})\cdot(\tens{\Lambda}{4}:\te{C}^2)] \hspace{1.5em}
				I_9 = \tr[(\tens{\Lambda}{4}:\te{C}^2)^2] \hspace{1.5em}
				I_{10} = \tr[\te{C}^2\cdot(\tens{\Lambda}{4}:\te{C})^2] \vspace{3pt}\newline
				I_{11} = \tr[(\tens{\Lambda}{4}:\te{C})^2\cdot(\tens{\Lambda}{4}:\te{C}^2)] \hspace{1.5em}
				I_{12} = \tr[\te{C}\cdot(\tens{\Lambda}{4}:\te{C}^2)^2] \hspace{1.5em}
				I_{13} = \tr[(\tens{\Lambda}{4}:\te{C})\cdot(\tens{\Lambda}{4}:\te{C}^2)^2] \vspace{3pt}\newline
				I_{14} = \tr[(\tens{\Lambda}{4}:\te{C}^2)^3]$ \\
				\midrule
				$\Group{T}_{d}$ / $\Group{A}_{11}$ &
				$I_1 = \tr[\te{C}] \hspace{1.5em}
				I_2 = \tr[\te{C}^2] \hspace{1.5em}
				I_3 = (\tens{\Sigma}{3}:\te{C})\cdot(\tens{\Sigma}{3}:\te{C}) \hspace{1.5em}
				I_4 = \tr[\te{C}^3] \hspace{1.5em}
				I_5 = (\tens{\Sigma}{3}:\te{C})\cdot\te{C}\cdot(\tens{\Sigma}{3}:\te{C}) \vspace{3pt}\newline
				I_6 = (\tens{\Sigma}{3}:\te{C})\cdot(\tens{\Sigma}{3}:\te{C}^2) \hspace{1.5em}
				I_7 = (\tens{\Sigma}{3}:\te{C}^2)\cdot(\tens{\Sigma}{3}:\te{C}^2)  \hspace{1.5em}
				I_8 = (\tens{\Sigma}{3}:\te{C})\cdot\te{C}^2\cdot(\tens{\Sigma}{3}:\te{C}) \vspace{3pt}\newline
				I_9 = (\tens{\Sigma}{3}:\te{C}^2)\cdot\te{C}\cdot(\tens{\Sigma}{3}:\te{C}^2)$ \\[36pt]
				$\Group{O}_{h}$ / $\Group{A}_{11}$ &
				$I_1 = \tr[\te{C}] \hspace{1.5em}
				I_2 = \tr[\te{C}^2] \hspace{1.5em}
				I_3 = \tr[(\tens{\Theta}{4}:\te{C})^2] \hspace{1.5em}
				I_4 = \tr[\te{C}^3] \hspace{1.5em}
				I_5 = \tr[(\tens{\Theta}{4}:\te{C})^3] \hspace{1.5em}
				I_6 = \tr[\te{C}^2\cdot(\tens{\Theta}{4}:\te{C})] \vspace{3pt}\newline
				I_7 = \tr[(\tens{\Theta}{4}:\te{C}^2)^2] \hspace{1.5em}
				I_8 = \tr[\te{C}^2\cdot(\tens{\Theta}{4}:\te{C})^2] \hspace{1.5em}
				I_9 = \tr[\te{C}\cdot(\tens{\Theta}{4}:\te{C}^2)^2]$ \\
				\midrule
				$\Group{C}_\infty$, $\Group{C}_{\infty h}$ / $\Group{A}_{12}$ &
				$I_1 = \tr[\te{C}] \hspace{1.5em}
				I_2 = \tr[\te{C}\cdot\te{G}_1^2] \hspace{1.5em}
				I_3 = \tr[\te{C}^2] \hspace{1.5em}
				I_4 = \tr[\te{C}^2\cdot\te{G}_1^2] \hspace{1.5em}
				I_5 = \tr[\te{C}^3] \hspace{1.5em}
				I_6 = \tr[\te{C}^2\cdot\te{G}_1^2\cdot\te{C}\cdot\te{G}_1]$ \\
				\midrule
				$\Group{C}_{\infty v}$, $\Group{D}_{\infty h}$ / $\Group{A}_{13}$ &
				$I_1 = \tr[\te{C}] \hspace{1.5em}
				I_2 = \tr[\te{C}\cdot\ve{a}_3\ve{a}_3] \hspace{1.5em}
				I_3 = \tr[\te{C}^2] \hspace{1.5em}
				I_4 = \tr[\te{C}^2\cdot\ve{a}_3\ve{a}_3] \hspace{1.5em}
				I_5 = \tr[\te{C}^3]$ \\
				\midrule
				$\Group{K}_h$ / $\Group{A}_{14}$ &
				$I_1 = \tr[\te{C}] \hspace{1.5em}
				I_2 = \tr[\te{C}^2] \hspace{1.5em}
				I_3 = \tr[\te{C}^3]$ \\
				\midrule
				$\Group{I}_h$ / $\Group{A}_{15}$ &
				$I_1 = \tr[\te{C}] \hspace{1.5em}
				I_2 = \tr[\te{C}^2] \hspace{1.5em}
				I_3 = \tr[\te{C}^3] \hspace{1.5em}
				I_4 = \tr[\te{C}\cdot(\tens{I}{6}_{h}::\te{C}\te{C})] \hspace{1.5em}
				I_5 = \tr[(\tens{I}{6}_{h}::\te{C}\te{C})^2] \vspace{3pt}\newline
				I_6 = \tr[\te{C}\cdot(\tens{I}{6}_{h}::\te{C}\te{C})^2] \hspace{1.5em}
				I_7 = \tr[\te{C}^2\cdot(\tens{I}{6}_{h}::\te{C}\te{C}^2)] \hspace{1.5em}
				I_8 = \tr[(\tens{I}{6}_{h}::\te{C}\te{C})^3] \hspace{1.5em}
				I_9 = \tr[(\tens{I}{6}_{h}::\te{C}\te{C}^2)^2] \vspace{3pt}\newline
				I_{10} = \tr[\te{C}\cdot(\tens{I}{6}_{h}::\te{C}\te{C}^2)^2]$ \\
				\bottomrule
			\end{tabularx}
		\end{small}
	\end{table}
	\paragraph{Cylindrical anisotropies}
	\begin{proposition}
		\label{prop:cylindrical_A13_minimal_integrity_basis}
		The invariants $I_1,\dots,I_5$ in Tab.~\ref{tab:integrtiy_bases_2} form a minimal integrity basis for the anisotropy $\Group{A}_{13}$, thereby $\PolRingE{R}{C_{11},C_{22},C_{33},C_{23},C_{13},C_{12}}{\Group{D}_{\infty h}} = \PolRing{R}{I_1,\dots,I_5}{}$.
	\end{proposition}
	\begin{proof}
		\label{proof:cylindrical_A13_minimal_integrity_basis}
		By selecting $[\ve{a}_3] = (0,0,1)^\top$ to construct the corresponding structural tensor $\ve{a}_3\ve{a}_3$ with respect to the cartesian basis $\{ \ve{e}_1, \ve{e}_2, \ve{e}_3\}$, we determine the polynomial relations $I_\alpha=I_\alpha(J_1,\dots,J_5)$ and $J_\beta=J_\beta(I_1,\dots,I_5)$, $\alpha,\beta\in\left\{1,2,3,4,5\right\}$ to the coordinate-dependent invariants $J_\beta$ of \citet{smith_transversely_1982}. These relations can be found in Sect.~2 of the supplementary material. Since $I_\alpha=I_\alpha(J_1,\dots,J_5)$, it is evident that all $I_\alpha$ are truly invariants. On the other hand, since the invariants $J_\beta$ of \citet{smith_transversely_1982} constitute an integrity basis, and $J_\beta=J_\beta(I_1,\dots,I_5)$, every invariant polynomial $f\in\PolRingE{R}{C_{11},C_{22},C_{33},C_{23},C_{13},C_{12}}{\Group{D}_{\infty h}}$ is expressible as a polynomial in $I_1,\dots,I_5$. Hence, $I_1,\dots,I_5$ form an integrity basis and $\PolRingE{R}{C_{11},C_{22},C_{33},C_{23},C_{13},C_{12}}{\Group{D}_{\infty h}} = \PolRing{R}{I_1,\dots,I_5}{}$. According to the rational form of the HS in Tab.~\ref{tab:molien_series} for $\Group{A}_{13}$ this is a Hironaka decomposition and $I_1,\dots,I_5$ are primary invariants. Thus, $I_1,\dots,I_5$ in Tab.~\ref{tab:integrtiy_bases_2} form a minimal integrity basis for the anisotropy $\Group{A}_{13}$.
	\end{proof}
	\begin{proposition}
		\label{prop:cylindrical_A12_minimal_integrity_basis}
		The invariants $I_1,\dots,I_6$ in Tab.~\ref{tab:integrtiy_bases_2} form a minimal integrity basis for the anisotropy $\Group{A}_{12}$, thereby $\PolRingE{R}{C_{11},C_{22},C_{33},C_{23},C_{13},C_{12}}{\Group{C}_{\infty h}} = \PolRing{R}{I_1,\dots,I_6}{}$.
	\end{proposition}
	\begin{proof}
		\label{proof:cylindrical_A12_minimal_integrity_basis}
		Analogously to the Proof of Proposition~\ref{prop:cylindrical_A13_minimal_integrity_basis} we first determine the polynomial relations between $I_1,\dots,I_6$ and the coordinate-dependent invariants of \citet{smith_transversely_1982} by setting $[\ve{a}_3] = (0,0,1)^\top$ with respect to the cartesian basis $\{ \ve{e}_1, \ve{e}_2, \ve{e}_3\}$ to construct the corresponding structural tensor $\tens{E}{3}\cdot\ve{a}_3$. These relations can be found in Sect.~2 of the supplementary material. Hence, $I_1,\dots,I_5$ form an integrity basis and $\PolRingE{R}{C_{11},C_{22},C_{33},C_{23},C_{13},C_{12}}{\Group{C}_{\infty h}} = \PolRing{R}{I_1,\dots,I_6}{}$. According to the rational form of the HS in Tab.~\ref{tab:molien_series} for $\Group{A}_{13}$ we next determine a corresponding Hironaka decomposition. By using the syzygy 
		\begin{equation}
			\begin{aligned}
				0 = &36I_6^2 + 4I_5^2 + 36I_4^3 + 90I_3I_4^2 + 72I_3^2I_4 + 18I_3^3 - 36I_2I_4I_5 - 24I_2I_3I_5 - 18I_2^2I_3I_4 - 9I_2^2I_3^2 + 12I_2^3I_5 \\
				&- 24I_1I_4I_5 - 24I_1I_3I_5 - 72I_1I_2I_4^2 - 108I_1I_2I_3I_4 - 36I_1I_2I_3^2 + 48I_1I_2^2I_5 - 54I_1^2I_4^2 - 72I_1^2I_3I_4 - 18I_1^2I_3^2 \\
				&+ 48I_1^2I_2I_5 + 54I_1^2I_2^2I_4 + 18I_1^2I_2^2I_3 + 16I_1^3I_5 + 72I_1^3I_2I_4 + 24I_1^3I_2I_3 - 12I_1^3I_2^3 + 24I_1^4I_4 + 6I_1^4I_3 - 21I_1^4I_2^2 \\
				&- 12I_1^5I_2 - 2I_1^6
			\end{aligned}
			\label{eq:syzygy_cylyndrical}
		\end{equation}
		and Lemma~\ref{lemma:example_monoclinic_hironaka_lemma} it directly follows, analogous to the Proof of Proposition~\ref{prop:example_monoclinic_hironaka}, that
		\begin{equation}
			\PolRingE{R}{C_{11},C_{22},C_{33},C_{23},C_{13},C_{12},}{\Group{G}} = \PolRingE{R}{I_1,I_2,I_3,I_4,I_5}{} \oplus I_6\PolRingE{R}{I_1,I_2,I_3,I_4,I_5} \point
		\end{equation}
		is a Hironaka decomposition. Thus $I_1,\dots,I_5$ are primary invariants and $I_6$ is an irreducible secondary invariant. With that $I_1,\dots,I_6$ constitute a minimal integrity basis.
	\end{proof}
	
	\paragraph{Icosahedral anisotropy}
	To demonstrate that the invariants for anisotropy $\Group{A}_{15}$ in Tab.~\ref{tab:integrtiy_bases_2} constitute a minimal integrity basis, we rely on a minimal integrity basis determined by the mathematics software Singular~\citep{singular}. The invariants of Singular are of coordinate-dependent form. Details are given in the Supplementary material, Sect.~3. Using these results, we formulate polynomial relationships between the invariants listed in Tab.~\ref{tab:integrtiy_bases_2} for anisotropy $\Group{A}_{15}$, and the invariants of Singular, thereby enabling the application of Proposition~\ref{prop:minimal_integrity_basis}. It is important to emphasize that this proof is computer-aided. A concise overview of the Singular calculation is provided in the supplementary material.
	The minimal integrity basis determined with Singular as well as the polynomial relations to the invariants in Tab.~\ref{tab:integrtiy_bases_2} for anisotropy $\Group{A}_{15}$
	are available at \url{https://github.com/NEFM-TUDresden/anisotropic_hyperelasticity_integrity_bases}.
	In summary, since all requirements of Proposition~\ref{prop:minimal_integrity_basis} are met, the proposed invariants for anisotropy $\Group{A}_{15}$ indeed form a minimal integrity basis.
	\section{Conclusions}
	\label{sec:conclusions}
	In this paper, we provide minimal integrity bases for all common anisotropies in hyperelasticity using the concept of structural tensors. Our work covers results for the 11 types of anisotropy that arise from the classical crystal systems, as well as findings for 4 additional non-crystal anisotropies derived from the cylindrical, spherical, and icosahedral symmetry systems.
	
	The article begins by summarizing selected fundamentals of continuum mechanics, such as various measures of deformation, strain, and stress, as well as basic concepts of constitutive modeling in hyperelasticity. Based on the principle of material symmetry, the point groups of crystal systems and the most important non-crystal systems relevant to material modeling are listed. Following this, two important concepts for modeling a scalar elastic potential using invariants are introduced: a coordinate-based description and a coordinate-free one using structural tensors. Subsequently, it is discussed how functional bases can be determined from the structural tensors using well-known information about isotropic tensor functions. 
	With the presented analytical-numerical approach, polynomial relations between invariants are identified, thereby reducing the number of invariants in the functional bases. To further show that the obtained functional bases are also minimal integrity bases, two approaches are presented: The first takes advantage of the fact that, for coordinate-based invariants, minimal integrity bases are already known for many relevant cases. The second uses the information provided by the associated HS regarding the number of invariants and the syzygies between them at each polynomial degree. Next, the proposed approach is carried out in detail exemplarily for the cases of monoclinic and cubic anisotropy. Finally, minimal integrity bases formulated with structural tensors are given for all common anisotropies and the icosahedral anisotropy in hyperelasticity.
	
	The provided invariant sets are of great importance for modeling anisotropic materials via the structural tensor concept using both classical approaches as well as modern techniques based on machine learning. 
	Thereby, this work is particularly relevant because it provides minimal integrity bases using the structural tensor concept for all common anisotropies in hyperelasticity.
	For further studies, the computed minimal integrity bases can be used for the formulation of classical material models as well as neural network-based models~\citep{linka_constitutive_2021, klein_polyconvex_2022, fuhg_learning_2022, tac_data-driven_2022, Linden2023,kalina_neural_2024}, and analyzed with regard to properties such as polyconvexity~\citep{ball_convexity_1976, Ebbing2010}. Furthermore, the provided invariants could be compared with those expressed in coordinate-based form~\citep{smith_1958,smith_further_1962} in the context of applications. A combination with algorithms for the automatic identification of structural tensors~\citep{kalina_neural_2024, patel_general_2025} would also be conceivable. Beyond the purely mechanical case, minimal integrity bases could also be computed for coupled field problems, such as magneto-mechanics. Lastly, based on the number of primary invariants indicated by the HS, one may further hypothesize that, for groups of the cylindrical non-crystal system, the number of primary invariants is one element fewer than the number of independent tensor coordinates of all constitutive variables.
	
	\section*{Acknowledgment}
	The authors thank the German Research Foundation (DFG) for the support within the Research Training Group GRK 2868 D${}^3$--Project Number 493401063. The authors also would like to thank Moritz~Flaschel and Dominik~K.~Klein for thoroughly reading initial drafts of this article and providing many very helpful comments. Finally, we would like to thank both reviewers for their extremely valuable comments, which ultimately led to the final version of the manuscript.
	
	\section*{Data availability}
	
	A Python code for determining group elements from generators and polynomial relations for the icosahedral symmetry group are available at \url{https://github.com/NEFM-TUDresden/anisotropic_hyperelasticity_integrity_bases}.
	
	\section*{Supplementary material}
	Supplementary material associated with this article can be found in the published version.
	
	\section*{CRediT authorship contribution statement}
	\textbf{Brain M. Riemer:} Conceptualization, Formal analysis, Investigation, Methodology, Software, Validation, Visualization, Writing -- original draft, Writing -- review \& editing.
	\textbf{J\"{o}rg Brummund:} Conceptualization, Formal analysis, Investigation, Methodology, Visualization, Supervision, Writing – review \& editing.
	\textbf{Karl A. Kalina:}  Conceptualization, Formal analysis, Methodology, Investigation, Visualization, Supervision, Writing -- original draft, Writing -- review \& editing, Funding acquisition.
	\textbf{Abel H. G. Milor:} Formal analysis, Writing -- original draft, Writing -- review \& editing.
	\textbf{Franz Dammaß:} Formal analysis, Methodology, Writing -- original draft, Writing -- review \& editing.
	\textbf{Markus Kästner:} Resources, Writing -- review \& editing, Funding acquisition.
	
	\section*{Usage of AI tools}
	In preparing this work, the authors partially used ChatGPT, a generative AI tool, to improve the readability and language of the manuscript. After using this tool, the authors reviewed and revised the content as necessary and take full responsibility for the content of the published article.
	
	\appendix
	\section{Glossary}
	\label{sec:glossar}
	
	Throughout this paper, we use several concepts from algebra. To ease the understanding for readers with an engineering background, we briefly recapitulate the most important definitions in the following.
	For further reading, we refer to textbooks on algebra, e.g., \citet{lang2005, gallian_contemporary_2017, derksen_computational_2002}.
	
	\subsection{Groups and generators}
	\begin{definition}[Group]
		\label{def:group}
		Let $\Group{S}$ a set and $\ast : \Group{S}\times \Group{S} \rightarrow \Group{S}$ a binary operation.
		The pair $(\Group{S}, \ast)$ is called a group if the following axioms are met:
		\item (G1)\quad $\left(a \ast b\right)\ast c = a\ast\left(b\ast c\right) \, \forall \, a, b, c \in \Group{S}$
		\item (G2)\quad $\exists\: u \in \Group{S}: \,  u\ast a = a \ast u = a \, \forall\, a \in \Group{S}$
		\item (G3)\quad $\forall a \in \Group{S} \; \exists \; a^{-1} \in \Group{S}: \,  a^{-1} \ast a = a\ast a^{-1} = u$
	\end{definition}
	
	\begin{remark}
		\label{rem:group_notation}
		To strengthen the connection to the Schoenflies notation, calligraphic letters, such as $\Group{G}$, are used for all groups relevant to this work.
		Moreover, we generally do not explicitly distinguish between algebraic structures such as groups, rings and vector spaces and their underlying sets, as this convention is also common in the literature.
	\end{remark}

	\begin{definition}[Generators of a group]
		\label{def:generators}
		Let $(\Group{G}, \ast)$ a group.
		A set $\Set{S}_{\textrm{gen}}\subseteq\Group{G}$ is called a generating set of $\Group{G}$, if, 
		for every $a \in \Group{G}$, for some $n \in \zset_{\geq 0}$, there exist $s_1, \dots, s_n$ such that for all $i \in \{1, \dots, n\}$, $s_i \in \Set{S}_{\textrm{gen}}$ or $(s_i)^{-1} \in \Set{S}_{\textrm{gen}}$ and $a = s_1 \ast \cdots \ast s_n$.
		The elements of a generating set $\Set{G}_{\textrm{gen}}$ are called group generators.   
		If $\Group{S}_{\textrm{gen}}$ is finite, $\Group{G}$ is called finitely generated.
	\end{definition}
	\begin{definition}[Minimal generating set of a group, cf. \citet{harper_maximal_2023}]
		\label{def:minimal_generating_set_group}
		Let $\Group{G}$ be a finite group and $\Set{S}\ttu{gen}$ one possible set of generators, then $\Set{S}\ttu{gen}$ is said to be a minimal generating set, if no generating set exists with fewer elements than~$\Set{S}\ttu{gen}$.
	\end{definition}

	\subsection{$\mathcal G$-invariant polynomials and integrity bases}
	
	\begin{definition}[Ring]
		\label{def:ring}
		Let $\Ring{R}$ be a set and let $+: \Ring{R} \times \Ring{R} \to \Ring{R}$ and $*: \Ring{R} \times \Ring{R} \to \Ring{R}$ binary operations. The triple $(\Ring{R},+,\ast)$ is called a ring if the following axioms are true:
		\item (R1)\quad $(\Ring{R}, +)$ is a group.
		\item (R2) \quad $a + b = b + a \, \forall \, a,b \in \Ring{R}$  
		\item (R3) \quad $a * (b * c) = (a * b ) *c \, \forall \, a,b,c \in \Ring{R}$
		\item (R4) \quad $a * (b+c) = a * b + a*c \quad \text{and} \quad  (a+b)*c = a * c + b*c   \, \forall \, a,b,c \in \Ring{R}$
	\end{definition}
	
	\begin{definition}[Polynomial ring]
		\label{def:polynomial_ring}
		Let $\Setnum{K}$ be a field, $n \in \zset_{>0}$ as well as $x_1, \dots x_n$ (formal) variables. We define: 
		\begin{itemize}
			\item a monomial $M$ in $x_1$, \dots, $x_n$ as a formal multiplication of the form:
			$$x_1^{\epsilon_1}\cdots x_n^{\epsilon_n}
			\qquad \text{with} \quad \epsilon_1, \dots, \epsilon_n \in \zset_{\geq0}$$
			The degree of the monomial $M$ is defined as $\text{deg}(M):= \epsilon_1+\dots+\epsilon_n$.
			\item a polynomial $P$ in $x_1$, ..., $x_n$ over $\Setnum{K}$ as a linear combination of $k \in \zset_{\geq 0}$ monomials $M_1, \dots, M_k$ in $x_1$, \dots,  $x_n$, of the form: 
			$$P= \lambda_1M_1+\dots+\lambda_kM_k
			\qquad \text{with} \quad\lambda_1, \dots, \lambda_k \in \Setnum{K}\setminus\{0\}$$
			The degree of a polynomial $P\neq0$ is defined as $\text{deg}(P):=\max(\{\text{deg}(M_1),  \dots, \text{deg}(M_k)  \})$. 
			
			\item a homogeneous polynomial of degree $q \in \zset_{\geq 0}$ as a polynomial whose monomials all have the degree $q$.
			\item the polynomial ring in $x_1, \dots, x_n$ over $\Setnum{K}$ as the ring defined by the set of all polynomials $P$ in $x_1, \dots, x_n$ over $\Setnum{K}$ together with the standard addition and multiplication. This ring is denoted by $\PolRing{\Setnum{K}}{x_1, \dots, x_{n}}{}$. 
		\end{itemize}
	\end{definition}
	Using the associative and distributive laws, it can be verified that $(\PolRing{\Setnum{K}}{x_1, \dots, x_{n}}{},+,\cdot)$ is indeed a ring fulfilling the axioms of Def.~\ref{def:ring}.
	\begin{definition}[Invariant polynomials, cf.~\citep{sturmfels_algorithms_2008}(p. 14)]
		\label{def:invariant_polynomial}
		Let $\M{Q}\in\Group{G} \subseteq \Group{GL}_n(\Setnum{K})$ with $n\in\Setnum{Z}_{>0}$ be an $n\times n$-matrix and $\V{X}\in\Setnum{R}^n$ a vector of formal variables $x_1,\dots,x_n$, then the ring of $\Group{G}$-invariant polynomials is defined as 
		\[\PolRing{\Setnum{K}}{x_1, \dots, x_{n}}{G} = \left\{ I \in \Setnum{K} \left[x_1, \dots, x_{n}\right] 
		\big|\: I\left( (\M{Q}\cdot\V{X})_1,\dots,(\M{Q}\cdot\V{X})_n \right) = I(x_1,\dots,x_n) \:\forall\: \M{Q} 
		\in\Group{G} \right\}  \,,\]
	\end{definition}
	We refer to the elements of $\PolRing{\Setnum{K}}{x_1, \dots, x_{n}}{G}$ as $\Set{G}$-invariant polynomials, $\Set{G}$-invariants or simply invariants.
	Again, using the associative and distributive laws, it can be verified that $\PolRing{\Setnum{K}}{x_1, \dots, x_{n}}{G} \subseteq \PolRing{\Setnum{K}}{x_1, \dots, x_{n}}{}$ is a subring, since it is closed under addition and multiplication and the zero polynomial is an element of $\PolRing{\Setnum{K}}{x_1, \dots, x_{n}}{G}$, cf.~\citet[Sec.~1.3]{sturmfels_algorithms_2008}.
	\begin{definition}[Fundamental invariant, cf.~\citep{sturmfels_algorithms_2008,Ebbing2010}]
		\label{def:fundamental_invariant}
		Let $I_1,\dots,I_m\subseteq \PolRing{\Setnum{K}}{x_1, \dots, x_{n}}{G}, \:n,m\in\Setnum{Z}_{\geq0},$ be homogeneous $\Group{G}$-invariant polynomials $I_\lambda$. 
		Furthermore, for all $k\in\{1,\dots,m\}$, let there exist no polynomial relation for $I_k$ such that $I_k = f_k(I_1,\dots,I_{k-1},I_{k+1},\dots,I_m)$.
		Then, the elements $I_\lambda$ are called fundamental or basic invariants of $\PolRing{\Setnum{K}}{x_1, \dots, x_{n}}{G}$ if 
		\begin{equation*}
			\PolRing{\Setnum{K}}{x_1, \dots, x_{n}}{G} = \PolRing{\Setnum{K}}{I_1,\dots,I_m}{} \commar
			\label{eq:ring_generation_condition}
		\end{equation*}
		i.e., $I_1,\dots,I_m$ generate the ring of $\Group{G}$-invariant polynomials.
	\end{definition}
	The important property that a set $\left\{I_1,\dots,I_m\right\}$ of fundamental invariants $I_\lambda$ is sufficient to generate the associated ring is also often denoted as polynomial completeness in engineering mechanics literature, cf. \citet{Apel2004, Ebbing2010}. Similarly, the property that no $I_\lambda\in\left\{I_1,\dots,I_m\right\}$ can be expressed as a polynomial in the other fundamental invariants is called irreducibility, cf. \citet{Apel2004, Ebbing2010}. 
	However, in mathematics, the concept of irreducibility is only well-defined for polynomials.
	
	\begin{definition}[Irreducible invariant, cf. \citet{froberg_introduction_1998}(p. 14)]
		\label{def:irreducibility}
		Let $I, J, K \in \PolRing{\Setnum{K}}{x_1, \dots, x_{n}}{G}$, where $\Setnum{K}$ is a field, $x_1, \dots, x_{n}$ are formal variables and $\Group{G}$ is a group.
		Furthermore, let $I$  homogeneous and its polynomial degree $\textrm{deg}(I)=i$.
		If there exist homogeneous $J, K \in \PolRing{\Setnum{K}}{x_1, \dots, x_{n}}{G}$ with degrees $\textrm{deg}(J)=j,\: \textrm{deg}(K)=k$ and $i>j\geq k \geq 1$ such that $I=J \cdot K$, $I$ is called reducible. Otherwise, $I$ is called irreducible.
	\end{definition}
	\begin{definition}[Homogeneous system of parameters, cf. \citet{derksen_computational_2002}(Def. 2.4.6)]
		\label{def:hsop}
		Let $n,p\in\Setnum{Z}_{>0}$, $p\leq n$, $\PolRing{\Setnum{K}}{x_1,\dots,x_n}{G}$ be a ring of $\Set{G}$-invariant polynomials and $\{P_1,\dots,P_p\}$ a finite set of homogeneous elements $P_\lambda\in\PolRing{\Setnum{K}}{x_1,\dots,x_n}{G}$. Then the set $\{P_1,\dots,P_p\}$ is called a homogeneous system of parameters (h.s.o.p.) if
		\begin{enumerate}
			\item $P_1,\dots,P_{p}$ are algebraically independent and
			\item $\PolRing{K}{x_1,\dots,x_n}{G}$ is a finitely generated module over $\PolRing{K}{P_1,\dots,P_{p}}{}$.
		\end{enumerate}
	\end{definition}
	The elements of a h.s.o.p. are called primary invariants. From this, it follows that every invariant $I_\lambda\in\PolRing{K}{x_1,\dots,x_n}{G}$ is the solution of a syzygy $\sum_{r=0}^sf_rI_\lambda^r=0, \:s\in\Setnum{Z}_{\geq0}$, where $f_r$ is a polynomial in the primary invariants. In general, primary invariants are not sufficient to generate the polynomial ring $\PolRing{K}{x_1,\dots,x_n}{G}$.
	\begin{definition}[Hironaka decomposition, cf. \citet{derksen_computational_2002}(Sect. 2.5)]
		\label{def:hironaka_decomposition}
		Let $\PolRing{\Setnum{K}}{x_1,\dots,x_n}{G}, \:n\in\Setnum{Z}_{\geq0},$ be a ring of $\Set{G}$-invariant polynomials over a field $\Setnum{K}$ and also Cohen-Macaulay, then
		\begin{equation*}
			\PolRing{\Setnum{K}}{x_1,\dots,x_n}{G} = \bigoplus\limits_{i=1}^q S_i \, \PolRing{K}{P_1,\dots,P_{p}}{}_i,\quad p,q\in\Setnum{Z}_{\geq0}
			\label{eq:hironaka_decomposition_def}
		\end{equation*}
		holds, where $P_1,\dots,P_{p}$ are primary invariants and $S_i$ so-called secondary invariants. The above representation of $\PolRing{\Setnum{K}}{x_1,\dots,x_n}{G}$ is refereed to as a Hironaka decomposition.
	\end{definition}
	Note that secondary invariants $S_i$ are also solutions of syzygies $\sum_{r=0}^sf_rS_i^r=0, \:s\in\Setnum{Z}_{\geq0}$, where $f_i$ is a polynomial in the primary invariants. A Hironaka decomposition is generally not unique, but can be very useful to show that a determined finite set of invariants is generating the polynomial ring $\PolRing{K}{x_1,\dots,x_n}{G}$.
	\begin{definition}[Integrity basis~\citep{smith_1963, smith_isotropic_1965, Ebbing2010}]
		\label{def:integrity_basis}
		Let $I_1,\dots,I_m\subseteq \PolRing{\Setnum{K}}{x_1,\dots,x_n}{G}, \:m,n\in\Setnum{Z}_{>0},$ be homogeneous $\Group{G}$-invariant polynomials that generate the polynomial ring $\PolRing{\Setnum{K}}{x_1,\dots,x_n}{G}$. Then this set is called an integrity basis.
		It may be possible to express some elements of $\left\{ I_1,\dots,I_m \right\}$ as a polynomial in   other elements of the set. 
		If this is not the case, i.e., if all the invariants of an integrity basis are polynomially independent, the set contains only the fundamental invariants and is referred to as a minimal integrity basis.
	\end{definition}
	
	\subsection{Separating invariants and functional basis}
	\begin{definition}[Separating invariants, cf. \citet{xiao_anisotropic_1997}]
		\label{def:separating_set}
		Let $\Set{N}$ be a set, $\Group{G}$ a group, $z\in\Set{N}$, $r\in\Setnum{Z}_{>0}$, $\alpha\in\left\{1,\dots,r\right\}$ and $I_\alpha : \Group{\Set{N}\to\Setnum{R}}$. The elements of the set $\left\{I_1,\dots,I_r\right\}$ are called separating invariants that separate the group orbits $\Set{M}:=\left\{ g*z\in\Set{N}\:|\:g\in\Group{G}  \right\}$ if, for all $x,y\in\Set{N}$ and all $\alpha\in\left\{1,\dots,r\right\}$, it holds
		\begin{equation*}
			I_\alpha(x) = I_\alpha(y) \quad\Leftrightarrow\quad
			\exists\:g\in\Group{G}: y = g * x \point
			\label{eq:functional_complete}
		\end{equation*}
	\end{definition}
	In summary, for a given $x\in\Set{N}$ and its corresponding group orbit $\Set{M}:=\left\{ g*x\in\Set{N}\:|\:g\in\Group{G}  \right\}$, all invariants are constant for every element in this orbit. This is because invariants, by definition, are invariant against group transformations. For an element $y\notin\Set{M}$, however, at least one of all the invariants must change its value with respect to $x$.
	Separable invariants capture this precisely: they take identical values for $x$ and $y$ if and only if both elements $x$ and $y$ lie in the same group orbit. Consequently, the group orbit can be uniquely determined using these separable invariants. Moreover, a specific element $z\in\Set{M}$, which is not necessarily the original $x$ and $y$, can be clearly identified. Once $z$ is known, the values of all possible invariants are known. Thus, the value of all possible invariants can be determined from the separable invariants. Hence, every invariant function can be expressed as a function, not necessarily polynomial~\citep{Boehler1977} and not even continuous~\citep[Sect. 5]{smith_isotropic_1971}, of the separable invariants.
	
	Analogous to minimal integrity bases, it seems reasonable to specify a set $\Set{S}$ of separating invariants, which contains as few elements as possible. To show that, for a given set $\Set{S}$, no proper subset $\Set{S}_{\textrm{sub}}\subsetneq\Set{S}$ still separates the group orbits, one can find $x\in\Set{N}$ that does not belong to the same group orbit, such that all separable invariants except one have the same value for $x$ and $y$. The invariant that has a different values is therefore necessary to distinguish the group orbits. If this is shown for all separable invariants, then the set $\Set{S}$ is minimal with respect to these invariants. Finally, please note that other authors, such as \citet{boehler_polynomial_1994, desmorat_generic_2019}, also introduce the concept of weak separating sets.
	\begin{definition}[Functional basis, cf.~\citep{xiao_anisotropic_1997}]
		\label{def:function_basis}
		A finite set of $\Group{G}$-invariants that separates the group orbits is termed a functional basis. Furthermore, every $\Group{G}$-invariant is a function of the separable invariants~\citep{wineman_material_1964}. If no proper subset of the functional basis is also a functional basis, then the functional basis is minimal.
	\end{definition}

	\section{Functional basis of Boehler}
	\label{app:boehler}
	The following Tab.~\ref{tab:boehler_table} is taken from \citet{Boehler1977} and allows constructing a functional basis for an arbitrary number of real-valued 1st order as well as real-valued symmetric and real-valued antisymmetric 2nd order tensors. Although originally developed for the isotropic case, this result can also be applied to various anisotropies through the structural tensor concept, as discussed in Sects.~\ref{sec:structural_tensor_concept}~and~\ref{sec:constructing_xiao}.
	\begin{table}[ht]
		\centering
		\caption{Isotropic invariants for any tuple $\Set{H} = \left( \ve{v}_1, \dots, \ve{v}_{a}, \te{A}_{1}, \dots, \te{A}_{b}, \te{W}_1, \dots, \te{W}_{c} \right)$ with $a,b,c\in\Setnum{Z}_{\geq0}$, $\ve{v}_{\alpha}\in\ts{1}{}$, $\te{A}_{\beta}\in\tss{2}$ and $\te{W}_{\gamma}\in\tsa{2}$ from the original article by~\cite{Boehler1977}. A functional basis is obtained by constructing all invariants from all unordered combinations of one, two, three, and four variables of $\Set{H}$.}
		\label{tab:boehler_table}
		\renewcommand{\arraystretch}{1.0}
		\begin{small}
			\begin{tabularx}{0.98\textwidth}{ll|llX}
				\toprule
				Variables & Isotropic invariants & Variables & Isotropic invariants\\
				\midrule
				$\te{A}$					& $\tr(\te{A})$,\:\: $\tr(\te{A}^2)$,\:\: $\tr(\te{A}^3)$ &
				$\te{A}$, $\te{W}$, $\ve{v}$		& $\te{A}\cdot\ve{v}\cdot\te{W}\cdot\ve{v}$,\:\: $\te{A}^2\cdot\ve{v}\cdot\te{W}\cdot\ve{v}$,\:\:
				$\te{A}\cdot\te{W}\cdot\ve{v}\cdot\te{W}^2\cdot\ve{v}$ \\
				$\te{W}$					& $\tr(\te{W}^2)$ &
				$\te{W}_1$, $\te{W}_2$, $\ve{v}$	& $\te{W}_1\cdot\ve{v}\cdot\te{W}_2\cdot\ve{v}$,\:\: $\te{W}_1^2\cdot\ve{v}\cdot\te{W}_2\cdot\ve{v}$,\:\:
				$\te{W}_1\cdot\ve{v}\cdot\te{W}_2^2\cdot\ve{v}$ \\
				$\ve{v}$					& $\ve{v}\cdot\ve{v}$ &
				$\te{W}$, $\ve{v}_1$, $\ve{v}_2$	& $\ve{v}_1\cdot\te{W}\cdot\ve{v}_2$,\:\: $\ve{v}_1\cdot\te{W}^2\cdot\ve{v}_2$ \\
				$\te{A}_1$, $\te{A}_2$		& $\tr(\te{A}_1\cdot\te{A}_2)$,\:\: $\tr(\te{A}_1^2\cdot\te{A}_2)$,\:\: $\tr(\te{A}_1\cdot\te{A}_2^2)$, &
				$\te{A}_1$, $\te{A}_2$, $\te{W}$	& $\tr(\te{A}_1\cdot\te{A}_2\cdot\te{W})$,\:\: $\tr(\te{A}_1^2\cdot\te{A}_2\cdot\te{W})$,\:\: 
				$\tr(\te{A}_1\cdot\te{A}_2^2\cdot\te{W})$, \\
				& $\tr(\te{A}_1^2\cdot\te{A}_2^2)$ &
				&  $\tr(\te{A}_1\cdot\te{W}^2\cdot\te{A}_2\cdot\te{W})$ \\
				$\te{A}$, $\te{W}$			& $\tr(\te{A}\cdot\te{W}^2)$,\:\: $\tr(\te{A}^2\cdot\te{W}^2)$, &
				$\te{A}$, $\te{W}_1$, $\te{W}_2$	& $\tr(\te{A}\cdot\te{W}_1\cdot\te{W}_2)$,\:\: $\tr(\te{A}\cdot\te{W}_1^2\cdot\te{W}_2)$, \\
				& $\tr(\te{A}^2\cdot\te{W}^2\cdot\te{A}\cdot\te{W})$ &
				& $\tr(\te{A}\cdot\te{W}_1\cdot\te{W}_2^2)$ \\
				$\te{A}$, $\ve{v}$			& $\ve{v}\cdot\te{A}\cdot\ve{v}$,\:\: $\ve{v}\cdot\te{A}^2\cdot\ve{v}$ &
				$\te{A}_1$, $\te{A}_2$, $\ve{v}$	& $\te{A}_1\cdot\ve{v}\cdot\te{A}_2\cdot\ve{v}$ \\
				$\te{W}_1$, $\te{W}_2$		& $\tr(\te{W}_1\cdot\te{W}_2)$ &
				$\te{A}$, $\ve{v}_1$, $\ve{v}_2$	& $\ve{v}_1\cdot\te{A}\cdot\ve{v}_2$,\:\: $\ve{v}_1\cdot\te{A}^2\cdot\ve{v}_2$ \\
				$\te{W}$, $\ve{v}$			& $\ve{v}\cdot\te{W}^2\cdot\ve{v}$ &
				$\te{A}_1$, $\te{A}_2$, $\ve{v}_1$, $\ve{v}_2$	& 
				$\te{A}_1\cdot\ve{v}_1\cdot\te{A}_2\cdot\ve{v}_2 - \te{A}_1\cdot\ve{v}_2\cdot\te{A}_2\cdot\ve{v}_1$ \\
				$\ve{v}_1$, $\ve{v}_2$		& $\ve{v}_1\cdot\ve{v}_2$ &
				$\te{A}$, $\te{W}$, $\ve{v}_1$, $\ve{v}_2$		& 
				$\te{A}\cdot\ve{v}_1\cdot\te{W}\cdot\ve{v}_2 - \te{A}\cdot\ve{v}_2\cdot\te{W}\cdot\ve{v}_1$ \\
				$\te{A}_1$, $\te{A}_2$, $\te{A}_3$	& $\tr(\te{A}_1\cdot\te{A}_2\cdot\te{A}_3)$ &
				$\te{W}_1$, $\te{W}_2$, $\ve{v}_1$, $\ve{v}_2$	& 
				$\te{W}_1\cdot\ve{v}_1\cdot\te{W}_2\cdot\ve{v}_2 - \te{W}_1\cdot\ve{v}_2\cdot\te{W}_2\cdot\ve{v}_1$ \\
				$\te{W}_1$, $\te{W}_2$, $\te{W}_3$	& $\tr(\te{W}_1\cdot\te{W}_2\cdot\te{W}_3)$ & & \\
				\bottomrule
			\end{tabularx}
		\end{small}
	\end{table}
	
	\section{Example: Hironaka decomposition and Hilbert series}
	\label{app:example_hs_derksen}
	To show the application of the Hilbert series (HS) in the context of determining integrity bases, we illustrate a common example, which can also be found in \citet[Example 3.8]{stanley_hilbert_1978} or \citet[Example 3.3.6 (b)]{derksen_computational_2002}. This example shows that the information provided by the rational form of the HS should be handled with care.
	
	Consider the complex numbers $\Setnum{C}$, $\Group{G_\textrm{exp.}}$ the abelian group with $\left|\Group{G}_\textrm{exp.}\right|=8$ generated by the diagonal matrices $\textrm{diag}(-1,-1,1)$ and $\textrm{diag}(1,1,i)$, where $i^2=-1$, and $\Group{G_{\textrm{exp.}}}$ acting on the polynomial ring $\PolRing{C}{x,y,z}{\Group{G}_\textrm{exp.}}$ via standard matrix vector multiplication on the coordinate vector $(x,y,z)^\top$. In this case, the matrix $\M{\varPhi}$ in Eq.~\eqref{eq:molien_series} is directly given by a group element $\M{Q}\in\Group{G}_\textrm{exp.}$. After determining all eight group elements from the two generators, one can calculate the HS via Eq.~\eqref{eq:molien_series} to
	\begin{equation}
		\Hs_{\Group{G}_\textrm{exp.}} = \frac{1}{\left(1-t^2\right)^3}
		\label{eq:hilbert_series_example_rational}
	\end{equation}
	and using Eq.~\eqref{eq:molien_series_taylor} to
	\begin{equation}
		\Hs_{\textrm{Tay},\Group{G}_\textrm{exp.}} = 1 + 3t^2 + 6t^4 + 10t^6 + 15t^8 + \textrm{HOT}
		\label{eq:hilbert_series_example_taylor}
	\end{equation}
	\paragraph{Information of Eq.~\eqref{eq:hilbert_series_example_taylor}}
	Now, let us consider the coefficients $b_\lambda$ in Eq.~\eqref{eq:hilbert_series_example_taylor}. Since $b_1=0$, there is no basis $\Setnum{C}[\Group{V}]_1^{\Group{G}_\textrm{exp.}}$ and therefore no non-constant invariant of polynomial degree one. Next, $b_2=3$, which means that a basis of linearly independent homogeneous invariants in polynomial degree two consists of exactly three elements or invariants. Furthermore, since there are no homogeneous invariants in polynomial degree one that are not the zero polynomial, these three invariants must be fundamental invariants. In fact, there are three fundamental invariants for this example~\citep{stanley_hilbert_1978}:
	\begin{equation}
		I_1 = x^2, \quad I_2=y^2 \quad \textrm{and} \quad I_3=xy \point
		\label{eq:stanley_example_deg1_invars}
	\end{equation}
	The next coefficient $b_3=0$ indicates that there are no non-constant homogeneous invariants in polynomial degree three. With $b_4=6$ we know that six polynomially independent invariants exist in polynomial degree four. A total of six products can be formed from the fundamental invariants~\eqref{eq:stanley_example_deg1_invars} in polynomial degree four:
	\begin{equation}
		I_1^2, \quad I_2^2, \quad I_3^2, \quad I_1 I_2, \quad I_1 I_3, \quad \textrm{and} \quad I_2I_3 \point
		\label{eq:stanley_products_dependent}
	\end{equation}
	But only five products are polynomially independent, because of $I_3^2=I_1I_2$. Please note the polynomial relations between such products are more complex in general. Due to $b_4=6$, we know that another polynomially independent invariant exists in polynomial degree four. Since this cannot be written as a homogeneous polynomial in the invariants $I_1, I_2$ and $I_3$, this missing invariant must also be a fundamental invariant. Indeed, this invariant $I_4=z^4$ does exist~\citep{stanley_hilbert_1978}. Therefore, this counting scheme is very helpful in checking whether the correct number of invariants has been determined up to a certain polynomial degree. For the higher polynomial degrees, no further fundamental invariants exist. Hence, a minimal integrity basis is given by $I_1$, $I_2$, $I_3$ and $I_4$. This counting procedure can be limited by using Noether's degree bound, which states that no polynomial degree of a fundamental invariant exceeds the order of a finite group, in this case eight.
	
	\paragraph{Contradiction of the rational form of the HS~\eqref{eq:hilbert_series_example_rational}}
	The HS in Eq.~\eqref{eq:hilbert_series_example_rational} already has a form that would be permissible according to Eq.~\eqref{eq:molien_series_prim_sek}. Then, the statement would be as follows: there exist exactly three primary invariants in polynomial degree two and one secondary invariant, namely the number $1$, which is always a secondary invariant.
	This immediately presents a contradiction, because invariant $I_4=z^4$ is missing. As elaborated in the last paragraph, this invariant with polynomial degree four can not be expressed in terms of invariants in lower degrees. Therefore, we will first take a closer look at the three invariants in Eq.~\eqref{eq:stanley_example_deg1_invars}. Although the number and polynomial degrees match, these three invariants are not algebraically independent and therefore can not form a homogeneous system of parameters (h.s.o.p.), see Def.~\ref{def:hsop}. At this point, it becomes clear that rational form given in Eq.~\eqref{eq:hilbert_series_example_rational} does not belong to any Hironaka decomposition. Therefore, one can not extract any information of the rational form of the HS~\eqref{eq:hilbert_series_example_rational}.
	
	\paragraph{A meaningful rational form of the HS}
	However, if the HS~\eqref{eq:hilbert_series_example_rational} is multiplied by $(1+t^2)/(1+t^2)$, another permissible form
	\begin{equation}
		\Hs_{\Group{G}_\textrm{exp}} = \frac{1+t^2}{\left(1-t^2\right)^2 \left(1-t^4\right)}
		\label{eq:hilbert_series_example_rational_correct}
	\end{equation}
	is obtained.\footnote{Note that both rational forms~\eqref{eq:hilbert_series_example_rational}~and~\eqref{eq:hilbert_series_example_rational_correct} lead to the same series expansion~\eqref{eq:hilbert_series_example_taylor}.} This provides now the following information: there are two primary invariants of degree two, one primary invariant of degree four, and, in addition to the secondary invariant 1, another secondary invariant of polynomial degree two. The form~\eqref{eq:hilbert_series_example_rational_correct} is already more promising, because now an invariant in polynomial degree four is also required. First, however, it must be checked whether three invariants in the corresponding polynomial degrees two and four actually form a h.s.o.p.
	To do this in the case of a finite matrix group $\Group{G}\subseteq\Group{GL}_n(\Setnum{K})$ for a field $\Setnum{K}$ with zero characteristic we have to check that primary invariants $P_1,\dots,P_{p}$ fulfill the four conditions:
	\begin{enumerate}[label=\Roman*.]
		\item $P_1,\dots,P_p$ are homogeneous polynomials,
		\item the count $p$ equals the number of coordinates, cf.~\citet[Proposition 2.1.1]{sturmfels_algorithms_2008},
		\item the jacobian $\frac{\partial P_\lambda}{\partial x_\mu}$ has full rank $\Leftrightarrow P_1,\dots,P_{p}$ are algebraically independent \citep{pandey_algebraic_2018}, and
		\item $\:P_\lambda(x_1,\dots,x_l)=0\:\forall\:\lambda\in\left\{1,\dots,p\right\} \Leftrightarrow x_1,\dots,x_l = 0$, cf. \citep[Lemma 2.4.5]{derksen_computational_2002}.
	\end{enumerate}
	This means for the current example and the invariants $I_1=x^2, I_2=y^2, I_3=xy$ and $I_4=z^4$ that $I_1, I_2$ and $I_4$ build a h.s.o.p and thus are primary invariants. Another choice of a h.s.o.p or primary invariants is not possible in this case. For instance, although $I_1, I_3$ and $I_4$ are also algebraically independent, they do not vanish only for $x,y,z=0$. In conclusion, since we found a set of primary invariants and secondary invariants, the invariant ring $\PolRing{C}{x,y,z}{\Group{G}_\textrm{exp}}$ is expressible as
	\begin{equation}
		\PolRing{C}{x,y,z}{\Group{G}_\textrm{exp}} = \PolRing{C}{I_1,I_2,I_4}{} \oplus I_3\PolRing{C}{I_1,I_2,I_4}{} \point
		\label{eq:hilbert_series_example_solution}
	\end{equation}

\bibliographystyle{apalike-ejor}
\bibliography{aniso.bib}

\end{document}